\documentclass[12pt]{article}
\usepackage{a4wide}
\usepackage{amssymb}
\usepackage{amsmath}
\usepackage{braket}
\usepackage{graphicx}
\usepackage{graphicx,caption}
\usepackage{soul}
\usepackage{subcaption}
\usepackage{hyperref,url}
\usepackage{cite}

\usepackage{tensor}

\usepackage{comment}

\usepackage[dvipsnames]{xcolor}

\newcommand\SB[1]{\textcolor{red}{\texttt{(SB: #1)}}}

\newcommand\ID[1]{\textcolor{purple}{\texttt{(ID: #1)}}}

\begin{document}
{\renewcommand{\thefootnote}{\fnsymbol{footnote}}
\begin{center}
{\LARGE Black holes in effective loop quantum gravity: \\Hawking radiation}\\
\vspace{1.5em}
Idrus Husin Belfaqih,$^1$\footnote{e-mail address: {\tt i.h.belfaqih@sms.ed.ac.uk}}
Martin Bojowald,$^2$\footnote{e-mail address: {\tt bojowald@psu.edu}}
Suddhasattwa Brahma$^{1,3}$\footnote{e-mail address: {\tt suddhasattwa.brahma@gmail.com}}
and Erick I.\ Duque$^2$\footnote{e-mail address: {\tt eqd5272@psu.edu}}
\\
\vspace{0.5em}
$^1$Higgs Centre for Theoretical Physics, School of Physics \& Astronomy,\\
University of Edinburgh, Edinburgh EH9 3FD, Scotland, UK\\
\vspace{0.5em}
$^2$ Institute for Gravitation and the Cosmos,\\
The Pennsylvania State
University,\\
104 Davey Lab, University Park, PA 16802, USA\\
\vspace{0.5em}
$^3$ Physics and Applied Mathematics Unit, Indian Statistical Institute,\\
203 B.T. Road, Kolkata 700108, India
\vspace{1.5em}
\end{center}
}

\setcounter{footnote}{0}

\begin{abstract}
  Emergent modified gravity provides a covariant framework for holonomy
  effects in models of loop quantum gravity with consistent black hole
  solutions coupled to a scalar field. Several independent studies of the
  Hawking thermal distribution are shown here to lead to the same final
  result. This internal consistency is a direct consequence of general
  covariance, which is analogous to the situation in classical general
  relativity but highly nontrivial in the context of modified canonical
  gravity.  Holonomy corrections to the evaporation rate enter through the
  greybody factor, slowing down the evaporation process when the holonomy modification function decreases
  monotonically. Accounting for backreaction, corrected covariant semi-classical stress-energy tensors are computed in various vacuum states. Thanks to these results, the new concept of a net stress-energy tensor makes it possible to compute evaporation rates
  directly from energy conservation laws.
\end{abstract}

\tableofcontents


\section{Introduction}
Black holes are among the most striking predictions of general relativity. As stationary vacuum solutions of Einstein equations, they appear -- within the purely classical theory -- to be eternal. In the 1970s, however, Hawking demonstrated that
quantum effects cause black holes to radiate thermally and gradually lose mass
\cite{Hawking}. This Hawking radiation raises a profound question about the
fate of information that initially fell into the black hole. If the emitted
quanta were to remain in a mixed state throughout the entire evaporation
process, the entropy of the external universe would keep increasing and the
information seemingly absorbed by the black hole would be lost. Page argued,
however, that if black-hole evaporation is unitary, this cannot persist
indefinitely \cite{Page1993Information}. After roughly half of the black hole’s initial Bekenstein–Hawking entropy has been emitted -- at the so-called Page time -- the remaining black hole is too small to purify the radiation on its own, so the newly emitted quanta must be entangled with the earlier radiation rather than with the interior.

In the classical Schwarzschild solution, a vacuum black hole contains a
central curvature singularity. If one were to assume that classical Einstein equations remain valid throughout the Hawking evaporation process,
then the black hole would eventually evaporate completely, leaving behind a
naked singularity \cite{HawkingExplosion}. Because the Hawking temperature
$T_{\rm H}$ scales as $T_{\rm H}\propto 1/M$, the black hole possesses a
negative heat capacity: As it loses energy through Hawking radiation, $M$
decreases and hence the temperature rises, making the evaporation
thermodynamically unstable.  It is commonly expected that
quantum-gravitational effects intervene to resolve the central singularity,
thereby modifying the black hole’s thermodynamic behavior at a characteristic
scale determined by the details of the quantum corrections. In particular, the
semiclassical backreaction of the quantum test field on spacetime should
become relevant before the black hole reaches the Planck regime, whereas
genuinely quantum-gravitational effects are expected to dominate only as it
approaches Planckian size.

A nonperturbative quantum correction to the classical Schwarzschild solution has been proposed recently in \cite{Bardaji2022, Idrus1}. The solution presented in \cite{Idrus1} is more general because it does not fix the dependence of a specific modification function $\lambda(x)$ on the areal radius and hence admits arbitrary profiles.
This effective solution can be used as a model of  holonomy corrections motivated by loop quantum gravity with $\lambda(x)$ playing the role of the holonomy length.
Previous proposals for finding holonomy-corrected Schwarzschild solutions
break covariance due to an incomplete analysis of the consequences of
introducing modifications after partial gauge fixings.  For example,
Refs. \cite{Campiglia,Edward-Ewing} fix the areal gauge for the angular
components of the metric and solve the radial diffeomorphism constraint prior
to introducing the holonomy modifications, leaving the modified Hamiltonian
constraint as the sole gauge generator; this remaining constraint can generate
time translations but no radial diffeomorphisms and hence covariance cannot be
analyzed or guaranteed.

Some modification of the classical space-time structure is expected from
  quantum theories of gravity, and therefore general covariance may not be
  required to hold in its standard form. However, general covariance is
  related not only to space-time structure but also to gauge consistency. In
  particular, breaking manifest covariance of the line element implies that
  the location of the horizon becomes gauge-dependent. Such gauge dependence
  is problematic because the horizon is central to black-hole thermodynamics:
  Quantities such as surface gravity, temperature, and entropy are defined at (or
  in terms of) the horizon and therefore should not depend on the choice of
  gauge or foliation.  As an example, in the tunneling picture
  \cite{Padmannabhan, Parikh-Wilczek}, the black hole's Boltzmann distribution
  $e^{-iS_{\rm O.S.}}=e^{-T^{-1}_{\rm H}\omega}$ is obtained by computing the
  tunneling probability through the WKB approximation, where $S_{\rm O.S.}$ is
  the on-shell Hamilton--Jacobi function. In the analysis of \cite{Padmannabhan} the
  derivation is based on the diagonal-gauge ($N^{x}=0$) while the approach of \cite{Parikh-Wilczek} is based on the Painlev\'e--Gullstrand
  gauge. Both are sensitive to the location of the horizon: The
  evaluation of the Hamilton--Jacobi function $S_{\rm O.S.}$ involves an
  integral with the poles located at the black hole horizon.  In the case of
  GR, the results of these two procedures are consistent.  When the line
  elements are not diffeomorphism invariant, the procedures yield inequivalent
  results due to the different locations of the horizon depending on the way
  the foliation is chosen.  Accordingly, failing to enforce covariance
  prevents a unique definition of black-hole thermal properties.
This universality can be achieved only when the theory underlying the
solutions is coordinate and slicing independent, or diffeomorphism covariant.

Keeping this in mind, the solution of \cite{Idrus1}, based on the recently introduced emergent
modified gravity (EMG) framework \cite{EMG,EMGCov}, solves these
problems. Rather than postulating an \textit{ad hoc} pointwise modification function
$K_{\varphi}\to f(K_{\varphi})$ of the angular extrinsic curvature
$K_{\varphi}$, usually referred to as a ``polymerization'' in this context,
EMG derives a consistent polymerization by requiring (i)
Hamiltonian-diffeomorphism constraint brackets that are off-shell closed
(anomaly free), and (ii) gauge transformations generated by these constraints
that are on-shell equivalent to coordinate transformations of the  physical
fields, such that the spacetime metric reconstructed from the canonical
variables obeys the tensor-transformation law. Consequently, admissible
holonomy modifications are derived rather than postulated. In this way, EMG
bypasses earlier no-go results
\cite{Bojowald-Brahma-Reyes,Bojowald-Brahma,Bojowald1,Bojowald2,Bojowald3}
that arise when a polymerization is imposed by hand without deriving a
compatible metric tensor. A distinctive feature of EMG is the separation
between the fundamental gravitational phase-space variables and the emergent
spacetime geometry: The intrinsic geometry defined by the line element is not
assumed \textit{a priori} but is reconstructed from the canonical data precisely to
satisfy covariance. In this way, classical covariance and its associated
  space-time structure, encoded by a specific dependence of the metric tensor
  on the canonical fields, is modified in a consistent manner that takes into
  account potential quantum space-time effects.

Hawking radiation of the black hole solution of \cite{Idrus1} and its
universal features were studied in \cite{Idrus-short thermal BH} with a
specific holonomy function $\lambda(x)$ associated with the $\bar{\mu}$-scheme
of loop quantum cosmology (LQC).  Here, we extend the results to more general
$\lambda(x)$ functions.  As we will show, the Boltzmann distribution is
insensitive to the function $\lambda(x)$, and hence the Hawking distribution
is exactly the same for any monotonically decreasing function $\lambda(x)$.
When $\lambda(x)$ is constant, we obtain the Hawking distribution up to an
overall constant, in agreement with \cite{Asier-radiation, Asier-gray, Menezes}.

Although holonomy effects do not provide corrections to the black hole thermal distribution to leading order, they leave their imprints through the scalar field potential. The impact of having a quantum-corrected potential results in deviation in the  quasinormal modes and greybody factors, to name a few. The quasinormal modes have been
studied in \cite{EMGscalarQNM}, where one of the main results was that the
imaginary part of some frequencies can change sign if the
scalar field is coupled nonminimally with the gravitational degrees of
freedom, leading to superradiant instability.  In the present paper, we will
study the evaporation aspect of the emergent black hole by including the
greybody factor as the central ingredient affected by holonomy corrections. As
we will see, the holonomy effect has significant impact on the evaporation
process when we choose the holonomy function associated with the
$\bar{\mu}$-scheme. Such a choice provides a way for the black-hole
evaporation process to slow down. Even though there are alternative proposals
for covariant solutions with holonomy modifications, this main result has so
far been achieved only within the EMG framework. On the other hand, the authors
\cite{Yongbin} have shown that the solution in \cite{Zhang1, Zhang2}, which is
also covariant and may be considered a special case of EMG, does not slow down
the evaporation process.

Our analysis is organized as follows.  In Section \ref{Sec: Emergent modified
  gravity in spherical symmetry}, we give a brief overview of emergent
modified gravity, focusing on the coupling of a scalar field to the
gravitational degrees of freedom.  Requiring anomaly freedom together with
the strong covariance conditions leaves some flexibility in how the scalar
field couples to gravity.  We examine both minimally and nonminimally coupled
scalar fields in this context.  In Section \ref{Sec: LQGBH Hawking
  distribution}, we analyze the thermal radiation of the emergent modified
gravity black hole using geometric optic approximation. The formalism
employ the static gauge, focussing on the exterior region of the
spacetime.  We find agreement with the results obtained in the stationary
gauge, where the horizon is crossed via the tunneling formalism discussed in
Section~\ref{Sec: The black hole tunneling rates}.  In that picture, pair
creation appears as quantum-mechanical tunneling: A particle escapes to the
exterior while its antiparticle falls into the black hole, decreasing the
black hole's mass parameter while preserving the ADM mass.  We argue
that the consistency between these different approaches stems from the
underlying covariance requirement of the model.  The holonomy effects enter
through the greybody factors, discussed in Section \ref{Sec: The greybody
  factors}, where we derive analytic expressions in the low-frequency regime.
Finally, we investigate the black-hole evaporation process using the thermal
spectrum corrected by these greybody factors.  When the holonomy function is
constant, the evaporation follows the same profile as in the classical case,
with the black hole evaporating without bound.  By contrast, a monotonically
decreasing holonomy parameter slows the evaporation process.



\section{Emergent modified gravity in spherical symmetry}\label{Sec: Emergent modified gravity in spherical symmetry}

Emergent modified gravity provides a systematic framework for constructing the
Hamiltonian constraint of a consistent gravitational theory, ensuring not only
anomaly freedom but also the emergence of a covariant spacetime structure
\cite{EMGCov}. In this section, we briefly review the coupling of scalar
matter, with a more detailed discussion available in \cite{Bojowald-Duque
  scalar}.

\subsection{Conditions for covariance in canonical gravity}
New classes of canonical gravity models can be formulated by modifying the classical Hamiltonian constraint $H$ of general relativity to a new Hamiltonian constraint $\tilde{H}$, such that the former can be attained in a specific limit of suitable parameters. If one considers a modified Hamiltonian constraint $\tilde{H}$ while maintaining the diffeomorphism constraint without
  modifications, as implied for instance by the regularization procedure of
LQG, then we must first ensure that the constraint brackets remain first class
for the constraints to vanish consistently in all gauges. In a minimal modification, one may assume that the phase space remains unchanged, with configuration variables $q_{ab}$ and momenta $p^{ab}$ for gravity. In the classical limit, the former is equal to the spatial metric while the latter is a combination of the spatial metric and extrinsic curvature components. 

Due to the modifications, in general the Poisson bracket between the constraints has anomalous terms if not chosen properly.
It is therefore necessary to require that the general form of hypersurface-deformation brackets be preserved
\begin{subequations}
\label{eq:Hypersurface deformation algebra - spherical - EMG}
\begin{eqnarray}\label{Eq: Constraints algebra}
    \{ \vec{H}[\vec{N}] , \vec{H}[\vec{M}] \} &=&  \vec{H} [\mathcal{L}_{\vec{N}}\vec{M}]
    \,, \\
    \{ \tilde{H} [N] , \vec{H} [\vec{N}] \} &=& - \tilde{H}[N^{b}\partial_{b}N] 
    \,,
    \label{eq:H,H_x bracket - spherical - EMG}
    \\
    \{ \tilde{H} [N] , \tilde{H}[M] \} &=& - \vec{H} \left[ \tilde{q}^{ab} \left( N\partial_{b}M-M\partial_{b}N \right)\right]
    \label{eq:H,H bracket - spherical - EMG}
    \,,
\end{eqnarray}
\end{subequations}
where the structure function $\tilde{q}^{ab}$ is determined by demanding that no additional terms appear on the right-hand sides. However, the specific structure function $\tilde{q}^{ab}$ need not equal the classical one, given by the inverse of the basic phase-space variable $q_{ab}$ (or an equivalent version in triad variables), on which the modified $\tilde{H}$ depends.
Accordingly, the geometrical structure must be adjusted by defining the emergent line element
\begin{eqnarray}\label{Eq: General expression for emergent line element}
    {\rm d}s^{2}=\tilde{g}_{\mu\nu}{\rm d}x^{\mu}{\rm d}x^{\nu}=-N^{2}{\rm d}t^{2}+\tilde{q}_{ab}\left({\rm d}x^{a}+N^{a}{\rm d}t\right)\left({\rm d}x^{b}+N^{b}{\rm d}t\right)
\end{eqnarray}
in ADM form.
The spatial components of the metric $\tilde{q}_{ab}$ are given by the
inverse of the structure function in regions in which $\tilde{q}^{ab}$ is
invertible. (If $\tilde{q}^{ab}$ is not invertible everywhere in spacetime
for a given solution, the modifications may give rise to signature change.)

The constraints generate the gauge transformation
$\delta_{\epsilon}f(q_{ab},p^{ab})=\{f,\tilde{H}[\epsilon]+\vec{H}[\vec{\epsilon}]\}$ for a phase-space function $f$. The gauge
transformation of the spacetime metric (\ref{Eq: General expression for
  emergent line element}) is required to be related
to the Lie derivative ${\cal L}_{\xi}\tilde{g}_{\mu\nu}$ such that
\begin{eqnarray}
    \delta_{\epsilon}\tilde{g}_{\mu\nu}\big|_{\rm
  O.S.}=\mathcal{L}_{\xi}\tilde{g}_{\mu\nu}\big|_{\rm O.S.}\,. 
\end{eqnarray}
This condition ensures that the gauge transformation of the metric components
is equivalent to an infinitesimal coordinate transformation of the spacetime metric
components, at least on-shell when the constraints and equations of motion are
satisfied (indicated by the subscript O.S.). The spacetime vector $\xi$
is related to the gauge parameters $\left(\epsilon^{0},\epsilon^{a}\right)$ by
\begin{eqnarray}
    \xi^{t}=\frac{\epsilon^{0}}{N}\quad,\quad \xi^{a}=\epsilon^{a}-\frac{\epsilon^{0}}{N}N^{a}
\end{eqnarray}
where the vector $\xi^{\mu}=\xi^{t}t^{\mu}+\xi^{a}s_{a}^{\mu}$ refers to the
gauge vector in spacetime components, while the gauge parameters $\epsilon^0$
and $\epsilon^a$ are components in a normal frame adjusted to the spacelike
hypersurfaces with unit normal $n^{\mu}$.  The lapse function $N$ and shift
vector $N^a$ appear in the relationship $t^{\mu}=N n^{\mu}+N^a s_a^{\mu}$.

It is possible to include matter field degrees of freedom in the modified
theory. For our purposes, it is sufficient to consider the simplest case of a
single-component scalar field $\phi$ with momentum $P_{\phi}$. For a canonical
theory with hypersurface deformation brackets given by (\ref{eq:Hypersurface deformation algebra - spherical - EMG}) for the combined constraints of gravitational and matter
contribution, $\tilde{H}_{\rm grav}[N]+\tilde{H}_{\rm matter}[N]$ and
$\vec{H}_{\rm grav}[\vec{N}]+\vec{H}_{\rm matter}[\vec{N}]$, we say that the
scalar field is covariant if
\begin{eqnarray}
    \delta_{\epsilon}\phi\big|_{\rm O.S.}=\mathcal{L}_{\xi}\phi\big|_{\rm O.S.}\,.
\end{eqnarray}
This condition can be shown to imply that no spatial derivatives of the scalar
momentum $P_{\phi}$ are allowed in the Hamiltonian constraint.


\subsection{Scalar field coupling in spherical symmetry}

The anomaly-freedom and covariance conditions allow for a variety of inequivalent couplings of scalar matter in spherically symmetric systems \cite{Bojowald-Duque scalar}.
Specifically, we consider two cases: The minimal coupling and the simplest nonminimally coupled model, which we define below.
In both cases, the diffeomorphism constraint of the full system takes the classical form
\begin{equation}
    H_{x}=E^{\varphi}K'_{\varphi}-K_{x}\left(E^{x}\right)'+P_{\phi}\phi'
\end{equation}
where $(K_\varphi,E^\varphi)$ and $(K_x,E^a)$ are the canonical pairs of the
gravitational degrees of freedom, while $\left(\phi,P_{\phi}\right)$ is the
canonical pair of the reduced two-dimensional scalar matter. (A factor of $\sqrt{4\pi}$ is absorbed in each matter variable.) The Hamiltonian constraint, on the other hand, takes the form 
\begin{eqnarray}\label{Eq: Modified Hamiltonian}
\tilde{H}\left(E^{x},E^{\varphi},P_{\phi};K_{x},K_\varphi,\phi\right)=\tilde{H}_{\rm grav}\left(E^{x},E^\varphi;K_{x},K_\varphi\right)+\tilde{H}_{\rm scalar}\left(E^{x},E^\varphi,P_{\phi};K_{x},K_\varphi,\phi\right)
\end{eqnarray}
where $\tilde{H}_{\rm grav}$ and $\tilde{H}_{\rm scalar}$ are the vacuum and the scalar matter contributions to the Hamiltonian constraint, respectively.
In the classical theory, the momenta $E^{x}$ and $E^{\varphi}$ are components of the densitized triad, while the configuration variables are related to the extrinsic curvature components $\mathcal{K}_{x}=2K_{x}$ and $\mathcal{K}_{\varphi}=K_{\varphi}$.

The modified Hamiltonian constraint (\ref{Eq: Modified Hamiltonian}) is
required to contain derivatives of the phase space variables of at most
second-order spatial derivatives. The anomaly freedom and covariance
conditions in vacuum then allow the presence of some arbitrary functions which can be
found in \cite{EMGCov}. Here, we will fix most of them to their classical
values and only maintain corrections that can be associated with holonomy
modifications as they commonly appear in models of loop quantum gravity:
\begin{eqnarray}
    \tilde{H}_{\rm grav} &=&
    - \chi \frac{\sqrt{E^x}}{2} \Bigg[
    E^\varphi \bigg( \frac{1}{E^x}  + \frac{1}{E^x} \frac{\sin^2 ( \lambda K_\varphi)}{\lambda^2}
    + 4 \left( K_\varphi \frac{\sin (2 \lambda K_\varphi)}{2 \lambda} - \frac{\sin^2 ( \lambda K_\varphi)}{\lambda^2} \right) \frac{\partial \ln \lambda}{\partial E^x} \bigg)
    \notag\\
    &&
    + 4 K_x \frac{\sin (2 \lambda K_\varphi)}{2 \lambda}
    - \frac{((E^x)')^2}{4 E^\varphi} \left( \frac{1}{E^x} \cos^2 (\lambda K_\varphi)
    - 4 \lambda^2 \left( \frac{K_x}{E^\varphi} + K_\varphi \frac{\partial \ln \lambda}{\partial E^x} \right) \frac{\sin (2 \lambda K_\varphi)}{2 \lambda} \right)
    \notag\\
    &&
    + \cos^2 (\lambda K_\varphi) \left( \frac{(E^x)' (E^\varphi)'}{(E^\varphi)^2} - \frac{(E^x)''}{E^\varphi} \right)
    \Bigg]
    \label{eq:Modified constraint - nonperiodic version - simple}
\end{eqnarray}
with the structure function
\begin{eqnarray}
    \tilde{q}^{x x} &=&
    \left( 1
    + \lambda^2 \left( \frac{(E^x)'}{2 E^\varphi} \right)^2
    \right)
    \cos^2 \left( \lambda K_\varphi \right) \chi^2 \frac{E^x}{(E^\varphi)^2}
    \nonumber\\
    &=&
    \left(1 + \lambda^2 \left(1 - \frac{2 \mathcal{M}}{\sqrt{E^x}} \right) \right) \chi^2 \frac{E^x}{(E^\varphi)^2}
    \,.
    \label{eq:Modified structure function - nonperiodic version - simple}
\end{eqnarray}
The phase-space function $\mathcal{M}$, given by
\begin{eqnarray}
    \mathcal{M} =
    \frac{\sqrt{E^x}}{2}
    \left(
    1 + \frac{\sin^2\left(\lambda K_{\varphi}\right)}{\lambda^2}
    - \cos^2 (\lambda K_\varphi) \left(\frac{(E^x)'}{2 E^\varphi}\right)^2 \right)
    \,,
    \label{eq:Weak observable in simple case - nonperiodic}
\end{eqnarray}
is a local Dirac observable---which plays the role of the mass term in the
vacuum---but not in the presence of matter. The classical expressions are
recovered in the limit $\lambda\to0$, $\chi\to1$.
The structure function defines the emergent line-element for the system,
\begin{eqnarray}\label{Eq: ADM spherical line element}
    {\rm d}s^{2}=-N^{2}{\rm d}t^{2}+\tilde{q}_{xx}\left({\rm d}x+N^{x}{\rm d}t\right)^{2}+q_{\theta\theta}{\rm d}\Omega^{2}
\end{eqnarray}
where $\tilde{q}_{xx}$ is the inverse of the structure function
(\ref{eq:Modified structure function - nonperiodic version - simple}).

The modification function $\lambda(E^x)$, designed to capture holonomy effects
from LQG, and the global factor $\chi(E^x)$ are arbitrary functions of
$E^x$. We impose the condition that the asymptotic value of the modification
function, $\lambda_\infty := \lim_{E^x \to \infty} \lambda(E^x)$, is a
constant. While there is no unique prescription for fixing the global function
$\chi(E^x)$, the requirement that the spacetime geometry be asymptotically
flat restricts the ambiguity in $\chi(E^x)$, leaving two simple choices. The first
option is to set it to a constant value, determined by the asymptotic behavior
of the modification function $\chi_{0}=1/\sqrt{1+\lambda_{\infty}^{2}}$. This
choice yields a nonEuclidean spatial structure when $M\to0$.  A different
choice of the global factor, namely $\chi=1/\sqrt{1+\lambda^{2}(E^x)}$, is
also allowed to recover asymptotic flatness, but it yields a
deviation from the classical Newtonian potential for nonrelativistic objects
\cite{Idrus1}. Given these conditions, the holonomy modification and
parameters such as $\lambda_{\infty}$ enter through the global
factor as well.

As mentioned earlier, coupling a real scalar field can be performed in
different inequivalent ways \cite{Bojowald-Duque scalar}. We consider the
following two simple scenarios: The minimal coupling results from substituting
the modified structure function $\tilde{q}^{xx}$ in place of the classical
$q^{xx}$ in the real, spherically symmetric Klein--Gordon contribution to the
Hamiltonian constraint:
\begin{equation}
    \tilde{H}^{\phi}_{\rm (MC)} = \frac{\sqrt{\tilde{q}^{xx}}}{E^x} \frac{P_\phi{}^2}{2}
    + \frac{E^x}{2} \sqrt{\tilde{q}^{xx}} (\phi')^2
    + \frac{E^x}{\sqrt{\tilde{q}^{xx}}} V
    \,.
    \label{eq:Scalar Hamiltonian constraint - real}
\end{equation}
For the nonminimal coupling, the contribution to the constraint  can be derived
as an alternative solution to the anomaly-freedom and covariance conditions,
starting with an ansatz
that depends on up to second-order derivatives
and to a quadratic dependence on first-order derivatives
\cite{Bojowald-Duque scalar}:
\begin{eqnarray}
    \tilde{H}^\phi_{\rm (NMC)} &=&
    \frac{\chi}{2E^\varphi\sqrt{E^x}} \left(1+\lambda^2 \left(\frac{(E^x)'}{2E^\varphi}\right)^2\right) \cos^2 (\lambda K_\varphi) P_\phi{}^2
    \nonumber\\
    &&
    + \chi \frac{(E^x)^{3/2}}{2E^\varphi} \left(\phi'\right)^2
    + \chi E^\varphi \sqrt{E^x} V
    \,.
    \label{eq:Hamiltonian constraint - scalar - real}
\end{eqnarray}

Generalizing to the complex scalar field is straightforward because the latter can be seen as the coupling of two real scalar fields $\phi_1$ and $\phi_2$ with respective conjugate momenta $P_1$ and $P_2$. In this case, we only have to substitute $\phi^2\to\phi_1^2+\phi_2^2$, $(\phi')^2\to(\phi_1')^2+(\phi_2')^2$, and $P^2\to P_1^2+P_2^2$ in the Hamiltonian constraints above and correspondingly in the diffeomorphism constraint.
The resulting constraints have an SO(2) symmetry in the scalar multiplet $\phi_i$ with $i=1,2$ (and the momenta $P_i$) which can be exchanged for a U(1) symmetry by instead using the complex field $\phi=(\phi_1+i\phi_2)/\sqrt{2}$ with complex momentum $P_\phi=(P_1-iP_2)/\sqrt{2}$.
This procedure results in the minimally coupled constraint, now for a complex field, as
\begin{equation}
    \tilde{H}^{\phi\mathbb{C}}_{\rm (MC)} = \frac{\sqrt{\tilde{q}^{xx}}}{E^x} |P_\phi|^2
    + E^x \sqrt{\tilde{q}^{xx}} (\phi^*)'\phi'
    + \frac{E^x}{\sqrt{\tilde{q}^{xx}}} V (|\phi|^2)\,,
    \label{eq:Scalar Hamiltonian constraint - MC - complex}
\end{equation}
and the nonminimally coupled constraint as
\begin{eqnarray}
    \tilde{H}^{\phi\mathbb{C}}_{\rm (NMC)} &=&
    \frac{\chi}{E^\varphi\sqrt{E^x}} \left(1+\lambda^2 \left(\frac{(E^x)'}{2E^\varphi}\right)^2\right) \cos^2 (\lambda K_\varphi) |P_\phi|^2
    \nonumber\\
    &&
    + \chi \frac{(E^x)^{3/2}}{E^\varphi} (\phi^*)'\phi'
    + \chi E^\varphi \sqrt{E^x} V (|\phi|^2)
    \,.
    \label{eq:Scalar Hamiltonian constraint - NMC - complex}
\end{eqnarray}

The diffeomorphism constraint is correspondingly extended to
\begin{equation}
    H_{x}=E^{\varphi}K'_{\varphi}-K_{x}\left(E^{x}\right)'+P_{\phi}\phi'+P_{\phi}^*(\phi^*)'\,,
\end{equation}
and the symplectic structure of the matter variables to
\begin{equation}\label{eq:Basic brackets - scalar}
    \{\phi(x),P_\phi(y)\}=\{\phi^*(x),P_\phi^*(y)\}=\delta(x-y)\,,
\end{equation}
with the other fundamental brackets vanishing.


\subsection{Dirac observables and the inner product}
Using either of the constraints results in an anomaly-free and covariant system with the same structure function (\ref{eq:Modified structure function - nonperiodic version - simple}).
It is important to note that the two constraints (\ref{eq:Scalar Hamiltonian constraint - MC - complex}) and (\ref{eq:Scalar Hamiltonian constraint - NMC - complex}) have the same symmetry generator
\begin{eqnarray}
    G^\phi [\Theta] = \int {\rm d} x\, \Theta i \left( \phi P_\phi - \phi^* P_\phi^* \right)
    \,,
    \label{eq:Symmetry generator - scalar - complex}
\end{eqnarray}
where $\Theta$ is a real constant.
This functional generates U(1) transformations of the scalar field and its momentum. Since the Hamiltonian and diffeomorphism constraints are invariant under U(1) transformations, (\ref{eq:Symmetry generator - scalar - complex}) commutes with the constraints up to boundary terms and is, therefore, a nonlocal Dirac observable.  Such a boundary term defines a conserved densitized current. The time components of the densitized currents are given by $J^t_{\rm (MC)}=G^\phi$ and $J^t_{\rm (NMC)}=G^\phi$, which are identical kinematically but may differ in dynamical solutions, while the radial components are defined by
\begin{eqnarray}\label{eq:Current - minimal scalar complex}
    \dot{J}^t_{\rm (MC)} &=& \{ G^\phi , \tilde{H}^{\rm (MC)} [N] + H_x^{\rm (MC)} [N^x]\}
    \nonumber\\
    &=& - \left(i N E^x \sqrt{\tilde{q}^{xx}} \left(\phi^* \phi'-\phi (\phi^*)'\right)
    + i N^x \left(P_\phi \phi-P_\phi^* \phi^*\right)\right)'
    \nonumber\\
    &=:& - (J^x_{\rm (MC)})'
\end{eqnarray}
for the minimal coupling and
\begin{eqnarray}\label{eq:Current - nonminimal scalar complex}
    \dot{J}_{\rm (NMC)}^t &=& \{ G^\phi , \tilde{H}^{\rm (NMC)} [N] + H_x^{\rm (NMC)} [N^x]\}
    \nonumber\\
    &=& - \left(i N \chi \frac{(E^x)^{3/2}}{E^\varphi} \left(\phi^* \phi'-\phi (\phi^*)'\right)
    + i N^x \left(P_\phi \phi-P_\phi^* \phi^*\right)\right)'
    \nonumber\\
    &=:& - (J^x_{\rm (NMC)})'
\end{eqnarray}
for the nonminimal coupling.
It follows that $\partial_\mu J^\mu_{\rm (MC)}=0$ and $\partial_\mu J^\mu_{\rm (NMC)}=0$, on shell, by definition. Recall that for a densitized vector field $\partial_\mu J^\mu=\nabla_\mu J^\mu$.

Having a densitized conserved current is essential for studying Hawking evaporation, which requires the quantization of the complex scalar field and hence an inner product. The latter is defined by the symmetry generator (\ref{eq:Symmetry generator - scalar - complex}), which is a constant of the motion.
Using Hamilton's equations of motion for the matter variables,
\begin{eqnarray}
    \dot{\phi} &=& N \frac{\sqrt{\tilde{q}^{xx}}}{E^x} P_\phi^* + N^x \phi'\,,\label{eq:MC dot phi}\\
    \dot{P}_\phi &=& \left(N E^x \sqrt{\tilde{q}^{xx}} (\phi^*)'\right)'
    - N \frac{E^x}{\sqrt{\tilde{q}^{xx}}} \frac{\partial V}{\partial \phi}
    + (N^xP_\phi)'\,,\label{eq:MC dot P phi}
\end{eqnarray}
for the minimal coupling,  and
\begin{eqnarray}
    \dot{\phi} &=&  N \frac{\chi P_\phi^*}{E^\varphi\sqrt{E^x}} \left(1+\lambda^2 \left(\frac{(E^x)'}{2E^\varphi}\right)^2\right) \cos^2 (\lambda K_\varphi)
    + N^x \phi'\,, \label{eq:NMC dot phi}\\
    \dot{P}_\phi &=& \left(N\chi \frac{(E^x)^{3/2}}{E^\varphi}(\phi^*)'\right)'
    - N \chi E^\varphi \sqrt{E^x} \frac{\partial V}{\partial \phi}
    + (N^xP_\phi)'\,,\label{eq:NMC dot P phi}
\end{eqnarray}
for the nonminimal coupling, we can find $P_\phi$ in terms of $\dot{\phi}^*$.
Substituting this into the symmetry generator (\ref{eq:Symmetry generator - scalar - complex}) results in
\begin{eqnarray}\label{Eq: Jt for MC}
    G^\phi_{\rm (MC)}
    &=& i E^x \sqrt{\tilde{q}_{xx}} \left[\phi^* \partial_0\phi-\phi \partial_0\phi^*\right]
    \\
    &=& i \frac{E^\varphi\sqrt{E^x}}{\chi} \left(1+\lambda^2 \left(\frac{(E^x)'}{2E^\varphi}\right)^2\right)^{-1/2} \sec (\lambda K_\varphi) \left(\phi^* \partial_0\phi-\phi \partial_0\phi^*\right)\,,\nonumber
\end{eqnarray}
for the minimal coupling, where $\partial_0=n^\mu \partial_\mu=N^{-1}(\partial_t-N^x\partial_x)$. Similarly, we find
\begin{eqnarray}\label{Eq: Jt for NMC}
    G^\phi_{\rm (NMC)} &=&i \chi \frac{(E^x)^{3/2}}{E^\varphi} \tilde{q}_{xx} \left(\phi^* \partial_0\phi-\phi \partial_0\phi^*\right)
    \nonumber\\
  &=& i \frac{E^\varphi\sqrt{E^x}}{\chi} \left(1+\lambda^2
      \left(\frac{(E^x)'}{2E^\varphi}\right)^2\right)^{-1} \sec^2 (\lambda
      K_\varphi) \left(\phi^* \partial_0\phi-\phi \partial_0\phi^*\right)  \,.
\end{eqnarray}
for the nonminimal coupling. (The overall global factor $\chi$ can be removed by choosing the smearing parameter $\Theta=1/\chi$ if $\chi$ is a constant.)

One can also use the expression for $P_\phi^*$ in terms of $\dot{\phi}$ to obtain the second-order evolution equation for the scalar field.
For the minimal coupling, it results in the complex Klein--Gordon equation,
\begin{equation}\label{Eq: Min coupling KG EMG}
    \frac{\partial_{\mu}\left[\sqrt{-\det\tilde{g}}\;\tilde{g}^{\mu\nu}\partial_{\nu}\phi\right]}{\sqrt{-\det\tilde{g}}} + \frac{\partial V}{\partial \phi^*} = \nabla^\mu \nabla_\mu \phi + \frac{\partial V}{\partial \phi^*} = 0\,,
\end{equation}
where $\nabla_\mu$ is the derivative operator compatible with the emergent metric.
For nonminimal coupling, the equation takes the form
\begin{eqnarray}\label{Eq: KG in NMC}
    \ddot{\phi} &=&  N \frac{E^\varphi}{(E^x)^{3/2}} \tilde{q}^{xx} \Bigg[\chi^{-1}\left(\chi N \frac{(E^x)^{3/2}}{E^\varphi}\phi'\right)'
    - N E^\varphi \sqrt{E^x} \frac{\partial V}{\partial \phi^*}\Bigg]
    \,,
\end{eqnarray}
for a static spacetime background in a gauge where $N^x=0$, which is our case of interest.
Both equations of motion imply that a free scalar field travels at the speed of null rays defined by the emergent line-element,
\begin{equation}
    \frac{{\rm d} x}{{\rm d} t} \bigg|_{\rm null} = \pm N \sqrt{\tilde{q}^{xx}}\,.
\end{equation}


\subsection{Spacetime structure}

Hamilton's equations of motion for the metric, generated by (\ref{eq:Modified constraint - nonperiodic version - simple}) in the vacuum, can be solved exactly in different gauges corresponding to different coordinate charts, all of which have been shown to be equivalent to each other via coordiante transformations. Sewing these solutions in regions of overlap one can discover the global spacetime structure \cite{Idrus1}. The exterior region $x>2M$ in the static Schwarzschild gauge is described by the line element
\begin{eqnarray}\label{Eq: Effective line-element in the full case}
    {\rm d}s^{2}=-\left(1-\frac{2M}{x}\right)\frac{{\rm d}t^2}{\alpha_{0}^{2}\chi^{2}}+\frac{{\rm d}x^{2}}{\chi^{2}\left(1-\frac{2M}{x}\right)\left(1+\lambda^{2}\left(1-\frac{2M}{x}\right)\right)}+x^{2}{\rm d}\Omega^2\,,
\end{eqnarray}
where $\alpha_{0}$ is an arbitrary constant and the global factor $\chi=\chi(E^x)$.

The line element has coordinate singularities at $x=2M$, corresponding to the Killing horizon, and at $x_{\lambda}$, defined as the solutions to the equation:
\begin{eqnarray}\label{Eq: condition for minimum radius}
    1+\lambda^{2}(x_{\lambda})\left(1-\frac{2M}{x_{\lambda}}\right)=0
\end{eqnarray}
associated with the location of a reflection symmetry surface. In what follows, we will regard this vacuum solution as a final product of the collapsing matter. 

The spacetime geometry (\ref{Eq: Effective line-element in the full case})
remains nontrivial in the zero-mass limit \( M \rightarrow 0 \), provided
that \( \lambda(x) \) is not constant. In particular, when \( \lambda(x) \) is
a monotonically decreasing function\footnote{This is the closest to what is
  known as the ``mu-bar'' scheme in homogeneous loop quantum cosmology
  models \cite{APS}. See \cite{Lessons} for a detailed comparison and a
  discussion of consistent holonomy modifications in the light of EMG.} 
the geometry retains its nontrivial structure at short distances. Two
possible choices in the zero mass limit have been proposed in
\cite{Idrus1}. The first one requires the spatial section to be Euclidean
while maintaining asymptotic flatness. This amounts to setting
$\chi(E^{x})=1/\sqrt{1+\lambda^{2}(E^x)}$ and
$\alpha_{0}=1/(1+\lambda^{2}_{\infty})$. More details of the spacetime with
this structure can be found in Appendix~\ref{Apendix}.

Instead of choosing flat Euclidean space in the zero-mass limit, one may  require a constant time component of the spacetime metric. This may be preferred due to its preservation of the classical Newtonian potential in the nonrelativistic limit. With this choice, the global factor is held constant with $\chi=\chi_{0}=1/\sqrt{1+\lambda_{\infty}^{2}}$ and $\alpha_{0}=\chi^{-1}$, which simplifies the line element (\ref{Eq: Effective line-element in the full case}) to
\begin{eqnarray}\label{Eq. Effective line-element}
    {\rm d}s^{2}=-\left(1-\frac{2M}{x}\right){\rm d}t^{2}+\frac{\left(1+\lambda^{2}_{\infty}\right){\rm d}x^{2}}{\left(1-\frac{2M}{x}\right)\left(1+\lambda^{2}(x)\left(1-\frac{2M}{x}\right)\right)}+x^{2}{\rm d}\Omega^{2}\,.
\end{eqnarray}

The four-acceleration required to keep a particle at rest at a constant radial position is given by
\begin{eqnarray}
a^{\mu}\partial_{\mu}=\chi_{0}^{2}\frac{M}{x^2}\left(1+\lambda^2\left(1-\frac{2M}{x}\right)\right)^{1/2}\partial_x\,,
\end{eqnarray}
The magnitude $a$ of the four-acceleration that implies the surface gravity, defined as
\begin{eqnarray}\label{Eq: General Surface Gravity}
    \kappa=\left(\sqrt{-g_{tt}} \ a\right)\big|_{x=2M}\,,
\end{eqnarray}
results in
\begin{eqnarray}\label{Eq: Surface gravity flat time}
    \kappa=\frac{\chi_{0}}{4M}\,.
\end{eqnarray}
It approaches its classical value when $\chi_{0}\to 1$. 

The global structure of the spacetime can then be obtained by performing the Kruskal transformation. Using null coordinates 
\begin{eqnarray}\label{Eq: Null coordinates definition for arbitrary mass}
u=\chi_{0}^{-1}\left(t-x_{*}\right)\quad\text{and}\quad v=\chi_{0}^{-1}\left(t+x_{*}\right)\,,
\end{eqnarray}
where the radial tortoise coordinate $x_*(x)$ may be kept formal without an
explicit evaluation of
\begin{eqnarray}\label{Eq: Radial EMG tortoise coordinates}
    {\rm d}x_{*}=\frac{\sqrt{1+\lambda^{2}_{\infty}}{\rm d}x}{\sqrt{1+\lambda^{2}(x)\left(1-2M/x\right)}\left(1-2M/x\right)}\,.
\end{eqnarray}
The line element in these coordinates is given by
\begin{eqnarray}\label{Eq: Null line-element of constant chi}
    {\rm d}s^{2}=-\chi_{0}^{2}\left(1-\frac{2M}{x}\right){\rm d}u{\rm d}v+x^{2}{\rm d}\Omega^{2}\,.
\end{eqnarray}
For latter purposes, we define the Kruskal null coordinates $\left(U,V\right)$ as
\begin{eqnarray}\label{Eq: Krsukal coordinates for flat space}
    U=-\frac{\chi_{0}}{\kappa} e^{-\kappa u}\quad{\rm and}\quad V=\frac{\chi_{0}}{\kappa}e^{\kappa v}\,.
\end{eqnarray}
In these coordinates, the line element becomes
\begin{eqnarray}\label{Eq: Metric in Kruskal coordinates for flat time}
    {\rm d}s^{2}=-\left(1-\frac{2M}{x}\right){\rm exp}\left(-\frac{2\kappa x_{*}}{\chi_{0}}\right){\rm d}U{\rm d}V+x^{2}{\rm d}\Omega^{2}\,,
\end{eqnarray}
where the conformal factor is similar to the classical Schwarzschild
solution. Here, however, the values of $U$ and $V$ can be extended beyond
  the classical limit imposed by the singularity at $x=0$.  




\subsection{Minimally coupled test scalar field}\label{sub: Minimally coupled test scalar field}
As mentioned earlier, one can minimally or nonminimally couple test scalar fields in the background vacuum spacetime given above \cite{Bojowald-Duque scalar, EMGscalarQNM}.
To maintain generality, we start with a generic spherically symmetric line-element
\begin{eqnarray}\label{Eq: General spherical line-element}
    {\rm d}s^2 &=& -f(x){\rm d}t^2+\frac{{\rm d}x^2}{f(x)h(x)}+x^2{\rm d}\Omega^2
    \nonumber\\
    &=:& f(x)\left(-{\rm d}t^{2}+{\rm d}x_{*}^{2}\right)+x^{2}(x_{*}){\rm d}\Omega^{2}\,.
\end{eqnarray}
For the minimal coupling, the test scalar field follows the classical Klein--Gordon (KG) equation, with the classical metric replaced by the emergent one, given by (\ref{Eq: Min coupling KG EMG}).
Under the decomposition
\begin{eqnarray}\label{Eq: Reduced 2d minimally coupled scalar}
\phi(x^\mu)=\sum_{l,m}\phi_{lm}(t,x)Y_{lm}(\theta,\varphi)\,,
\end{eqnarray}
the KG equation reduces to
\begin{eqnarray}\label{Eq: General KG in radial coordinates}
    -f^{-1}(x)\partial_{t}^{2}\phi_{lm}+\frac{\sqrt{h}}{x^2}\partial_{x}\left[x^{2}f(x)\sqrt{h(x)}\partial_{x}\phi_{lm}\right]-\frac{l(l+1)}{x^{2}}\phi_{lm}=0
\end{eqnarray}
for a free scalar field. Rescaling the field as $\phi_{lm}(t,x)=\Psi_{lm}(t,x)/x$, the equation simplifies in the tortoise radial coordinates as
\begin{eqnarray}\label{Eq: General KG in tortoise coordinates}
    \left[-\partial_{t}^{2}+\partial_{x_{*}}^{2}-\left(\frac{f\sqrt{h}}{x}\partial_{x}\left(f\sqrt{h}\right)+\frac{l(l+1)}{x^{2}}\right)\right]\Psi_{lm}=0\,.
\end{eqnarray}
Using the Fourier transform
\begin{eqnarray}\label{Eq: Stationary Fourier transform}
    \Psi_{lm}(t,x)=\frac{1}{2\pi}\int_{-\infty}^{\infty}{\rm d}\omega\, \psi_{lm}(\omega,x)e^{i\omega t}\,,
\end{eqnarray}
we obtain a time-independent Schrodinger-like equation
\begin{eqnarray}\label{Eq: General Schrodinger in tortoise coordinates}
    \frac{{\rm d^2}\psi_{lm}}{{\rm d}x_{*}^{2}}+\left(\omega^{2}-V_{l}(x)\right)\psi_{lm}=0\,,
\end{eqnarray}
where
\begin{eqnarray}\label{Eq: Generic scalar field potential}
   V_{l}(x)= \frac{f\sqrt{h}}{x}\partial_{x}\left(f\sqrt{h}\right)+\frac{l(l+1)}{x^{2}}
\end{eqnarray}
is the effective potential for the multipole $l$.

For the classical Schwarzschild solution we have $f(x)=1-2M/x$ and $h(x)=1$, in which case Eq.~(\ref{Eq: General KG in radial coordinates}) reduces to
\begin{eqnarray}
   0&=& -\left(1-\frac{2M}{x}\right)\partial_{t}^{2}\phi_{lm}+\frac{1}{x^{2}}\partial_{x}\left[x^{2}\left(1-\frac{2M}{x}\right)\partial_{x}\phi_{lm}\right]-\frac{l(l+1)}{x^{2}}\phi_{lm} \label{Eq: Reduce classical KG-equation in radial coordinates 1} \\
   &=&-\left(1-\frac{2M}{x}\right)^{-1}\ddot{\phi}_{lm}+\frac{2}{x}\left(1-\frac{M}{x}\right)\partial_{x}\phi_{lm}+\left(1-\frac{2M}{x}\right)\partial_{x}^{2}\phi_{lm}-\frac{l(l+1)}{x^{2}}\phi_{lm} \label{Eq: Reduce classical KG-equation in radial coordinates 2}.
\end{eqnarray}
while Eq.~(\ref{Eq: General KG in tortoise coordinates}) simplifies to
\begin{eqnarray}
    \left(-\partial_{t}^{2}+\partial_{x_{*}}^{2}-V_{l}(x)\right)\Psi_{lm}=0
\end{eqnarray}
with the potential
\begin{eqnarray}\label{classical potential}
    V^{({\rm cl})}_{l}(x)=\left(1-\frac{2M}{x}\right)\left(\frac{l(l+1)}{x^{2}}+\frac{2M}{x^{3}}\right)\,,
\end{eqnarray}
which vanishes near the horizon as well as asymptotically.

The minimally coupled scalar field is determined by replacing the classical
metric with the emergent one $\tilde{g}_{\mu\nu}$. However, as we have seen,
the emergent line element is not uniquely defined: depending on whether
Euclidean spatial geometry or a constant time-time component is imposed in the
zero-mass limit. Here we will discuss the latter case while the former can be found in Appendix~\ref{Appendix Scalar Field Coupling}.

In the case when the spacetime has a nontrivial geometry when $M\to 0$, namely the line-element described in Eq.~(\ref{Eq. Effective line-element}), the Klein--Gordon equation (\ref{Eq: General KG in radial coordinates}) reduces to
\begin{eqnarray}
    -\left(1-\frac{2M}{x}\right)\ddot{\phi}_{lm}+\frac{\sqrt{\beta}}{x^{2}}\partial_{x}\left[x^{2}\sqrt{\beta}\left(1-\frac{2M}{x}\right)\partial_{x}\phi_{lm}\right]-\frac{l(l+1)}{x^{2}}\phi_{lm}=0\, .
\end{eqnarray}
where we defined 
\begin{eqnarray}\label{beta}
    \beta(x)=\chi_{0}^{2}\left(1+\lambda^2\left(1-\frac{2M}{x}\right)\right)
\end{eqnarray}
for the sake of simplicity. The equation can be further simplified if we work with the tortoise radial coordinates (\ref{Eq: Radial EMG tortoise coordinates}), as before. The dynamics reduce to the wave equation
\begin{eqnarray}\label{Reduce KG dynamics}
    \left[-\partial_{t}^{2}+\partial_{x_{*}}^{2}-V_{l}(x(x_{*}))\right]\Psi_{lm}=0\,,
\end{eqnarray}
where
\begin{eqnarray}\label{General holonomy potential}
    \tilde{V}_{l}(x)=\left(1-\frac{2M}{x}\right)\left[\frac{2M}{x^3}\chi^{2}+\frac{l(l+1)}{x^2}+\chi_{0}^{2}\left(1-\frac{2M}{x}\right)\left(\frac{3M\lambda^2}{x^3}+\frac{\lambda\lambda'}{x}\left(1-\frac{2M}{x}\right)\right)\right]\,.
\end{eqnarray}
Here, $x$ is implicitly a function of $ x_*$. Notably, the potential vanishes at the horizon as $x\to 2M$ as well as asymptotically as $x\to\infty$, just as in the classical case.

Since both cases imply the same behavior of the potential as the classical
system, near the horizon as well as asymptotically, the vacuum states at
$ \mathcal{J}^{-}$ and $ \mathcal{J}^{+}$ resemble their classical
counterparts. Using the Fourier transform (\ref{Eq: Stationary Fourier
  transform}), we end up with
\begin{eqnarray}
    \frac{{\rm d}^{2}\Psi_{lm}}{{\rm d}x_{*}^{2}}+\left[\omega^{2}-V_{l}(x)\right]\Psi_{lm}=0\,,
\end{eqnarray}
which is a time-independent Schrodinger-like equation for the potentials
(\ref{General holonomy potential}). (As shown in Appendix~\ref{Appendix:
  Covariant tensor components}, this equation takes the same form in the case
of the Euclidean spatial geometry, but with a different potential.)


In this work, inspired by the discretization scheme in loop quantum cosmology
(LQC), we study two cases of interest for the modification function
$\lambda(x)$. The first one is the constant modification function
$\lambda(x)=\tilde{\lambda}$ with
$\chi_{0}=\left(1+\tilde{\lambda}^{2}\right)^{-1/2}$ associated with the original $\mu_{0}$-scheme in LQC. The line-element here reads
\begin{eqnarray}\label{Eq: Asier-Brizuela line-element}
    {\rm d}s^{2}=-\left(1-\frac{2M}{x}\right){\rm d}t^2+\frac{{\rm d}x^{2}}{\left(1-2M/x\right)\left(1-x_{\tilde{\lambda}}/x\right)}+x^{2}{\rm d}\Omega^{2}
\end{eqnarray}
where
\begin{eqnarray}\label{Eq: Minimum radius for constant holonomy}
x_{\tilde{\lambda}}=\frac{2M\tilde{\lambda}^{2}}{1+\tilde{\lambda}^{2}}
\end{eqnarray}
is the solution of (\ref{Eq: condition for minimum radius}), defining the minimal radius of the system. 
In this former case, the discretization does not refine on large scales and consequently it will impact the large scale physics \cite{Idrus1}.
The scalar field potential for this choice reads
\begin{eqnarray}\label{Eq: Potential for constant holonomy function}
    V_{l}^{(\tilde{\lambda})}=V_{l}^{(\rm cl)}+\frac{x_{\tilde{\lambda}}}{2x^{3}}\left(1-\frac{2M}{x}\right)\left(1-\frac{6M}{x}\right)
\end{eqnarray}
where $V_l^{({\rm cl})}$ is the classical potential given in \eqref{classical potential}.

A refined discretization is introduced in LQC by making the modification
function scale dependent, a specific case being known as the
$\bar{\mu}$-scheme. In this scheme the modification function is monotonically
decreasing as $\lambda(x)=\sqrt{\Delta}/x$ with $\chi_0=1$ where
$\Delta$ is a positive constant which may be related to the so-called `area gap' of
loop quantum gravity, $8\pi \gamma \ell_P^2$. The line
element now reads
\begin{eqnarray}\label{Eq: Line-element decreasing holonomy function}
    {\rm d}s^{2}&=&-\left(1-\frac{2M}{x}\right){\rm d}t^{2}+\frac{{\rm d}x^{2}}{\left(1-\frac{2M}{x}\right)\left(1+\frac{\Delta}{x^{2}}\left(1-\frac{2M}{x}\right)\right)}+x^{2}{\rm d}\Omega^{2} \nonumber \\
    &=&\left(1-\frac{2M}{x}\right)\left(-{\rm d}t^{2}+{\rm d}x_{*}^{2}\right)+x^{2}(x_{*}){\rm d}\Omega^{2}\,.
\end{eqnarray}
As in the previous case, the spacetime also has a minimal radius associated to the solution of (\ref{Eq: condition for minimum radius}), which gives \cite{Idrus1} 
\begin{eqnarray}
x_\Delta = (\Delta M)^{1/3} \,\,\frac{\left(1 + \sqrt{1+ \Delta/(27 M^2)}\right)^{2/3} - \left(\Delta/(27 M^2)\right)^{1/3}}{\left(1 + \sqrt{1+ \Delta/(27 M^2)}\right)^{1/3}}\, .
\end{eqnarray}
The scalar field potential for this scheme is given by
\begin{eqnarray}
    V_{l}^{(\Delta)}=V_{l}^{(\rm cl)}-\frac{\Delta}{x^{4}}\left(1-\frac{2M}{x}\right)^{2}\left(1-\frac{5M}{x}\right)\,.
\end{eqnarray}


\subsection{Non-minimally coupled test scalar field}
The equation of motion for nonminimal coupling (\ref{Eq: KG in NMC}) can be rewritten as
\begin{eqnarray}\label{Eq: Generic NMC Coupled scalar field}
    \ddot{\phi}_{lm}=\frac{N E^\varphi \tilde{q}^{xx}}{\left(E^{x}\right)^{3/2}}\left(\frac{N\left(E^{x}\right)^{3/2}}{E^\varphi}\phi_{lm}'\right)'-N^{2}\tilde{q}^{xx}\phi_{lm}'\left(\ln \chi\right)'+N^{2}\frac{\left(E^\varphi\right)^2\tilde{q}^{xx}}{E^x}\partial_{\phi_{lm}}V(\phi_{lm})
\end{eqnarray}
where we have used the decomposition (\ref{Eq: Reduced 2d minimally coupled scalar}).
The second term vanishes for a constant global factor $\chi=\chi_0$. In this case, the equation reduces to
\begin{eqnarray}
    0&=& -\beta^{-1}\left(1-\frac{2M}{x}\right)^{-1}\ddot{\phi}_{lm}+\frac{1}{x^{2}}\partial_{x}\left[x^{2}\left(1-\frac{2M}{x}\right)\partial_{x}\phi_{lm}\right]-\frac{l(l+1)}{x^{2}}\phi_{lm} \label{Eq: Reduce nonminimal KG-equation in radial coordinates 1} \\
    &=&-\beta^{-1}\left(1-\frac{2M}{x}\right)^{-1}\ddot{\phi}_{lm}+\frac{2}{x}\left(1-\frac{M}{x}\right)\partial_{x}\phi_{lm}+\left(1-\frac{2M}{x}\right)\partial_{x}^{2}\phi_{lm}-\frac{l(l+1)}{x^{2}}\phi_{lm} \nonumber
\end{eqnarray}
in the background vacuum solution.

Notice that with the exception of the $\beta^{-1}$ in the first term, this is very similar to its classical counterparts (\ref{Eq: Reduce classical KG-equation in radial coordinates 1}) and (\ref{Eq: Reduce classical KG-equation in radial coordinates 2}).
As $x \to \infty$, the equation of motion reduces to that of flat space, ensuring that the vacuum state at $\mathcal{J}^{+}$ closely resembles the classical one. In contrast, taking the limit $M \to 0$ leads to a modified Klein--Gordon equation, arising from the nontrivial behavior of $\beta$ in this limit. This modification reflects the fact that the spacetime described by Eq.~\eqref{Eq. Effective line-element} remains non-Euclidean even in the zero-mass limit.

The equation (\ref{Eq: Reduce nonminimal KG-equation in radial coordinates 1}) can be further simplified by rewriting it in terms of the radial tortoise coordinates (\ref{Eq: Radial EMG tortoise coordinates}). Defining
\begin{eqnarray}
    \phi_{lm}(t,x)=\frac{\Psi_{lm}(t,x)}{x}\,,
\end{eqnarray}
equation (\ref{Eq: Reduce nonminimal KG-equation in radial coordinates 1}) reduces to
\begin{eqnarray}\label{Eq: nonminimally coupled reduced scalar field}
    \left[-\partial_{t}^{2}+\partial_{x_{*}}^{2}+2\zeta(x)\partial_{x_{*}}-V_{l}(x)\right]\Psi_{lm}=0
\end{eqnarray}
where 
\begin{equation}
    V_{l}(x)=\frac{\beta(x)}{x^{2}}\left(1-\frac{2M}{x}\right)\left(l(l+1)+\frac{2M}{x}\right)\label{Eq:
      nonminimally coupling potential}
  \end{equation}
  and
  \begin{equation}
    \zeta(x)=-\frac{\chi_{0}^{2}\lambda^{2}(x)}{2\sqrt{\beta(x)}}\left(1-\frac{2M}{x}\right)\left(\frac{2M}{x^{2}}+\left(1-\frac{2M}{x}\right)\frac{\partial \ln\lambda^{2}(x)}{\partial x}\right)\label{Eq: nonminimally coupling damping term}
\end{equation}
The nonminimal coupling introduces an (anti)damping term through $\zeta(x)$ and an overall multiplicative correction to $V_{l}$. Using the Fourier transform (\ref{Eq: Stationary Fourier transform}), the equation of motion can be rewritten as the Schrodinger-like equation
\begin{eqnarray}\label{Eq: nonminimal Schrodinger-like coupling in tortoise coordinates} 
    \frac{{\rm d}^{2}\Psi_{lm}}{{\rm d}x_{*}^{2}}+2\zeta\frac{{\rm d}\Psi_{lm}}{{\rm d}x_{*}}+\left[\omega^{2}-V_{l}\right]\Psi_{lm}=0\,.
\end{eqnarray}
Similar dynamics are obtained in the case of nonconstant $\chi$, which can be found in Appendix~\ref{Appendix Scalar Field Coupling}.

\subsection{Classical action functionals and conservation laws: Net stress-energy tensor}

The functionals,
\begin{equation}\label{eq:EH functional}
    S_{\rm EH} [\tilde{g}]
    = \int {\rm d}^4 x \sqrt{-\det \tilde{g}} R
\end{equation}
and
\begin{equation}\label{eq:KG functional}
    S_{\rm KG}[\tilde{g},\phi]
    = \int {\rm d}^4 x \sqrt{-\det \tilde{g}} \left(\frac{1}{2} \tilde{g}^{\mu \nu} (\partial_\mu \phi) (\partial_\nu \phi) + V (\phi) \right)
    \,,
\end{equation}
where $R$ is the Ricci scalar associated to the emergent metric, are not
necessarily action contributions in EMG because they are not guaranteed to
generate the correct equations of motion.
However, as functionals, and neglecting boundary terms, they are invariant under spacetime diffeomorphisms and hence also under the gauge transformations of EMG:
\begin{equation}\label{eq:SEH invariance}
    \delta_\epsilon S_{\rm EH} [\tilde{g}] = \int {\rm d}^4 x \partial_\mu\left[\xi^\mu \sqrt{-\det \tilde{g}} R\right] = 0
  \end{equation}
  and
\begin{equation}\label{eq:SKG invariance}
    \delta_\epsilon S_{\rm KG} [\phi,\tilde{g}] = \int {\rm d}^4 x \partial_\mu\left[\xi^\mu \sqrt{-\det \tilde{g}} \left(\frac{1}{2} \tilde{g}^{\mu \nu} (\partial_\mu \phi) (\partial_\nu \phi) + V (\phi) \right)\right] = 0\,.
\end{equation}

The explicit gauge transformations of these functionals, neglecting boundary terms, are given by
\begin{eqnarray}\label{eq:SEH gauge transf}
    \delta_\epsilon S_{\rm EH} [\tilde{g}]
    &=& \int {\rm d}^4 x\sqrt{-\det \tilde{g}} G_{\mu\nu} \delta_\epsilon \tilde{g}^{\mu\nu}
    \,,
\end{eqnarray}
and
\begin{eqnarray}\label{eq:SKG gauge transf}
    \delta_\epsilon S_{\rm KG} [\phi,\tilde{g}]
    &=& \int {\rm d}^4 x \sqrt{-\det \tilde{g}} \left( - T_{\mu\nu} \delta_\epsilon \tilde{g}^{\mu\nu} - {\cal E} \delta_\epsilon \phi \right)
    \,,
\end{eqnarray}
respectively, where $G_{\mu\nu}$ is the Einstein tensor associated to the emergent metric,
\begin{equation}
    T_{\mu\nu} = (\partial_\mu \phi) (\partial_\nu \phi)
    - \frac{1}{2} \tilde{g}_{\mu\nu} \tilde{g}^{\alpha\beta} (\partial_\alpha \phi) (\partial_\beta \phi)\,,
\end{equation}
is the scalar field's stress-energy tensor, and
\begin{equation}
    {\cal E} = \tilde{g}^{\mu \nu} \nabla_\mu \nabla_\nu \phi - \frac{\partial V}{\partial \phi}
\end{equation}
implies the Klein--Gordon equation when set equal to zero.

Using the on-shell covariance conditions $\delta_\epsilon \tilde{g}^{\mu \nu} = \mathcal{L}_\xi \tilde{g}^{\mu \nu}=2 \nabla^{(\mu}\xi^{\nu)}$ and $\delta_\epsilon \phi^I=\mathcal{L}_\xi \phi^I=\xi^\mu \nabla_\mu \phi^I$, performing integrations by parts, and neglecting boundary terms, equations (\ref{eq:SEH gauge transf}) and (\ref{eq:SKG gauge transf}) become
\begin{equation}
    \delta_\epsilon S_{\rm EH} [\tilde{g}]
    = - 2 \int {\rm d}^4 x \sqrt{-\det\tilde{g}} \xi^\nu \nabla^\mu G_{\mu\nu}
    \ .
\end{equation}
and
\begin{equation}
    \delta_\epsilon S_{\rm KG} [\tilde{g},\phi]
    = \int {\rm d}^4 x \sqrt{-\det\tilde{g}} \xi^\nu \left( 2 \nabla^\mu T_{\mu\nu}
    - E \nabla_\nu \phi \right)
    \,,
\end{equation}
respectively.
Using (\ref{eq:SEH invariance}) and (\ref{eq:SKG invariance}), we obtain
\begin{equation}\label{eq:Einstein tensor conservation}
    \nabla^\mu G_{\mu\nu} = 0\,,
\end{equation}
as expected from the Bianchi identities, and
\begin{equation}\label{eq:KG conservation}
    \nabla^\mu T_{\mu\nu} = \frac{1}{2} {\cal E} \nabla_\nu \phi\,.
\end{equation}

We conclude that the covariant divergence of the Einstein tensor evaluated on solutions of the equations of motion is always preserved in
EMG, while that of the Klein--Gordon stress-energy tensor is preserved only in
the case of minimal coupling where ${\cal E}=0$, but not in the EMG version of
nonminimal
coupling.
However, the models we analyze here turn out to reproduce the classical matter equations of motion asymptotically and at the horizon,
\begin{equation}\label{eq:KG conservation - asymptotically}
    \nabla^\mu T_{\mu\nu} \big|_{x \to 2M, \infty} = \frac{1}{2} {\cal E} \nabla_\nu \phi \big|_{x \to 2M,\infty} = 0\,,
\end{equation}
and hence the Klein--Gordon stress-energy is indeed conserved near the horizon
and in the asymptotic region, which are the relevant regions for the
energy-conservation arguments to be employed in our discussion of  the black hole's evaporation.

Finally, a relevant quantity is the net stress-energy tensor, defined in
\cite{Idrus1} as $\bar{T}_{\mu\nu}=T_{\mu\nu}-G_{\mu\nu}/8\pi$, which vanishes in
GR because of Einstein's equations but not necessarily in EMG.  However, it is
important to note that the covariant divergence of $G_{\mu\nu}$ and
$T_{\mu\nu}$ (asymptotically or at the horizon) vanish independently of each other.  This
property, combined with the fact that the classical action functionals do not
generate the equations of motion, implies an ambiguity in the definition of a
net-stress energy tensor: This concept may be generalized to
\begin{equation}\label{eq:Net stress-energy tensor}
    \bar{T}_{\mu\nu}=T_{\mu\nu}-\frac{\alpha}{8\pi} G_{\mu\nu}\,,
  \end{equation}
where $\alpha$ is a free constant whose classical limit is unity. As will be shown later, 
$\alpha$ acts as the distinguishing parameter between minimal and nonminimal coupling. 


\section{Hawking distribution: Geometric optics approach}\label{Sec: LQGBH Hawking distribution}
Particle creation in the vicinity of a black hole can be interpreted as arising due to different notions of vacua. Namely, if a black hole is considered as a final product of collapsing matter, then the initial Minkowski vacuum will nonunitarily transform to a different vacuum state. The two vacua will have different timelike Killing vector fields and hence have different positive-frequency modes. Accordingly, the particle number operator associated to the initial Minkowski vacuum is nontrivial when acting in the late-time vacuum state. We will see this in what follows through two approaches: (1) the original derivation by Hawking using the Bogoliubov transformation \cite{Hawking} and (2) the CFT approach \cite{Fabbri-Salas-Olmo}. We will focus on the former in this section, while the latter can be found in Appendix~\ref{Appendix B}.


\subsection{Scalar field mode expansion}
Let us consider a massless scalar field $\phi$ propagating in the background of (\ref{Eq. Effective line-element}).
At an early time $t\to -\infty$, the black hole has not formed and the geometry can be described by the line-element (\ref{Eq: Effective line-element in the full case}) in the $M\to0$ limit, which is a nontrivial geometry.
In this limit, the metric (\ref{Eq. Effective line-element}) becomes
\begin{eqnarray}
  \lim_{M\to 0}  {\rm d}s^{2}=-{\rm d}t^{2}+\frac{1+\lambda^{2}_{\infty}}{1+\lambda^{2}(x)}{\rm d}x^{2}+x^{2}{\rm d}x^{2}=-{\rm d}t^{2}+dx_{*}^{2}+x^2(x_*){\rm d}\Omega^{2}\,,
\end{eqnarray}
which reduces to the flat geometry if the modification function is constant $\lambda(x)=\tilde{\lambda}=\lambda_\infty$.

In the derivation of Hawking radiation, the early modes are prepared from $\mathcal{J}^{-}$, namely at asymptotic spatial coordinates  $x\rightarrow\infty$.
Consequently, if $\lambda(x)$ is asymptotically constant or a decreasing function, the vacuum at $\mathcal{J}^{-}$ can be considered a Minkowski vacuum.
We will be particularly interested in these two cases in this work.

The Klein--Gordon field $\phi$ can be expanded in terms of early-time modes
\begin{eqnarray}
\phi(x)= \int{\rm d}\omega \left(a_{\omega}f_{\omega}+a^{\dagger}_{\omega}f^{*}_{\omega}\right)\,,
\end{eqnarray}
where $\{f_{\omega},f^{*}_{\omega}\}$ is a complete set of basis at the null past $\mathcal{J}^{-}$. The complete set of bases in the asymptotic future must also include $\mathcal{H}^{+}$, the future horizon. Accordingly, the field expansion in the asymptotic future must be written as
\begin{eqnarray}
\phi(x)=\int{\rm d}\Omega \left(b_\Omega g_\Omega(x)+c_{\Omega}h_{\Omega}(x)+b^{\dagger}_{\Omega}g^{*}_{\Omega}(x)+c^{\dagger}_{\Omega}h^{*}_{\Omega}(x)\right)\,,
\end{eqnarray}
where the set $\{g_{\Omega},g^{*}_{\Omega}\}$ and $\{h_{\Omega},h^{*}_{\Omega}\}$ are the bases on $\mathcal{J}^{+}$ and $\mathcal{H}^{+}$, respectively. Therefore, if we truncate the state at $\mathcal{J}^{+}$ in the far future $\mathcal{S}=\mathcal{J}^{+}\cup \mathcal{H}^{+}$, then this leads to a different notion of vacuum than $\mathcal{J}^{-}$.

Due to completeness, the modes $g_{\Omega}$ can be expanded in terms of the modes in $\mathcal{J}^{-}$:
\begin{eqnarray}\label{Bogolyubov transformation modes}
     g_{\Omega}&=&\int_{0}^{\infty} {\rm d}\omega \left(A_{\Omega\omega}f_\omega+B_{\Omega\omega}f^{*}_\omega\right)\,,
\end{eqnarray}
with Bogoliubov coefficients $A_{\Omega\omega}$ and $B_{\Omega\omega}$.
If $B_{\Omega\omega}\neq0$, then the transformation is not unitary.
In order to isolate the coefficient $B_{\Omega\omega}$, one needs to define the inner product of the basis modes. 


\subsection{Symplectic current and inner product}
The inner product can be defined through the symplectic structure \cite{Wald-QFT Book}
\begin{eqnarray}\label{Eq: inner-product general}
\tilde{\Omega}[(\phi_{1},P_{\phi_{2}}),(\phi_{2}^{*},P_{\phi_{1}}^{*})]\big|_{\Sigma}=\frac{1}{2}\int_{\Sigma}{\rm d}\Sigma_{a}J^{a}[(\phi_{1},P_{\phi_{2}}),(\phi_{2}^{*},P_{\phi_{1}}^{*})]\,,
\end{eqnarray}
where $J^{a}$ is the symplectic current, and the integration is done on the spatial hypersurface $\Sigma$. The current can be defined from the $U(1)$ conserved current given by (\ref{eq:Current - minimal scalar complex}) for the minimal coupling and by (\ref{eq:Current - nonminimal scalar complex}) for the nonminimal coupling. The former is given by
\begin{eqnarray}
    J^{t}_{{\rm MC}}[(\phi_{1},P_{\phi_{2}}),(\phi_{2}^{*},P_{\phi_{1}}^{*})]=G^\phi_{\rm MC}[(\phi_{1},P_{\phi_{2}}),(\phi_{2}^{*},P_{\phi_{1}}^{*})] &=& i\left(\phi_{1}P_{\phi_{2}}-\phi_{2}^{*}P^{*}_{\phi_{1}}\right) \nonumber \\
&=& i\frac{E^{x}}{\sqrt{\tilde{q}^{xx}}}\left[\phi_{1}\partial_{0}\phi_{2}^{*}-\phi_{2}^{*}\partial_{0}\phi_{1}\right]\,,
\end{eqnarray}
with $\partial_0 = n^{\mu} \partial_{\mu}$. Conservation can be demonstrated analogously to the derivation of Eq.~(\ref{eq:Current - minimal scalar complex}), namely:
\begin{eqnarray}\label{Eq: Conservation MC}
    \dot{J}^{t}_{{\rm MC}}[(\phi_{1},P_{\phi_{2}}),(\phi_{2}^{*},P_{\phi_{1}}^{*})]=-i\left[NE^{x}\sqrt{\tilde{q}^{xx}}\left(\phi^{*}_{2}\phi'_{1}-\phi_{1}\phi_{2}^{*'}\right)+N^{x}\left(P_{\phi_{2}}\phi_{1}-P^{*}_{\phi_{1}}\phi^{*}_{2}\right)\right]'\,,
\end{eqnarray}
which gives $\partial_{\mu}J^{\mu}=0$, with the radial component being
\begin{eqnarray}
    J^{x}_{{\rm MC}}[(\phi_{1},P_{\phi_{2}}),(\phi_{2}^{*},P_{\phi_{1}}^{*})]=i\left[NE^{x}\sqrt{\tilde{q}^{xx}}\left(\phi_{2}^{*}\phi_{1}'-\phi_{2}^{*'}\phi_{1}\right)+N^{x}\left(P_{\phi_{2}}\phi_{1}-P^{*}_{\phi_{1}}\phi^{*}_{2}\right)\right]\,.
\end{eqnarray}
The inner product on the space of the solution can hence be written as
\begin{eqnarray}\label{Eq: Inner product MC}
\tilde{\Omega}_{\rm MC}\left[(\phi_{1},P_{\phi_{2}}),(\phi_{2}^{*},P_{\phi_{1}}^{*})\right]&=&\frac{i}{2}\int_{\Sigma}{\rm d}\Sigma_{\mu}J^{\mu}_{\rm MC}\left[(\phi_{1},P_{\phi_{2}}),(\phi_{2}^{*},P_{\phi_{1}}^{*})\right] \,.
\end{eqnarray}
Due to the conservation and by applying Stokes' theorem, it follows that the symplectic structure (\ref{Eq: Conservation MC}) is independent of the choice of surface $\Sigma$.

The inner product for the nonminimally coupled scalar field is defined similarly through Eq.~(\ref{Eq: inner-product general}), with the symplectic current 
\begin{eqnarray}
    J^{t}_{{\rm NMC}}[(\phi_{1},P_{\phi_{2}}),(\phi_{2}^{*},P_{\phi_{1}}^{*})] = G^\phi_{\rm NMC}[(\phi_{1},P_{\phi_{2}}),(\phi_{2}^{*},P_{\phi_{1}}^{*})] &=& i\left(\phi_{1}P_{\phi_{2}}-\phi_{2}^{*}P^{*}_{\phi_{1}}\right) \\
&=&i \chi \frac{(E^x)^{3/2}}{E^\varphi} \tilde{q}_{xx} \left[\phi_{1}\partial_{0}\phi_{2}^{*}-\phi_{2}^{*}\partial_{0}\phi_{1}\right]\,.\nonumber
\end{eqnarray}
Conservation then follows in a way similar to  Eq.~(\ref{eq:Current - nonminimal scalar complex}):
\begin{eqnarray}\label{Eq: Conservation NMC}
    \dot{J}_{\rm NMC}^{t}= - i\left(N \chi \frac{(E^x)^{3/2}}{E^\varphi} \left(\phi_{1}\phi_{2}^{*'}-\phi^{*}_{2} \phi'_{1}\right)
    +  N^x \left(P_{\phi_{2}} \phi_{1}-P_{\phi_{1}}^{*} \phi_{2}^{*}\right)\right)'\,,
\end{eqnarray}
which gives $\partial_{\mu}J^{\mu}=0$ with 
\begin{eqnarray}
    J^{x}_{\rm NMC}\left[\left(\phi_{1},P_{\phi_{2}}\right),\left(\phi_{2}^{*},P_{\phi_1}^{*}\right)\right]=i\left(N \chi \frac{(E^x)^{3/2}}{E^\varphi} \left(\phi^{*}_{2} \phi'_{1}-\phi_{1}\phi_{2}^{*'}\right)
    +  N^x \left(P_{\phi_{1}}^{*} \phi_{2}^{*}-P_{\phi_{2}} \phi_{1}\right)\right)
\end{eqnarray}
as the radial component. The inner product, as before, is given by
\begin{eqnarray}
    \tilde{\Omega}_{\rm NMC}\left[(\phi_{1},P_{\phi_{2}}),(\phi_{2}^{*},P_{\phi_{1}}^{*})\right]&=&\frac{i}{2}\int_{\Sigma}{\rm d}\Sigma_{\mu}J^{\mu}_{\rm NMC}\left[(\phi_{1},P_{\phi_{2}}),(\phi_{2}^{*},P_{\phi_{1}}^{*})\right]\,.
\end{eqnarray}
which is again independent of the chosen surface $\Sigma$ due to the conservation of the current.



In deriving the black hole thermal distribution, we only need to work in the exterior region of the horizon. Therefore, we choose to work in the Schwarzschild gauge where $N^{x}=0$,
\begin{eqnarray}
E^{x}=x^{2}\,,\quad{\rm and}\quad E^\varphi=\frac{x}{\sqrt{1-\frac{2M}{x}}}\,,
\end{eqnarray}
and the lapse function is given by $N = \sqrt{1 - \frac{2M}{x}}$. Since we are considering a massless scalar field, it is more natural to define the inner product on a null surface. The null coordinates are defined as in Eq.~(\ref{Eq: Null coordinates definition for arbitrary mass}).

 

The corresponding four-density component $J^{\mu}$ in the $x^{'\tilde{\mu}}=\left(u,v\right)$ coordinates can be obtained by the coordinate transformation
\begin{eqnarray}
    J^{\tilde{\mu}}=\left|\frac{\partial x}{\partial x'}\right|^{-1}\frac{\partial x^{'\tilde{\mu}}}{\partial x^{\nu}}J^{\nu}\,,
\end{eqnarray}
with $\left|\partial x/\partial x'\right|$ defining the Jacobian determinant. This gives
\begin{eqnarray}
    J^{u}_{\rm MC}\left(\phi_{1},\phi_{2}\right)&=&\frac{2ix^{2}}{\chi_{0}^{3}\left(1-\frac{2M}{x}\right)\left(1+\lambda^2\left(1-2M/x\right)\right)^{1/2}}\\ &&\times\left[\frac{\phi_{1}\dot{\phi}_{2}^{*}-\phi_{2}^{*}\dot{\phi}_{1}}{\chi_{0}}-\frac{1}{\chi_{0}}\left(\phi_{2}^{*}\partial_{x_{*}}\phi_{1}-\phi_{1}\partial_{x_{*}}\phi_{2}^{*}\right)\right]\,,\nonumber \\
    J^{v}_{\rm MC}\left(\phi_{1},\phi_{2}\right)&=&\frac{2ix^{2}}{\chi_{0}^{3}\left(1-\frac{2M}{x}\right)\left(1+\lambda^2\left(1-2M/x\right)\right)^{1/2}} \nonumber \\ &&\times\left[\frac{\phi_{1}\dot{\phi}_{2}^{*}-\phi_{2}^{*}\dot{\phi}_{1}}{\chi_{0}} +\frac{1}{\chi_{0}}\left(\phi_{2}^{*}\partial_{x_{*}}\phi_{1}-\phi_{1}\partial_{x_{*}}\phi_{2}^{*}\right)\right]\,,\nonumber 
\end{eqnarray}
for the minimal coupling, and
\begin{eqnarray}
    J^{u}_{\rm NMC}\left(\phi_{1},\phi_{2}\right)&=&\frac{2i x^{2}}{\chi_{0}^{4}\left(1-\frac{2M}{x}\right)\left(1+\lambda^{2}\left(1-2M/x\right)\right)^{1/2}} \left[\phi_{1}\dot{\phi}_{2}^{*}-\phi_{2}^{*}\dot{\phi}_{1}-\left(\phi_{2}^{*}\partial_{x_{*}}\phi_{1}-\phi_{1}^{*}\partial_{x_{*}}\phi_{2}\right)\right]\,, \nonumber \\
    J^{v}_{\rm NMC}\left(\phi_{1},\phi_{2}\right)&=&\frac{2i x^{2}}{\chi_{0}^{4}\left(1-\frac{2M}{x}\right)\left(1+\lambda^{2}\left(1-2M/x\right)\right)^{1/2}} \left[\phi_{1}\dot{\phi}_{2}^{*}-\phi_{2}^{*}\dot{\phi}_{1}+\left(\phi_{2}^{*}\partial_{x_{*}}\phi_{1}-\phi_{1}^{*}\partial_{x_{*}}\phi_{2}\right)\right]\,,\nonumber \\
\end{eqnarray}
for the nonminimal coupling. In both cases, the dependence on the scalar momentum has been removed by expressing the momentum in terms of the velocity.
As mentioned earlier, the inner product~(\ref{Eq: inner-product general}) is independent of the choice of the surface $\Sigma$. 

In what follows, we will define the inner product in the traditional way, namely on the past null infinity $\mathcal{J}^{-}$. We use $\left(u_{\rm in},v_{\rm in}\right)$  to denote null coordinates on this surface. On this surface, $x\to\infty$ and hence the symplectic currents simplify to
\begin{eqnarray}
    J^{u_{\rm in}}_{\rm MC}\left(\phi_{1},\phi_{2}\right) \bigg|_{\mathcal{J}^{-}}&=&\frac{4ix^{2}}{\chi_{0}^{4}}\left[\phi_{1}\partial_{v_{\rm in}}\phi^{*}_{2}-\phi_{2}^{*}\partial_{v_{\rm in}}\phi_{1}\right]\,, \nonumber \\
    J^{v_{\rm in}}_{\rm MC}\left(\phi_{1},\phi_{2}\right) \bigg|_{\mathcal{J}^{-}}&=&\frac{4ix^{2}}{\chi_{0}^{4}}\left[\phi_{1}\partial_{u_{\rm in}}\phi_{2}^{*}-\phi_{2}^{*}\partial_{u_{\rm in}}\phi_{1}\right]\,,
\end{eqnarray}
for the minimal coupling and
\begin{eqnarray}
   J^{u_{\rm in}}_{\rm NMC}\left(\phi_{1},\phi_{2}\right) \bigg|_{\mathcal{J}^{-}}&=&\frac{4ix^{2}}{\chi_{0}^{3}}\left[\phi_{1}\partial_{v_{\rm in}}\phi^{*}_{2}-\phi_{2}^{*}\partial_{v_{\rm in}}\phi_{1}\right]\,, \nonumber \\
    J^{v_{\rm in}}_{\rm NMC}\left(\phi_{1},\phi_{2}\right) \bigg|_{\mathcal{J}^{-}}&=&\frac{4ix^{2}}{\chi_{0}^{3}}\left[\phi_{1}\partial_{u_{\rm in}}\phi_{2}^{*}-\phi_{2}^{*}\partial_{u_{\rm in}}\phi_{1}\right]\,,
\end{eqnarray}
for the non-minimal coupling. They both gives the same expression for the decreasing holonomy function.
Thus, the symplectic current defined on an arbitrary null surface $\mathcal{N}^{-}$ of constant $u$ is defined as
\begin{eqnarray}
\tilde{\Omega}\left[\left(\phi_{1},P_{\phi_{2}}\right),(\phi_{2}^{*},P_{\phi_{1}}^{*})\right]=\frac{1}{2}\int_{\mathcal{N}^{-}}{\rm d}\Sigma_{\mu}J^{\mu}(\phi_{1},\phi_{2})\,,
\end{eqnarray}
whereas, for the null surface, the volume measure is given by \cite{Possion-book}
\begin{eqnarray}
    {\rm d}\Sigma_{\mu}=-n_{\mu}{\rm d}\Omega{\rm d}v\,,
\end{eqnarray}
and thus, choosing $\mathcal{N}^{-}=\mathcal{J}^{-}$, we have 
\begin{eqnarray}\label{Eq: Inner product EMG}
    \tilde{\Omega}_{\rm MC}(\phi_{1},\phi_{2})&=&-\frac{2i}{\chi_{0}^{4}}\int_{\mathcal{J}^{-}}{\rm d}v_{\rm in}{\rm d}\Omega x^{2}\left(\phi_{1}\partial_{v}\phi_{2}^{*}-\phi_{2}^{*}\partial_{v}\phi_{1}\right)\,, \\
    \tilde{\Omega}_{\rm NMC}(\phi_{1},\phi_{2})&=&-\frac{2i}{\chi_{0}^{3}}\int_{\mathcal{J}^{-}}{\rm d}v_{\rm in}{\rm d}\Omega x^{2}\left(\phi_{1}\partial_{v}\phi_{2}^{*}-\phi_{2}^{*}\partial_{v}\phi_{1}\right)\, \label{Eq: Inner product EMG 2}
\end{eqnarray}
as the inner product.


\subsection{Bogoliubov transformation and occupation numbers}\label{Subsection: Bogoliubov}
The dynamical process of collapse generates different notions of vacua at late and early times. At early times, the \textit{in}-vacuum $\ket{0_{\rm in}}$, annihilated by $a_{\omega}$, contains \textit{out}-particles since the annihilation operator $b_{\Omega}$ is a linear combination of $\left(a_{\omega},a^{\dagger}_{\omega}\right)$. To justify this, we need to define the orthonormal set of basis.

We define the positive frequency early modes as
\begin{eqnarray}
    f_{lm\omega}(v_{\rm in},x,\theta,\varphi)&=&f_{\omega}(v_{\rm in},x)Y_{lm}(\theta,\varphi)=\frac{\chi_{0}^{n/2}}{\sqrt{8\pi \omega}}e^{-i\omega v_{\rm in}}\frac{Y_{lm}(\theta,\varphi)}{x}\,, \nonumber \\
f^{*}_{lm\omega}(v_{\rm in},x,\theta,\varphi)&=&f^{*}_{\omega}(v_{\rm in},x)Y^{*}_{lm}(\theta,\varphi) =\frac{\chi_{0}^{n/2}}{\sqrt{8\pi \omega}}e^{i\omega v_{\rm in}}\frac{Y^{*}_{lm}(\theta,\varphi)}{x}\,,
\end{eqnarray}
as the complete set for an orthonormal basis with respect to the inner product \eqref{Eq: Inner product EMG} and \eqref{Eq: Inner product EMG 2}, where $n=4$ for minimally coupled and $n=3$ for the non-minimally coupled. The orthonormal relation reads in this basis:
\begin{eqnarray}
    \tilde{\Omega}(f_{\omega lm},f_{\omega'l'm'})&=&\delta_{ll'}\delta_{mm'}\delta\left(\omega-\omega'\right)\,, \\
    \tilde{\Omega}(f^{*}_{\omega lm},f^{*}_{\omega'l'm'})&=&-\delta_{ll'}\delta_{mm'}\delta\left(\omega-\omega'\right)\,, \\
    \tilde{\Omega}(f_{\omega lm},f^{*}_{\omega'l'm'})&=& \tilde{\Omega}(f^{*}_{\omega lm},f_{\omega'l'm'})=0 \,.
\end{eqnarray}
The annihilation operator $b_{\Omega}$ can therefore be written as
\begin{eqnarray}
b_{\Omega}&=&\tilde{\Omega}\left(\phi,g_{\Omega lm}\right) \nonumber \\
    &=&\sum_{l'm'}\int {\rm d}\omega{\rm d}\tilde{\omega}\tilde{\Omega}\left(a_{\omega} f_{\omega l'm'}+a_{\omega}^{\dagger}f_{\omega l'm'}^{*},A_{\Omega \tilde{\omega}}f_{\tilde{\omega}lm}+B_{\Omega\tilde{\omega}}f_{\tilde{\omega}lm}^{*}\right) \,, \nonumber \\
    &=&\int {\rm d}\omega\left(A_{\Omega\omega}a_{\omega}-B_{\Omega\omega}a^{\dagger}_{\omega}\right)\,,
\end{eqnarray}
where we have omitted $\delta_{ll'}$ and $\delta_{mm'}$ to simplify the notation. When $B_{\Omega\omega}$ is nonvanishing, the late-time state will be an excited one with particles. In fact, the number of particles can be computed to be given by
\begin{eqnarray}\label{Eq: Occupation numbers definition}
\braket{N_{\Omega}}:=\braket{0_{\rm in}\big|b_{\Omega}^{\dagger}b_{\Omega}\big|0_{\rm in}}=\int _{0}^{\infty}{\rm d}\omega \big|B_{\Omega\omega}\big|^{2}\,.
\end{eqnarray}
The Bogoliubov coefficient $B_{\Omega\omega}$ can be similarly derived by utilizing the inner-product as
\begin{eqnarray}\label{Eq: B Bogoliubov coefficient}
    B_{\Omega\omega}&=&-\tilde{\Omega}\left(g_{\Omega},f^{*}_{\omega}\right) = \frac{2i}{\chi_{0}^{n}}\int_{\mathcal{J}^{-}}{\rm d}v_{\rm in}   x^{2}\left(g_{\Omega}\partial_{v}f_{ \omega}-f_{\omega}\partial_{v}g_{\omega}\right)\,,
\end{eqnarray}
which requires the basis modes on $\mathcal{I}^{+}$ to be defined. 

The mode expansion at late-times consists of those reaching $\mathcal{J}^{+}$ and $\mathcal{H}^{+}$. We define $\left(u,v\right)$ as the null coordinates on the latter.
The set of orthonormal basis modes on $\mathcal{J}^{+}$ can be defined in a similar fashion as
\begin{eqnarray}\label{Eq: Late time basis}
    g_{\Omega lm}(u,x,\theta,\varphi)&=&g_{\Omega}(u,x)Y_{lm}(\theta,\varphi)=\frac{\mathcal{T}_{l}(\Omega) \chi_{0}^{n/2}}{\sqrt{8\pi\Omega}}\frac{e^{-i\Omega u}}{x}Y_{lm}(\theta,\varphi) \nonumber \\
    g^{*}_{\Omega lm}(u,x,\theta,\varphi)&=&g^{*}_{\Omega}(u,x)Y_{lm}^{*}(\theta,\varphi)=\frac{\mathcal{T}_{l}^{*}(\Omega) \chi_{0}^{n/2}}{\sqrt{8\pi\Omega}}\frac{e^{i\Omega u}}{x}Y^{*}_{lm}(\theta,\varphi)
\end{eqnarray}
where $\mathcal{T}_{l}(\Omega)$ is the transmission coefficient.
Note that the orthonormality condition for $g_{\Omega}$ reads
\begin{eqnarray}\label{Eq: Orthonormality condition of g}
    \tilde{\Omega}(g_{\Omega lm},g_{\Omega'l'm'})&=&\mathcal{T}_{l}(\Omega)\mathcal{T}^{*}_{l'}(\Omega')\delta_{ll'}\delta_{mm'}\delta\left(\Omega-\Omega'\right),
\end{eqnarray}
Our line element (\ref{Eq: Effective line-element in the full case}) does not
displace the location of the event horizon (a property related  to its covariant nature), hence the typical logarithmic relation between late-time and early-time coordinates remains valid,
\begin{eqnarray}
    v&=&v_{\rm in}\,,\label{Eq: Trivial relation} \\
    u&=&v_{\rm H}-\kappa^{-1}\ln \kappa\left(v_{\rm in}-v_{\rm H}\right)=v_{\rm H}-\kappa^{-1}\ln \kappa\left(v-v_{\rm H}\right)\,, \label{Eq: Typical logarithmic relation}
\end{eqnarray}
where $v_{\rm H}$ represents the last ray from the $\mathcal{J}^{-}$ which generates the future horizon $\mathcal{H}^{+}$.
Using this relation, the orthonormal basis (\ref{Eq: Late time basis}), and (\ref{Eq: Trivial relation}), the coefficient (\ref{Eq: B Bogoliubov coefficient}) can be written as
\begin{eqnarray}
    B_{\Omega\omega}&=&\frac{4i}{\chi_{0}^{n}}\int_{\mathcal{J}^{-}}{\rm d}v x^{2}g_{\Omega} \partial_{v}f_{\omega} \nonumber \\
    &=&-\frac{\mathcal{T}_l(\Omega)}{2\pi}\sqrt{\frac{\omega}{\Omega}}e^{-i\Omega v_{\rm H}}\int_{-\infty}^{v_{\rm H}}{\rm d}v \kappa^{i\Omega/ \kappa}\left(v_{\rm H}-v\right)^{i\Omega/\kappa}e^{-i\omega v} \nonumber \\
    &=&\frac{\mathcal{T}_{l}(\Omega)}{2\pi}\sqrt{\frac{\omega}{\Omega}}\kappa^{i\Omega/\kappa}e^{-i\left(\omega+\Omega\right)v_{\rm H}}\int_{0}^{\infty}{\rm d}z z^{i\Omega/\kappa}e^{i\omega z-\epsilon z}\,.
\end{eqnarray}
In the final step we have defined $z=v_{\rm H}-v$ and introduced the $\epsilon$-prescription for damping high $z$-oscillations to ensure convergence.
This integral can be performed, yielding
\begin{eqnarray}
    B_{\Omega\omega}=-\frac{\mathcal{T}_{l}(\Omega)}{2\pi}\sqrt{\frac{\omega}{\Omega}}\kappa^{i\Omega/\kappa}e^{-i(\omega+\Omega)v_{\rm H}}\frac{\Gamma(1+i\Omega \kappa^{-1})}{\left(\epsilon-i\omega\right)^{1+i\Omega\kappa^{-1}}}\,.
\end{eqnarray}
The factor $n$ drops out, and the resulting coefficient is therefore independent of the way the scalar field couples to the gravitational degrees of freedom. Using a similar procedure, the coefficient $A_{\Omega\omega}$ is given by
\begin{eqnarray}
A_{\Omega\omega}&=&\tilde{\Omega}(g_{\Omega},f_{\omega}) \nonumber \\ 
&=&-\frac{\mathcal{T}_{l}(\Omega)}{2\pi}\sqrt{\frac{\omega}{\Omega}}\kappa^{i\Omega/\kappa}e^{-i(-\omega+\Omega)v_{\rm H}}\frac{\Gamma(1+i\Omega \kappa^{-1})}{\left(\epsilon+i\omega\right)^{1+i\Omega\kappa^{-1}}}\,.
\end{eqnarray}
By rewriting
\begin{eqnarray}
    \frac{1}{\left(i\omega +\epsilon\right)^{1+i\Omega \kappa^{-1}}}={\rm exp}{\left[\left(-1-i\Omega\kappa^{-1}\right)\ln\left(\epsilon+i\Omega\right)\right]}
\end{eqnarray}
and using
\begin{eqnarray}
    \ln\left(-i\epsilon-\omega\right)=-i\pi+\ln\omega\,,
\end{eqnarray}
we find the typical relation between $A$ and $B$,
\begin{eqnarray}\label{Eq: Typical A and B relation}
    \left|A_{\Omega\omega}\right|=e^{\pi\Omega \kappa^{-1}}\left|B_{\Omega\omega}\right|\,.
\end{eqnarray}

The orthonormality condition can be written as
\begin{eqnarray}
\tilde{\Omega}\left(g_{\Omega},g_{\Omega}\right)&=&\int_{0}^{\infty}\left(A_{\Omega\omega}f_{\omega}+B_{\Omega\omega}f_{\omega}^{*},A_{\Omega\omega'}f_{\omega'}+B_{\Omega\omega'}f_{\omega'}^{*}\right)\nonumber \\
&=&\int_{0}^{\infty}{\rm d}\omega \left(\left|A_{\Omega\omega}\right|^{2}-\left|B_{\Omega\omega}\right|^{2}\right)\,;
\end{eqnarray}
comparing this with (\ref{Eq: Orthonormality condition of g}) and using relation (\ref{Eq: Typical A and B relation}), the occupation number distribution (\ref{Eq: Occupation numbers definition}) takes the standard form
\begin{eqnarray}\label{Eq: Thermal distribution}
    \int_{0}^{\infty}{\rm d}\omega \left|B_{\Omega\omega}\right|^{2}=\frac{\left|\mathcal{T}_{l}(\Omega)\right|^{2}}{e^{2\pi \Omega \kappa^{-1}}-1}\,.
\end{eqnarray}



\section{The black hole tunneling rates}\label{Sec: The black hole tunneling rates}
The semi-classical analysis performed in the previous subsection was derived under the assumption of negligible backreaction. However, as shown in \cite{Kraus-Wilczek}, when modeling radiation as outgoing shells with energy \( \omega \), the shell follows a geodesic in a black hole spacetime, while the mass parameter \( M \) decreases by \( M - \omega \). Building on this concept, the authors of \cite{Parikh-Wilczek} proposed a straightforward derivation of the black hole's thermal spectrum that incorporates backreaction by analyzing the tunneling process of an \( s \)-wave near the horizon.  

In this section, we apply the tunneling formalism to study Hawking radiation. This method was first applied in \cite{Padmannabhan} using the WKB approximation to study the lowest order solution of the propagating scalar field. Another approach is through the $s$-wave approximation of \cite{Parikh-Wilczek}. A key aspect of this computation is that one needs to hold the ADM mass fixed, \textit{i.e.} the total energy of the black hole and the outgoing shell, \( M+\omega \), remains constant. However, for a quantum-corrected black hole,  the ADM mass \( M_{\rm ADM} \) and local Dirac observable denoted by the mass parameter \( M \) need not necessarily be identical, such as when \( \lambda_\infty \neq 0 \) \cite{Idrus1}, unlike in the classical case. As we will show, recovering the Bekenstein--Hawking entropy asymptotically requires the holonomy modification function to asymptotically vanish.


\subsection{Hamilton--Jacobi analysis of the massless $\phi$-field}
In this subsection we will study the propagation of a massless scalar test field on the emergent metric background. We separate the discussion between minimal and nonminimal coupling. To begin with, let us consider a generic spherically symmetric line element:
\begin{eqnarray}\label{Eq: Generic diagonal metric}
   {\rm d}s^{2}=-f(x){\rm d}t^{2}+\frac{{\rm d}x^{2}}{f(x)h(x)}+x^{2}{\rm d}{\Omega}^{2} \,,
\end{eqnarray}
where we have chosen to work with the diagonal and areal gauge.
In particular, our line element (\ref{Eq: Effective line-element in the full case}) can be cast into the above form with
\begin{eqnarray}\label{eq:f def}
    f(x)&=&1-\frac{2M}{x}\,,\\
    h(x)&=&\chi_{0}^{2}\left(1+\lambda^{2}\left(1-\frac{2M}{x}\right)\right)\,.\label{eq:h def}
\end{eqnarray}


\subsubsection{Minimally coupled scalar}
When the scalar field $\phi$ is minimally-coupled to the metric (\ref{Eq: Generic diagonal metric}), the Klein--Gordon equation is given by Eq.~(\ref{Eq: General KG in radial coordinates}). The nonangular component $\phi_{lm}(t,x)$ can be determined perturbatively by making the following ansatz:
\begin{eqnarray}\label{Eq: WKB Expansion1}
    \phi_{lm}(t,x)={\rm exp}\left(i\tilde{\epsilon} S(t,x)\right)\,,
\end{eqnarray}
where we introduced $\tilde{\epsilon}$ for the purpose of labeling the perturbative expansion. 
The function $S(t,x)$ has the usual series expansion of the form
\begin{eqnarray}\label{Eq: Series expansion on WKB}
    S(t,x)=S_{0}(t,x)+\tilde{\epsilon} S_{1}(t,x)+\tilde{\epsilon}^{2}S_{2}(t,x)+ \ldots\,.
\end{eqnarray}
In terms of $S$, the Klein--Gordon equation (\ref{Eq: General KG in radial coordinates}) reads
\begin{eqnarray}
    \left(\frac{1}{f}\dot{S}^{2}-fh\left(S'\right)^{2}\right)-\tilde{\epsilon}\left(-\frac{i \ddot{S}}{f}+\frac{i\sqrt{h}}{x^2}\left(x^{2}f\sqrt{h}\right)'S'-ifhS''\right)-\frac{\tilde{\epsilon}^2 l(l+1)}{x^2}=0
\end{eqnarray}
and we have
\begin{eqnarray}\label{Eq: Wave equation for S0}
    \left(\frac{\partial S_{0}}{\partial t}\right)^2-f^{2}h\left(\frac{\partial S_{0}}{\partial x}\right)^2=\left(\frac{\partial S_{0}}{\partial t}\right)^2-\left(\frac{\partial S_{0}}{\partial x_{*}}\right)^2=0
\end{eqnarray}
to lowest perturbative order, where $x_*= \int {\rm d}x/(f\sqrt{h})$ denotes the
tortoise coordinate. This is just the wave equation with the well-known
general solution
\begin{eqnarray}
    S_{0}(t,x)=F_{1}(t-x_{*})+F_{2}(t+x_{*})\,,
\end{eqnarray}
where $F_{1}$ and $F_{2}$ correspond to the outgoing and ingoing waves, respectively.
Since the solution depends on the radial coordinates $x$ only through the tortoise coordinates, it is independent of the choice of $\chi$.
For plane waves, the outgoing and ingoing solutions have the  form
\begin{eqnarray}
    F_{1}(t-x_{*})&=&\omega\left(t-
\int^{x}\frac{{\rm d}\tilde{x}}{\left(1-\frac{2M}{\tilde{x}}\right)\sqrt{1+\lambda^{2}\left(1-\frac{2M}{\tilde{x}}\right)}}\right)\,, \label{Eq: Outgoing F1}\\
F_{2}(t+x_{*})&=&\omega\left(t+
\int^{x}\frac{{\rm d}\tilde{x}}{\left(1-\frac{2M}{\tilde{x}}\right)\sqrt{1+\lambda^{2}\left(1-\frac{2M}{\tilde{x}}\right)}}\right) \,,\label{Eq: Ingoing F2}
\end{eqnarray}
where the constant $\omega$ has to be identified with the field's particle energy and we have substituted the expressions of $f$ and $h$, (\ref{eq:f def}) and (\ref{eq:h def}), that are relevant for our emergent metric (\ref{Eq: Effective line-element in the full case}).

The philosophy taken in \cite{Padmannabhan} was that the scalar field
$\phi_{lm}$ is to be interpreted as the wave-function, a construction commonly
known as the first quantization. In this case, the modulus square of the
solution (\ref{Eq: WKB Expansion1}) can be associated with the probability,
and hence to leading order
\begin{eqnarray}
    P_{\rm E}&\approx& {\rm exp}\left(i\tilde{\epsilon}\left(F_{1}-F_{1}^{*}\right)\right)\,,\label{Eq: Probability emission} \\
     P_{\rm A}&\approx& {\rm exp}\left(i\tilde{\epsilon}\left(F_{2}+F_{2}^{*}\right)\right)\,,\label{Eq: Probability Absorption}
\end{eqnarray}
where the subscripts E and A stand for emission and absorption, respectively. Due to the appearance of the energy $\omega$ in the exponent, we can interpret this as thermal radiation if the relationship \cite{Padmannabhan}
\begin{eqnarray}\label{Eq: Probability Emission and Absorption relation}
    P_{\rm E}={\rm exp}\left(-\frac{\omega}{T_{\rm H}}\right)P_{\rm A}
\end{eqnarray}
holds, where $T_{\rm H}$ is the Hawking temperature.

First, consider the tunneling probability from the interior to the exterior region, $P_{\rm E}$.
The relevant radial integration is given by
\begin{eqnarray}
    I_{E}=-\omega\int_{2M-\epsilon}^{2M+\epsilon}\frac{{\rm d}x'}{\left(1-\frac{2M}{x'}\right)\sqrt{1+\lambda^{2}(x')\left(1-\frac{2M}{x'}\right)}}\,.
\end{eqnarray}
The reverse process, when an anti-particle with energy $-\omega$ tunnels into the black hole with probability $P_{\rm A}$, requires the integral
\begin{eqnarray}
I_{A}=-\omega\int_{2M+\epsilon}^{2M-\epsilon}\frac{{\rm d}x'}{\left(1-\frac{2M}{x'}\right)\sqrt{1+\lambda^{2}(x')\left(1-\frac{2M}{x'}\right)}}\,.
\end{eqnarray}
These integrals can be performed analytically using the Feynman $i\epsilon$-prescription, which will be done in Subsection~\ref{Subsection Tunneling picture}. Using this method, one finds
\begin{eqnarray}
    I_{\rm E}&=&-\frac{\omega}{\chi_{0}}4\pi M i=-\frac{i \omega}{\kappa}\,,\\
     I_{\rm A}&=&\frac{\omega}{\chi_{0}}4\pi M i=\frac{i \omega}{\kappa}\,.
\end{eqnarray}
Using these results, the relations (\ref{Eq: Probability emission}) and (\ref{Eq: Probability Absorption}) become
\begin{eqnarray}
    P_{\rm E}&\approx& {\rm exp}\left(-\frac{2\pi\omega}{\kappa}\right)\,, \\
    P_{\rm A}&\approx& {\rm exp}\left(\frac{2\pi\omega}{\kappa}\right)\,,
\end{eqnarray}
and hence they indeed satisfy the thermality condition (\ref{Eq: Probability Emission and Absorption relation}) with $T_{\rm H}=\kappa/2\pi$.

\subsubsection{Non-minimally coupled $\phi$-field}
For the nonminimally coupled case, the scalar field follows  (\ref{Eq: Reduce nonminimal KG-equation in radial coordinates 1}) independent of the chosen $\chi$ function. Using the same ansatz (\ref{Eq: WKB Expansion1}), this equation of motion reduces to
\begin{eqnarray}
    -\beta^{-1}\left(1-\frac{2M}{x}\right)^{-1}\left(\frac{i}{\hbar}\ddot{S}-\frac{1}{\hbar^{2}}\dot{S}^{2}\right)+\frac{2i}{\hbar}\frac{1}{x}\left(1-\frac{M}{x}\right)S'\nonumber\\
    +\left(1-\frac{2M}{x}\right)\left(\frac{i}{\hbar}S''-\frac{\left(S'\right)^2}{\hbar^{2}}\right)-\frac{l(l+1)}{x^{2}}=0\,,
\end{eqnarray}
and it implies (\ref{Eq: Wave equation for S0}) to leading order.
Therefore, the semiclassical expansion is identical for both minimal and nonminimal coupling to leading-order.
Since the thermal distribution is determined by this contribution, both cases yield the same temperature.


\subsection{The ADM mass}
In order to apply the Parikh--Wilczek method, one needs to first define the ADM
mass for the LQG black hole. One way to compute the ADM mass is through the
Brown--York quasilocal energy, followed by taking the limit in which the
2-dimensional surface goes to infinity. The Brown--York quasilocal energy as
seen by observers along a worldline with tangent vector $\hat{t}^\mu=\hat{N}n^\mu+\hat{N}^xs^\mu_x$ is given  by
\begin{equation}
    E = - \frac{1}{8 \pi} \int {\rm d}^{2} z\ \hat{N} \left( \sqrt{\det \sigma} \,\mathcal{K}^{(S)} - \sqrt{\det \bar{\sigma}}\, \bar{\mathcal{K}}^{(S)} \right)
    \,,
    \label{eq:Brown-York quasilocal energy}
\end{equation}
where the integration is over a 2-dimensional surface with coordinates $z$ and
the induced 2-metric $\sigma$, and $\mathcal{K}^{(S)}$ is the trace of
extrinsic curvature on the 2-sphere. The barred quantities are usually evaluated in the ground state of classical general relativity, namely flat Minkowski spacetime. However, we should point out here that this is no longer the case for our LQG model. We will, therefore, choose our zero-mass solution as the relevant ground state.

We are interested in the net quasilocal energy defined by 2-spheres enclosing the black hole. Each symmetric 2-sphere has a normal vector
\begin{equation}
    r^\mu \partial_\mu = \sqrt{\tilde{q}^{xx}} \,\partial_x
    \ ,
\end{equation}
in space, such that $\tilde{g}_{\mu \nu} r^\mu r^\nu=1$. With this information, we can compute the extrinsic-curvature tensor of the spheres and find
\begin{eqnarray}
    \mathcal{K}^{(S)}_{\mu \nu} {\rm d} x^\mu {\rm d} x^\nu &:=& \left(\frac{1}{2} \mathcal{L}_r \tilde{q}_{\mu \nu}\right)  {\rm d} x^\mu {\rm d} x^\nu
    = \sqrt{\tilde{q}^{xx}}\, x \, {\rm d} \Omega^2
    \nonumber\\
    &=&
    \chi_{0}\, x \ \sqrt{1 + \lambda^2 \left( 1 - \frac{2 M}{x} \right)} \ \ \sqrt{1 - \frac{2 M}{x}} {\rm d} \Omega^2
    \ ,
\end{eqnarray}
with its trace given by
\begin{equation}
    \mathcal{K}^{(S)} = \frac{2 \chi_{0}}{x} \sqrt{1 + \lambda^2 \left( 1 - \frac{2 M}{x} \right)} \  \ \sqrt{1 - \frac{2 M}{x}}
    \  \label{eq:Trace of 2-sphere extrinsic curvature - modified}
\end{equation}
and hence
\begin{equation}
    \mathcal{\bar{K}}^{(S)} = \frac{2 \chi_{0}}{x} \sqrt{1 + \lambda^2}
    \  \label{eq:Trace of 2-sphere extrinsic curvature - modified ground state}
\end{equation}
Thus, using this expression to compute the net quasilocal energy in the holonomy-modified LQG  black hole, for observers with $\hat{N}=1$ and $\hat{N}^x=0$, gives
\begin{eqnarray}
    E (x) &=& x \chi_0 \left( \sqrt{1 + \lambda^2} - \sqrt{1 - \frac{2 M}{x}}
              \sqrt{1 + \lambda^2 \left( 1 - \frac{2 M}{x} \right)} \right)
    \,.
    \label{eq:Brown-York quasilocal energy - modified}
  \end{eqnarray}
We can now compute the ADM mass or total energy by taking the limit
\begin{eqnarray}\label{Eq: ADM mass in general}
    M_{\rm ADM}=\lim_{x\rightarrow\infty}E(x)=M\frac{1+2\lambda_{\infty}^{2}}{1+\lambda_{\infty}^{2}}.
\end{eqnarray}

It is worth noting that the ADM mass is not necessarily the same as the black hole mass parameter $M$. They are measuring the same quantities only when $\lambda(x)$ is monotonically decreasing, and thus the $\bar{\mu}$-scheme is within this class. However, the $\mu_{0}$-scheme modification function yields different notions of these two quantities \cite{Bardaji2022}.

\subsection{The Parikh--Wilczek approach}\label{Subsection Tunneling picture}
In this approach, the black-hole radiation can be understood as a tunneling process in which particles escape from just inside the horizon. If a particle pair is created near the horizon, one particle can tunnel outward, reducing the black-hole mass \( M \). However, it is crucial to emphasize that the ADM mass, \( M_{\rm ADM} \), must remain constant throughout the process. The total energy of the system, represented by the ADM mass, is given by
\begin{eqnarray}
    {\cal H}=M_{\rm ADM}-\tilde{\omega}\,.
\end{eqnarray}
This can be associated with the decrease in the black hole mass parameter $M$ to $M-\omega$, where 
\begin{eqnarray}
\omega=\frac{1+\lambda_{\infty}^{2}}{1+2\lambda_{\infty}^{2}}\tilde{\omega}\,.
\end{eqnarray}
due to the inequality $M_{\rm ADM}\neq M$ for arbitrary $\lambda$. Consequently, the horizon decreases by an amount
\begin{eqnarray}
    \Delta x=2\left(M-\omega\right)\,.
\end{eqnarray}

The tunneling rates $\Gamma\simeq {\rm exp}\left(-2{\rm Im}S_{\rm o.s.}\right)$ can be computed through the WKB approximation by computing the on-shell action of a null free particle,
\begin{eqnarray}
    S_{\rm o.s.}&=&\int_{2M-\epsilon}^{2(M-\omega)+\epsilon}\int_{0}^{p_{x}}{\rm d}p_{x}{\rm d}x\,.
\end{eqnarray}
Using Hamilton's equation $\dot{x}={\rm d}{\cal H}/{\rm d}p_{x}=-{\rm d}\tilde{\omega}/{\rm d}p_x$, where $p_{x}$ is the momentum of the shell gives
\begin{eqnarray}\label{Eq: On-shell action}
    S_{\rm o.s.}&=&\int_{2M-\epsilon}^{2(M-\omega)+\epsilon}\int_{0}^{p_{x}}\frac{{\rm d}{\cal H}}{\dot{x}}{\rm d}x\nonumber \\
    &=&-\frac{1+2\lambda_{\infty}^{2}}{1+\lambda_{\infty}^{2}}\int_{0}^{\omega}{\rm d}\omega'\int_{2M-\epsilon}^{2\left(M-\omega\right)+\epsilon}\frac{{\rm d}x}{\dot{x}}\,.
\end{eqnarray}
The radial velocity $\dot{x}={\rm d}x/{\rm d}t$ of the null shell can be
computed using the null condition ${\rm d}s^{2}=0$ for the radial
motion. According to the spherically symmetric ADM line element (\ref{Eq: ADM
  spherical line element}), the result is
\begin{eqnarray}\label{Eq: Three velocity}
    \dot{x}=\frac{{\rm d}x}{{\rm d}t}  = N\sqrt{\tilde{q}^{xx}}\left[1-\sqrt{\tilde{q}_{xx}}\frac{N^{x}}{N}\right]\,.
\end{eqnarray}

The coordinates in (\ref{Eq. Effective line-element}) are not suitable for evaluating (\ref{Eq: Three velocity}) because the metric components are not well-behaved at the horizon. Instead, we use the PG gauge, by setting \( N = 1 \). In this gauge, the line element (\ref{Eq. Effective line-element}) becomes
\begin{eqnarray}
    {\rm d} s^2_{\rm GP}
    &=&
    - {\rm d} t_{\rm GP}^2
    + \frac{1}{\chi_0^{2}} \left( 1 + \lambda^2 \left( 1 - \frac{2M}{x} \right)
        \right)^{-1}\left( {\rm d} x 
    + \chi_0 \sqrt{\frac{2M}{x}} \sqrt{1 + \lambda^2 \left( 1 - \frac{2M}{x} \right)} {\rm d} t_{\rm GP} \right)^2\nonumber\\
    & & \,\,\,\, + x^2 {\rm d} \Omega^2
    \label{eq:Spacetime metric - modified - GP simplified}
\end{eqnarray}
Thus, by evaluating Eq.~(\ref{Eq: Three velocity}) in the PG gauge
(\ref{eq:Spacetime metric - modified - GP simplified}), the on-shell action (\ref{Eq: On-shell action}) reduces to
\begin{eqnarray}
S_{\rm o.s.}&=&\frac{1+2\lambda_{\infty}^{2}}{\sqrt{1+\lambda_{\infty}^{2}}}\int_{0}^{\omega}\int_{2\left(M-\omega\right)+\epsilon}^{2M-\epsilon}\frac{{\rm d}x}{\left(1-\sqrt{\frac{2\left(M-\omega'\right)}{x}}\right)\left(1+\lambda^{2}(x)\left(1-\frac{2\left(M-\omega'\right)}{x}\right)\right)}{\rm d}\omega'\nonumber \\
&=&\frac{1+2\lambda_{\infty}^{2}}{\sqrt{1+\lambda_{\infty}^{2}}}\int_{0}^{\omega}{\rm d}\omega'\int_{u_{\rm out}}^{u_{\rm in}}\frac{2u^{4}{\rm d}u}{\left(u-\sqrt{2\left(M-\omega'\right)}\right)\left(u^{2}+\lambda^{2}(u)\left(u^{2}-2\left(M-\omega'\right)\right)\right)}\,,\nonumber\\
& &
\end{eqnarray}
where we have performed the change of variable to $u=\sqrt{x}$. The integral has a pole at the horizon $u=\sqrt{2\left(M-\omega\right)}=:u_{0}$. The integral can be computed by using the Feynman $i\epsilon$-prescription where we relocate the pole at $u_{0}\rightarrow u_{0}+i\epsilon$ for $\epsilon>0$, covering the lower complex plane. The direction of the semicircle $C_{\epsilon}$ is counter-clockwise from $u_{\rm out}=u_{0}+\epsilon$ to $u_{\rm in}=\sqrt{2M}-\epsilon$ with radius $\epsilon$ and enclose the pole at $u_{0}=\sqrt{2\left(M-\omega\right)}+i\epsilon$. Therefore, the on-shell action can be written as
\begin{eqnarray}
    & &\int_{u_{\rm in}}^{u_{\rm out}}\frac{2u^{4}{\rm d}u}{\left(u-u_{0}\right)\left(u^{2}+\lambda^{2}(u)\left(u-u_{0}\right)\left(u+u_{0}\right)\right)} \nonumber\\
    &=&4\pi i{\rm Res}\left[\frac{u^{4}}{\left(u-u_{0}\right)\left(u^{2}+\lambda^{2}(u)\left(u-u_{0}\right)\left(u+u_{0}\right)\right)}\right]\nonumber\\
&&    -\int_{C_{\epsilon}}\frac{2u^{4}}{\left(u-u_{0}\right)\left(u^{2}+\lambda^{2}(u)\left(u-u_{0}\right)\left(u+u_{0}\right)\right)}\nonumber \\
    &=& 8\pi i \left(M-\omega'\right) - \int_{C_{\epsilon}}\frac{2u^{4}}{\left(u-u_{0}\right)\left(u^{2}+\lambda^{2}(u)\left(u-u_{0}\right)\left(u+u_{0}\right)\right)}\,.
\end{eqnarray}
In the last integral, we take the limit $\epsilon\rightarrow0^{+}$ which gives $4\pi i \left(M-\omega'\right)$, and therefore
\begin{eqnarray}
    S_{\rm o.s.}&=&\frac{1+2\lambda_{\infty}^{2}}{\sqrt{1+\lambda_{\infty}^{2}}}\int_{0}^{\omega}{\rm d}\omega'4\pi i \left(M-\omega'\right)\nonumber \\
    &=&\frac{1+2\lambda_{\infty}^{2}}{\sqrt{1+\lambda_{\infty}^{2}}}4\pi i \omega\left(M-\frac{\omega}{2}\right)\,.
\end{eqnarray}
The on-shell action above is identical to the classical result for any decreasing modification function with $\lambda_{\infty}= 0$.

A similar computation can be done for the ingoing pair. The decrease in the black hole mass parameter is then $M-(-\omega)=M+\omega$. By using instead the ingoing radial null geodesics in (\ref{Eq: Three velocity}), and computing the on-shell action by setting the limit of $\omega$ integration from zero to $-\omega$, one ends up with the same expression. The tunneling rates, which are defined as $\Gamma\simeq {\rm exp}\left(-2{\rm Im} S_{\rm o.s.}\right)$, are then evaluated as 
\begin{eqnarray}\label{Eq: Tunneling rates asymptotic relation}
    \Gamma\simeq{\rm exp}\left(-8\pi \frac{1+2\lambda_{\infty}^{2}}{\sqrt{1+\lambda_{\infty}^{2}}}\omega \left(M-\frac{\omega}{2}\right)\right)={\rm exp}\left(-\frac{8\pi\tilde{\omega}}{\chi_{0}}\left(M-\frac{\omega}{2}\right) \right)\,.
\end{eqnarray}
The last expression can be thought of as the Boltzmann factor, with the temperature 
\begin{eqnarray}
    T_{\rm H}=\frac{\chi_0}{8\pi\left(M-\frac{\omega}{2}\right)}\,,
\end{eqnarray}
which is precisely the effective Hawking temperature of the black hole with mass $M-\omega/2$. On the other hand, the argument inside the exponent of the first term can be cast as the change in the black hole entropy ($\Delta S_{\infty}$), as seen by the asymptotic observer,  with
\begin{eqnarray}
    S_{\infty}=\frac{1+2\lambda_{\infty}^{2}}{1+\lambda^{2}_{\infty}}S_{\rm BH}(M)
\end{eqnarray}
where $S_{\rm BH}(M)=A_{\rm H}(M)/4$ is the Bekenstein--Hawking entropy. The above relation is in agreement with the result in \cite{Idrus1}.

According to (\ref{Eq: Tunneling rates asymptotic relation}), the most probable mode corresponds to \( \omega = M \) with tunneling rate
\begin{eqnarray}
    \Gamma\simeq {\rm exp}\left(-4\pi \frac{1+2\lambda_{\infty}^{2}}{\sqrt{1+\lambda_{\infty}^{2}}}M^{2} \right)={\rm exp}(-S_{\infty}(M))=\frac{e^{S_{f_{\infty}}}}{e^{S_{i_{\infty}}}}\,.
\end{eqnarray}
A statistical-mechanics interpretation implies that there are \( \exp(S_{\infty}) \) possible configurations for this specific macroscopic state of outgoing modes with \( \omega = M \).


\section{The greybody factors}\label{Sec: The greybody factors}
The appearance of the transmission coefficient $\mathcal{T}_{l}(\omega)$ in (\ref{Eq: Thermal distribution}) shows that the black hole is not really black because some of the incoming modes are reflected back to $\mathcal{J}^{+}$. It is not known how to analytically compute the greybody factors for arbitrary frequency. However, analytical estimates can be attained if one considers rather extreme energy ranges for the incoming particles.
For low frequency, one uses matching solutions in three regions \cite{Unruh2}, while for high frequency one usually refers to the monodromy matching technique \cite{Neitzke, Motl}.
We will focus on the former case in this work.

The discussion of the low-frequency case originates from Unruh's work \cite{Unruh2} concerning scalar and fermion fields in a classical Schwarzschild background. This analysis was later extended to encompass $d$-dimensional asymptotically flat black holes \cite{Mathur}, concluding that, for any asymptotically flat black hole in an arbitrary number of dimensions, the transmission coefficient is always proportional to the horizon area.
This finding establishes the universality of black hole absorption rates across all dimensions, provided the black hole is asymptotically flat. This concept of universality was further extended to include dS and AdS black holes in \cite{Harmark}.

In this approach, the energy of the incoming scalar particles was assumed to be below the background thermal energy
\begin{eqnarray}\label{low1}
    \omega\ll T_{H}
\end{eqnarray}
since the bath temperature is proportional to the inverse of the black-hole energy $T_{H}\sim \frac{1}{x_{\rm H}}\sim \frac{1}{2M}$, condition (\ref{low1}) can be written as
\begin{eqnarray}\label{low2}
    x_{\rm H}\omega \ll 1 \quad\rightarrow\quad M\omega\ll1.
\end{eqnarray}
We have shown that the temperature expression is identical to the classical one; therefore, the conditions (\ref{low1}) and (\ref{low2}) remain valid. 

In what follows we will define three regions \cite{Harmark}: 
\begin{itemize}
    \item Region I: The region near the event horizon $x\approx 2M$ and $V(x)\ll \omega^2$.
    \item Region II: The intermediate region, between the horizon and the asymptotic region. Here the peak of the potential dominates $V(x)\gg\omega^2$.
    \item Region III: The asymptotic region as $x\rightarrow \infty$. Here again the potential goes flat; however, as the wave propagates to infinity, it carries the information of region I and II.
\end{itemize}
We will then require that the solution is matched continuously at the boundary
between regions. In this section, we will, in particular, be interested in the nontrivial spatial component in the $M\to0$ limit, namely by choosing $\chi=\chi_{0}$.


\subsection{The minimally coupled scalar field}
We begin by examining the minimally coupled scalar field, studying the schemes
introduced in Sec.~\ref{sub: Minimally coupled test scalar field}: The constant
holonomy function ($\mu_{0}$-scheme) and monotonically decreasing holnomy
function ($\bar{\mu}$-scheme). As we will demonstrate, an analytical solution can be obtained only in the \(\mu_0\)-scheme. In contrast, for the \(\bar\mu\)-scheme, the integration must be handled perturbatively.

The scalar field propagation (\ref{Eq: General Schrodinger in tortoise coordinates}) can be cast into a Schrodinger-like equation, in terms of the tortoise coordinates, by considering a stationary state, which then gives 
\begin{eqnarray}\label{Eq: Reduced Schrodinger equation in tortoise coordinates}
    \left[\frac{\rm d^{2}}{{\rm d}x_{*}^{2}}+U_{l}\right]\psi_{lm}=0\,,
\end{eqnarray}
where
\begin{eqnarray}\label{U potential}
    U_{l}(x)=\omega^{2}-V_{l}(x)\,.
\end{eqnarray}
The potential $V_l(x)$ is given by Eq.~(\ref{Eq: Generic scalar field potential}).

Alternatively, one could also rewrite this expression in terms of the radial coordinate $x$ as 
\begin{eqnarray}\label{Eq: Reduced Schrodinger equation in radial coordinates}
    \sqrt{\beta(x)}\frac{{\rm d}}{{\rm d}x}\left[x^{2}\left(1-\frac{2M}{x}\right)\sqrt{\beta}\frac{{\rm d}\left(\psi_{lm}/x\right)}{{\rm d}x}\right]+\left(\frac{\omega^{2}x^{2}}{1-\frac{2M}{x}}-l(l+1)\right)\frac{\psi_{lm}(x)}{x}=0\,,
\end{eqnarray}
where $\beta$ has been defined in \eqref{beta}. In the low-frequency regime, the dominant multipole is the lowest one; therefore, we specifically consider the \(s\)-wave. This is because the peak of the effective potential increases with higher multipoles, causing most incoming low-frequency particles to be reflected. 


\subsubsection{Constant holonomy function}
In this subsection, we assume the modification function $\lambda(x)=\tilde{\lambda}$. The spacetime line element can be written in this scheme as
\begin{eqnarray}\label{Eq: Constant holonomy line element}
    {
\rm d}s^{2}&=&-\left(1-\frac{2M}{x}\right){\rm d}t^{2}+\frac{{\rm d}x^{2}}{\left(1-\frac{2M}{x}\right)\left(1-\frac{x_{\tilde{\lambda}}}{x}\right)}+x^{2}{\rm d}\Omega^{2} \nonumber \\
&=&\left(1-\frac{2M}{x}\right)\left(-{\rm d}t^{2}+{\rm d}x_{*}^{2}\right)+x^{2}(x_{*}){\rm d}\Omega^{2}
\end{eqnarray}
where $x_{\tilde{\lambda}}=2M\tilde{\lambda}^{2}/(1+\tilde{\lambda}^{2})$.
The tortoise radial coordinate $x_{*}$ has an analytical expression in this particular scheme, which is given by
\begin{eqnarray}\label{Eq: Tortoise radial coordinates constant holonomy}
    x_{*}&=&x \, \sqrt{1-\frac{x_{\tilde{\lambda}}}{x}}+\frac{2M}{\chi_{0}}\ln\left(\frac{x-2M}{M\left(x-\frac{\tilde{\lambda}^{2}\chi^{2}}{2}\left(x+2M\right)+x\chi \sqrt{1-\frac{x_{\tilde{\lambda}}}{x}}\right)}\right)\nonumber \\&&+2M\chi_{0}^{2}\left(1+\frac{3\tilde{\lambda}^{2}}{2}\right)\ln\left(\frac{x}{4}\left(1+\sqrt{1-\frac{x_{\tilde{\lambda}}}{x}}\right)^{2}\right)\,.
\end{eqnarray}
The limit $\tilde{\lambda}\to 0$ gives
\begin{eqnarray}
    \lim_{\tilde{\lambda}\rightarrow 0}x_{*}=x+2M\left(\frac{x-2M}{2M}\right)
\end{eqnarray}
which reduces to its classical expression.
The potential has the form
\begin{eqnarray}\label{Eq: Potential for constant holonomy function}
    V_{l}
    ^{(\tilde{\lambda})}=V_{l}^{(\rm C)}+\frac{x_{\tilde{\lambda}}}{2x^{3}}\left(1-\frac{2M}{x}\right)\left(1-\frac{6M}{x}\right)\,,
\end{eqnarray}
 where
\begin{eqnarray}
    V^{({\rm C})}=\left(1-\frac{2M}{x}\right)\left(\frac{l(l+1)}{x^{2}}+\frac{2M}{x^{3}}\right)\,.
\end{eqnarray}

The regime we are interested in is the near the horizon, $x\sim 2M$, which gives
\begin{eqnarray}\label{Eq: tortoise coordinates near the horizon}
    x_{*}\simeq2M\gamma_{\tilde{\lambda}}(M)+\frac{2M}{\chi_0}\ln\left(\frac{x-2M}{M\left(1+\chi^{2}\left(1-\tilde{\lambda}^{2}\right)\right)}\right)
\end{eqnarray}
with
\begin{eqnarray}\label{Eq: Definition for gamma constant holonomy}
    \gamma_{\tilde{\lambda}}(M)=\chi_0+\chi_{0}^{2}\left(1+\frac{3\tilde{\lambda}^{2}}{2}\right)\ln\left(\frac{M}{2}\left(1+\chi_{0}^{2}\right)^{2}\right)-\frac{1}{\chi_{0}}\ln (2M)\,.
\end{eqnarray}
As we can see, the logarithmic term (\ref{Eq: tortoise coordinates near the horizon}) is dominant near the horizon $x\simeq 2M$. However, we will keep track of the small correction term $\gamma_{\tilde{\lambda}}(M)$, since it will play a nontrivial role later on.

\begin{figure}[h]
    \centering
\includegraphics[width=10cm]{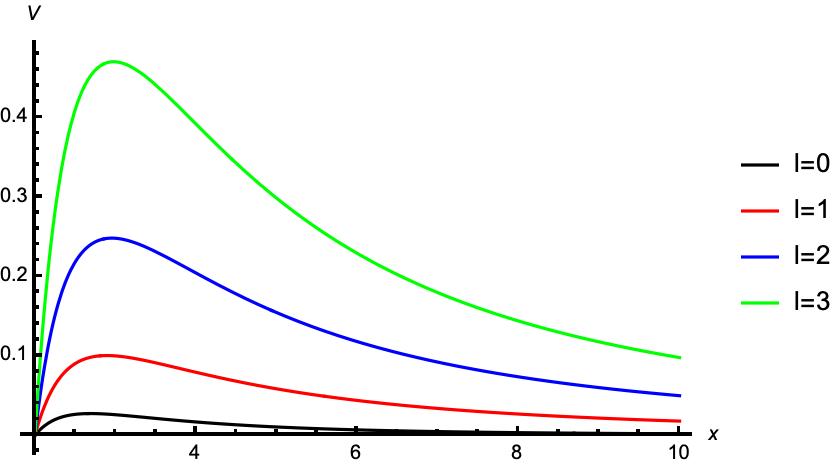}
    \caption{Potential vs.\ radial coordinates $x$ for constant holonomy function with $\tilde{\lambda}=0.3$ and $M=1$. The potential vanishes at the horizon and flattens asymptotically.}
    \label{Fig: Constant holonomy potential}
\end{figure}

We analyze the propagation of the minimally coupled scalar field in this
spacetime through the time-independent Schrodinger-like equation for
$u_{lm}(x)$, namely (\ref{Eq: Reduced Schrodinger equation in tortoise
  coordinates}) in tortoise coordinates and (\ref{Eq: Reduced Schrodinger
  equation in radial coordinates}) in radial coordinates, with the potential
$V_{l}^{(\tilde{\lambda})}$ given by Eq.(\ref{Eq: Potential for constant
  holonomy function}). In Fig.~\ref{Fig: Constant holonomy potential}, we see
that the lower the multipole number $l$, the lower the peak of the potential. Accordingly, by studying an incoming low-energy scalar field from $\mathcal{J}^{-}$, the transmitted modes will be dominated by the $s$-wave modes. Consequently, it is justified to use the $s$-wave approximation, by considering the $l=0$ case.

Near the horizon (region I), the potential $V_{0}\approx 0$ and, thus, $U_{0}\approx \omega^{2}$ from \eqref{U potential}. The solution to Eq.~(\ref{Eq: Reduced Schrodinger equation in tortoise coordinates}) is, therefore, a plane wave
\begin{eqnarray}\label{Eq: Region I Near the horizon}
    \psi_{0}^{(
    I)}(x)&\simeq& A_{I}e^{i\omega x_{*}} \nonumber \\
    &\approx&A_{I}\left(1+\frac{2iM\omega}{\chi_{0}}\left(\chi_0\gamma_{\tilde{\lambda}}(M)+\ln\left(\frac{x-2M}{M\left(1+\chi_{0}^{2}\left(1-\tilde{\lambda}^{2}\right)\right)}\right)\right)\right)\,.
\end{eqnarray}
Notice that this is an incoming solution near the horizon because $x_{*}$ is
negative definite when $x\sim 2M$. The intermediate region, by contrast, is dominated by the potential $V_{0}\gg\omega^{2}$, which, according to Eq.~(\ref{Eq: Reduced Schrodinger equation in radial coordinates}), implies
\begin{eqnarray}
    \frac{{\rm d}}{{\rm d}x}\left[x^{2}\left(1-\frac{2M}{x}\right)\sqrt{1-\frac{x_{\tilde{\lambda}}}{x}}\frac{{\rm d}\psi^{II}_{0}}{{\rm d}x}\right]=0\,,
\end{eqnarray}
with the solution
\begin{eqnarray}\label{Eq: Region II General}
    \psi_{0}^{(II)}&=&\tilde{A}_{II}+B_{II}\int\frac{{\rm d}x}{x^{2}\left(1-\frac{2M}{x}\right)\sqrt{1-\frac{x_{\tilde{\lambda}}}{x}}}\nonumber \\
    &=&A_{II}+\frac{B_{II}}{2M\chi_0} \ln\left(\frac{8M\left(x-2M\right)}{4Mx\left(1+\chi_0\sqrt{1-\frac{x_{\tilde{\lambda}}}{x}}\right)-x_{\tilde{\lambda}}\left(x+2M\right)}\right)\,,
\end{eqnarray}
where $\tilde{A}_{II}=A_{II}+B_{II}\ln(8M/x_{\tilde{\lambda}})/(2M\chi_0) $ has been included as an appropriate constant so as to recover the proper classical limit. Near the horizon, the solution can be approximated as
\begin{eqnarray}\label{Eq: Region II Near the horizon}
    \psi_{0}^{(II)}\simeq A_{II}+B_{II} \ \ln\left(\frac{x-2M}{M\left(1+\chi_{0}^{2}\left(1-\tilde{\lambda}^{2}\right)\right)}\right)\,.
\end{eqnarray}
This must be matched with the solution in region I, which gives the conditions
\begin{eqnarray}
    A_{II}&=&A_{I}\left(1+2iM\omega \gamma_{\tilde{\lambda}}(M)\right) \nonumber \\
    B_{II}&=&4i M^{2}\omega\,.
\end{eqnarray}
The asymptotic form of  (\ref{Eq: Region II General}) on the other hand is given by 
\begin{eqnarray}\label{Eq: Region III Asymptotic}
    \psi_{0}^{(II)}&\simeq&A_{II}+\frac{B_{II}}{2M\chi_{0}}\ln\left(\frac{2}{1+\chi_{0}\left(1-\frac{\tilde{\lambda}^{2}}{2}\right)}\right)+\frac{B_{II}}{2M\chi_{0}}\ln\left(1-\frac{2M}{x}\right) \nonumber \\
    &\approx&A_{I}\left[1+\frac{2iM\omega}{\chi_0}\left(\chi_{0}\gamma_{\tilde{\lambda}}+\ln\left(\frac{2}{1+\chi_{0}\left(1-\frac{\chi_{0}\tilde{\lambda}^{2}}{2}\right)}\right)\right)-\frac{4iM^{2}\omega^{2}}{\chi_{0}\rho}\right]
\end{eqnarray}
where we have defined $\rho:=\omega x$.

In the asymptotic region (region III), one can then approximate
\begin{eqnarray}\label{Eq: Asymptotic approximation}
    1-\frac{2M}{x}\rightarrow 1\quad {\rm and}\quad
    1-
\frac{x_{\tilde{\lambda}}}{x}\rightarrow 1\,.
\end{eqnarray}
This simplifies the equation (\ref{Eq: Reduced Schrodinger equation in radial coordinates}) to
\begin{eqnarray}
    \frac{{\rm d}}{{\rm d}\rho}\left[\rho^{2}\frac{{\rm d}\psi_{0}^{(III)}}{{\rm d}\rho}\right]+\rho^{2}\psi_{0}^{(III)} = 0\,,
\end{eqnarray}
which has the solution
\begin{eqnarray}\label{Eq: Region III}
    \psi_{0}^{(III)}=\frac{e^{-i\rho}A_{III}}{\rho}-\frac{ie^{i\rho}B_{III}}{2\rho}\,,
\end{eqnarray}
where the first and second terms correspond to the incoming and outgoing solutions, respectively. To determine $A_{III}$ and $B_{III}$ in terms of $A_{I}$, we require that this solution match with the solution in region II. Near region II, it is then safe to assume that $\rho<1$, and, thus, we can approximate
\begin{eqnarray}\label{Eq: Region III match with II}
    \psi_{0}^{(III)}\simeq -i\left(A_{III}+\frac{iB_{III}}{2}\right)+\frac{1}{\rho}\left(A_{III}-\frac{iB_{III}}{2}\right)\,.
\end{eqnarray}
Matching this solution to Eq.~(\ref{Eq: Region III Asymptotic}) gives
\begin{eqnarray}\label{Eq: Asymptotic coefficients MC constant holonomy}
    A_{III}&=&\frac{iA_{I}}{2\chi_{0}}\left[\chi_{0}-4M^{2}\omega^{2}+2iM\omega\left(\gamma_{\tilde{\lambda}}(M)\chi_{0}+\ln\left(\frac{2}{1+\chi_{0}\left(1-\frac{\tilde{\lambda}^{2}\chi_{0}}{2}\right)}\right)\right)\right]\,,\\
     B_{III}&=&\frac{A_{I}}{\chi_{0}}\left[\chi_{0}+4M^{2}\omega^{2}+2iM\omega\left(\gamma_{\tilde{\lambda}}(M)\chi_{0}+\ln\left(\frac{2}{1+\chi_{0}\left(1-\frac{\tilde{\lambda}^{2}\chi_{0}}{2}\right)}\right)\right)\right]\,.
\end{eqnarray}
Note that the term inside the round parentheses is exactly the same for $A_{III}$ and $B_{III}$, which is a crucial condition for the conservation of the current flux. To see this, we need to compute the total flux near the horizon. The current flux in general is given by (\ref{eq:Current - minimal scalar complex}), which in the Schwarzschild gauge reads
\begin{eqnarray}\label{Eq: Current MC Schawrzschild gauge Constant holonomy}
    J^{x}_{{\rm MC}}=i N E^x \sqrt{\tilde{q}^{xx}} \left(\psi^{*}_{0} \psi'_{0}-\psi_{0} (\psi^{*}_{0})'\right)\,.
\end{eqnarray}
Near the horizon, the above relation gives
\begin{eqnarray}\label{Eq: Current MC at the horizon}
    J_{\mathcal{H}}&=&i\left(1-\frac{2M}{x}\right)x^{2}\left(\psi_{0}^{(I)*}\frac{{\rm d}\psi_{(0)}^{(I)}}{{\rm d}x_{*}}\frac{{\rm d}x_{*}}{{\rm d}x}-\psi_{0}^{(I)}\frac{{\rm d}\psi_{(0)}^{(I)*}}{{\rm d}x_{*}}\frac{{\rm d}x_{*}}{{\rm d}x}\right)
    \nonumber \\
    &=&-8 M^{2}\omega \left|A_{I}\right|^{2}\,.
\end{eqnarray}
On the other hand, the asymptotic waves consist of ingoing and outgoing modes (\ref{Eq: Region III}), with the total flux
\begin{eqnarray}
    J_{\infty}&=&\frac{2}{\omega}\left(\left|A_{III}\right|^{2}-\frac{\left|B_{III}\right|^{2}}{4}\right)=J_{\mathcal{J}^{-}}-J_{\mathcal{J}^{+}}\nonumber \\
    &=&-8M^{2}\omega\left|A_{I}\right|^{2}   =J_{\mathcal{H}}
   \nonumber \\
\end{eqnarray}
where $J_{\mathcal{J}^{+}}$ is the outgoing modes reaching future null infinity, while $J_{\mathcal{J}^{-}}$ is the ingoing modes from past null infinity. The first line is obtained by utilizing (\ref{Eq: Asymptotic coefficients MC constant holonomy}). As can be seen, the conservation relation is satisfied,
\begin{eqnarray}
    J_{\mathcal{J}^{-}}&=&J_{\mathcal{J}^{+}}+J_{\mathcal{H}}\,, \nonumber \\
    1&=&\mathcal{R}_{0}+\mathcal{T}_{0}\,,
\end{eqnarray}
which shows that the current flux coming in from $\mathcal{J}^{-}$ is partly
reflected back to $\mathcal{J}^{+}$ and partly transmitted to the horizon
$\mathcal{H}$ with a relative amount of $\mathcal{R}_{0}$ and
$\mathcal{T}_{0}$, respectively. The subscript ``0'' is used to indicate that
this result has been computed for the $s$-wave ($l=0$). 

The transmission coefficient $\mathcal{T}_{0}(\omega)$  appears as a greybody
factor in  Eq.~(\ref{Eq: Thermal distribution}). For the lowest multipole, we
have 
\begin{eqnarray}\label{Eq: Minimally coupled transmission coefficient constant holonomy}
    \mathcal{T}_{0}(\omega)=\frac{16M^2\omega^2}{\chi_{0}^{-2}\left(\chi_{0}^{2}-4M^2\omega^2\right)^2+M^2\omega^2\alpha_{\tilde{\lambda}}^2(M)}\,,
\end{eqnarray}
where we implicitly defined
\begin{eqnarray}\label{Eq: Definition for alpha constant holonomy}
    \alpha_{\tilde{\lambda}}(M):=2\left(\gamma_{\tilde{\lambda}}(M)+\chi_{0}^{-1}\ln\left(\frac{2}{1+\chi_{0}\left(1-\frac{\chi_{0}^{2}\tilde{\lambda}}{2}\right)}\right)\right)\,.
\end{eqnarray}
To leading order, we regain the universality of black hole absorption
\cite{Mathur} because (\ref{Eq: Minimally coupled transmission coefficient constant holonomy}) is proportional to the horizon area. In the classical limit $\tilde{\lambda}\rightarrow0$ and $\chi_{0}\rightarrow1$ we have $\gamma_{\tilde{\lambda}}(M)\rightarrow1$
 and $\alpha_{\tilde{\lambda}}(M)\rightarrow2$, and thus their values are
 independent of the mass. The limit, therefore, results in the correct classical result, 
 \begin{eqnarray}
\lim_{\tilde{\lambda}\rightarrow0,\,\chi\rightarrow1}\mathcal{T}_{0}\rightarrow \frac{16M^{2}\omega^{2}}{1-4M^{2}\omega^{2}+16M^{4}\omega^{4}}\,.
\end{eqnarray}



\subsubsection{Decreasing holonomy function}
For a decreasing holonomy function analogous to the $\bar{\mu}$-scheme, the
line element is given by Eq.~(\ref{Eq: Line-element decreasing holonomy
  function}), and hence the tortoise coordinate is defined by
\begin{eqnarray}\label{Eq: Tortoise integration decreasing holonomy function}
    {\rm d}x_{*}=\frac{{\rm d}x}{\left(1-\frac{2M}{x}\right)\sqrt{1+\frac{\Delta}{x^{2}}\left(1-\frac{2M}{x}\right)}}\,.
\end{eqnarray}
The integral cannot be performed analytically, but there are suitable
approximations in specific regions. Near the horizon, one can expand the
right-hand side of (\ref{Eq: Tortoise integration decreasing holonomy
  function}), resulting in  the near-horizon tortoise coordinate
\begin{eqnarray}\label{Eq: Tortoise coordinates decreasing holonomy function}
    x_{*}(x)\simeq x+2M\ln\left(\frac{x-2M}{2M}\right)+\frac{\Delta}{2x}+\frac{3M-2x}{16x^{4}}\Delta^{2}\,.
\end{eqnarray}

We analyze the time-independent Schr\"odinger-like equation in both the
tortoise coordinate, (\ref{Eq: Reduced Schrodinger equation in tortoise
  coordinates}), and the radial coordinate, (\ref{Eq: Reduced Schrodinger
  equation in radial coordinates}), similar to the approach used for the
constant holonomy function. As shown in Fig.~\ref{Fig: Decreasing holonomy
  potential}, the qualitative profile of the potential remains similar to
that of the constant holonomy function. Notably, for the lowest multipole,
$l=0$, the peak remains the smallest among all modes. Given that the
lowest multipole dominates in the low-frequency regime, we
restrict our analysis to the \( s \)-wave.
\begin{figure}[h]
    \centering
\includegraphics[width=10cm]{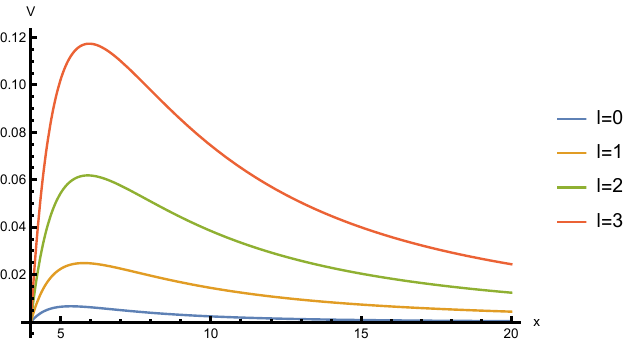}
    \caption{Potential as a function of the radial coordinate $x$ for decreasing holonomy function with $\Delta=1$ and $M=2$. The potential is vanishing at the horizon and flattens asymptotically.}
    \label{Fig: Decreasing holonomy potential}
\end{figure}

The near horizon solution (region I) is, as before, given by an ingoing plane wave  in terms of the tortoise coordinates in \eqref{Eq: Tortoise coordinates decreasing holonomy function}
\begin{eqnarray}\label{Eq: Region I decreasing holonomy function}
    \psi_{0}^{(I)}&\simeq& A_{I}e^{i\omega x_{*}}\nonumber \\
    &\approx&A_{I}\left[1+2iM\omega\left(1+\frac{\Delta}{8M^{2}}-\frac{\Delta^{2}}{512M^{4}}+\ln\left(\frac{x-2M}{2M}\right)\right)\right]\,.
\end{eqnarray}
The intermediate region (region II) remains dominated by the potential $V_{0}\gg\omega^{2}$, and, therefore, Eq.~(\ref{Eq: Reduced Schrodinger equation in radial coordinates}) is reduced to
\begin{eqnarray}\label{Eq: Region II decreasing holonomy function}
     \psi_{0}^{(II)}&=&A_{II}+B_{II}\int\frac{{\rm d}x}{x^{2}\left(1-\frac{2M}{x}\right)\sqrt{1+\frac{\Delta}{x^{2}}\left(1-\frac{2M}{x}\right)}}\,.
\end{eqnarray}
Evaluated near the horizon, this integral gives
\begin{eqnarray}
   \psi_{0}^{(II)}\simeq A_{II}+\frac{\Delta B_{II}}{7680M^{5}}\left(160M^{2}-3\Delta\right)+\frac{B_{II}}{2M}\ln\left(\frac{x-2M}{2M}\right)\,
\end{eqnarray}
which must match the solution in region I, namely Eq.~(\ref{Eq: Region I decreasing holonomy function}), yielding
\begin{eqnarray}
A_{II}&=&A_{I}\left(1+2iM\omega\left(1+\frac{\Delta}{12M^{2}}-\frac{3\Delta^{2}}{2560M^{4}}\right)\right)\,, \\
B_{II}&=&4iM^{2}\omega A_{I}\,.
\end{eqnarray}
Asymptotically, the solution (\ref{Eq: Region II decreasing holonomy function}) can be approximated as
\begin{eqnarray}\label{Eq: Region II asymptotically decreasing holonomy function}
    \psi_{0}^{(II)}&\simeq& A_{II}-\frac{B_{II}}{x}-\frac{M}{x^{2}}\left(1+\frac{1}{2}\lambda_{\infty}^{2}\chi^{2}\right)\approx     A_{II}-\frac{B_{II}}{x}\nonumber \\
&=&A_{I}\left(1+2iM\omega\left(1+\frac{\Delta}{12M^{2}}-\frac{3\Delta^{2}}{2560M^{4}}\right)-\frac{4iM^{2}\omega^{2}}{\rho}\right)\,,
\end{eqnarray}
where we have again used the new radial coordinate $\rho=\omega x$. This must also be matched with the solution in region III, where the exact solution is given by
\begin{eqnarray}\label{Eq: Region III decreasing holonomy function}
    \psi_{0}^{(III)}=\frac{e^{-i\rho}A_{III}}{\rho}-\frac{e^{i\rho}B_{III}}{2\rho}\,.
\end{eqnarray}
To ensure consistency with the solution in Eq.~(\ref{Eq: Region II asymptotically decreasing holonomy function}), we expand the above relation for \( \rho \ll 1 \). This expansion is the same as the one which was obtained in Eq.~(\ref{Eq: Region III match with II}). Consequently, we arrive at the following expressions for the coefficients:
\begin{eqnarray}\label{Eq: Coefficient of Region III in terms of I}
    A_{III}&=&\frac{iA_{I}}{2}\left[1+2iM\omega-4M^{2}\omega^{2}-iM\omega\left(-\frac{\Delta}{12M^{2}}+\frac{3\Delta^{2}}{2560M^{4}}\right)\right]\,, \nonumber \\
    B_{III}&=&A_{I}\left[1+2iM\omega-4M^{2}\omega^{2}+iM\omega\left(\frac{\Delta}{6M^{2}}-\frac{3\Delta^{2}}{1280M^{4}}\right)\right]\,.
\end{eqnarray}
Thus, the expressions for the coefficients in regions II and III are now fully determined in terms of those in region I.

Given the solutions in all three regions, we can evaluate the greybody
factor. This expression is defined through the current flux, Eq.~(\ref{eq:Current - minimal scalar complex}), which in the Schwarzschild gauge is reduced to Eq.~(\ref{Eq: Current MC Schawrzschild gauge Constant holonomy}). At the horizon, the relation simplifies to (\ref{Eq: Current MC at the horizon}). The asymptotic solution (\ref{Eq: Region III decreasing holonomy function}) consists of incoming and outgoing modes, and hence the current is given by
\begin{eqnarray}
    J_{\infty}=\frac{1}{2\omega}\left(4\left|A_{III}\right|^{2}-\left|B_{III}\right|^{2}\right)=J_{\mathcal{J}^{+}}-J_{\mathcal{J}^{-}}\,.
\end{eqnarray}
Using (\ref{Eq: Coefficient of Region III in terms of I}), one can show that
\begin{eqnarray}\label{Eq: Reflection plus transmission for decreasing holonomy MC}
J_{\infty}=-8M^{2}\omega\left|A_{I}\right|^{2} \,,
\end{eqnarray}
such that $1=\mathcal{R}_{0}(\omega)+\mathcal{T}_{0}(\omega)$: As expected,
the current is conserved also in this case. The transmission coefficient is given by
\begin{eqnarray}\label{Eq: Transmission coefficient exact decreasing holonomy function}
    \mathcal{T}_{0}(\omega)&=&\frac{16M^{2}\omega^{2}}{\left(1-4M^{2}\omega^{2}\right)^{2}+4M^{2}\omega^{2}\left(1+\frac{\Delta}{12M^{2}}-\frac{3\Delta^{2}}{2560M^{4}}\right)^{2}}\,,
\end{eqnarray}
from which we can infer the greybody factor. Note that the holonomy contribution reduces the magnitude of the greybody factor, which can be seen by approximating
\begin{eqnarray}
    \mathcal{T}_{0}(\omega)&\simeq& \frac{16M^{2}\omega^{2}}{1-8M^{2}\omega^{2}\left(1-\frac{1}{2}\left(1+\zeta_{\Delta}(M)\right)\right)}\nonumber\\
    &\approx& 16M^{2}\omega^{2}\left(1+8M^{2}\omega^{2}\left(1-\frac{1}{2}\left(1+\zeta_{\Delta}(M)\right)^{2}\right)\right)
\end{eqnarray}
and noticing that
\begin{eqnarray}
    \zeta_{\Delta}(M)=1+\frac{\Delta}{12M^{2}}-\frac{3\Delta^{2}}{2560M^{4}}
\end{eqnarray}
has positive leading-order terms.
In the classical limit, Eq.~(\ref{Eq: Transmission coefficient exact decreasing holonomy function}) gives
\begin{eqnarray}
    \lim_{\Delta\rightarrow0}\mathcal{T}_{0}=\frac{16M^{2}\omega^{2}}{1-4M^{2}\omega^{2}+16M^4\omega^4}\,.
\end{eqnarray}
Therefore, the holonomy correction \( \zeta_{\Delta}(M) \) plays a role in reducing the amount of radiation absorbed by the black hole. As we will see in the next section, this reduction influences the backreaction of Hawking radiation on the black hole's evaporation. 

\subsection{The non-minimally coupled scalar field}
The scalar field in EMG can be coupled in a nontrivial manner while still
preserving covariance. The dynamics of a scalar field coupled nonminimally to
an LQG black hole, expressed in terms of the radial
coordinates in Eq.~(\ref{Eq: Reduce nonminimal KG-equation in radial
  coordinates 1}), is similar to the classical case (\ref{Eq: Reduce classical KG-equation in
  radial coordinates 2}), the only
modification being in the first  term through $\beta^{-1}$. Consequently, the
modification will appear only through the intermediate region (region I),
while changes to the rest follow from imposing the matching conditions.

As in the previous subsection, the near-horizon structure is easier to solve
in the radial tortoise coordinate, in which the dynamics are given by
Eq.~(\ref{Eq: nonminimal Schrodinger-like coupling in tortoise
  coordinates}). Notice that for an arbitrary function $\lambda(x)$, both the damping and potential terms are negligible near the horizon\footnote{Both these terms are multiplied by $(1-2M/x)$.}. This means that the quantum effect appears only through the tortoise coordinate $x_{*}$, and the details of the damping term as well as the potential are negligible in this case.


\subsubsection{Constant holonomy function}
The line element for this case is given by Eq.~(\ref{Eq: Constant holonomy line element}), where the tortoise coordinates are exact Eq.~(\ref{Eq: Tortoise radial coordinates constant holonomy}). The potential (\ref{Eq: nonminimally coupling potential}) and damping factor (\ref{Eq: nonminimally coupling damping term}) simplify to
\begin{eqnarray}
    V_{l}(x)&=&\left(1+\tilde{\lambda}^{2}\left(1-\frac{2M}{x}\right)\right)\left(1-\frac{2M}{x}\right)\left(\frac{l(l+1)}{x^{2}}+\frac{2M}{x^{3}}\right)\,, \\
    \zeta(x)&=&-\frac{\chi^{2}\tilde{\lambda}^{2}}{\sqrt{1+\tilde{\lambda}^{2}\left(1-\frac{2M}{x}\right)}}\left(1-\frac{2M}{x}\right)\frac{M}{x^{2}}\,.
\end{eqnarray}
It can be easily shown that the peak of the potential decreases as the multipole number decreases, which means that the dominant contribution in the low-frequency regime is $l=0$.

Near the horizon (region I), we use the Eq.~(\ref{Eq: nonminimal Schrodinger-like coupling in tortoise coordinates}). Since both the potential and damping terms are negligible, the equation reduces to
\begin{eqnarray}
   \left(\frac{{\rm d}^{2}}{{\rm d}x_{*}^{2}}+\omega^{2}\right)x\psi_{0}^{(I)}=0\,,
\end{eqnarray}
with the solution
\begin{eqnarray}\label{Eq: Region I Near the horizon nonminimal coupling}
    \psi_{0}^{(I)}&\simeq& A_{I}e^{i\omega x_{*}} \nonumber \\
    &\approx&A_{I}\left(1+\frac{2iM\omega}{\chi_{0}}\left(\chi_0\gamma_{\tilde{\lambda}}(M)+\ln\left(\frac{x-2M}{M\left(1+\chi_{0}^{2}\left(1-\tilde{\lambda}^{2}\right)\right)}\right)\right)\right)\,.
\end{eqnarray}
In the intermediate region (region II), the potential and damping terms are dominant, and hence it is easier to tackle the Schr\"odinger equation written in terms of the radial coordinate (\ref{Eq: Reduce nonminimal KG-equation in radial coordinates 1}), which for $l=0$ reduces to
\begin{eqnarray}
    \partial_{x}\left[x^{2}\left(1-\frac{2M}{x}\right)\partial_{x}\psi_{0}^{(II)}\right]=0\,.
\end{eqnarray}
Since the time-derivatives are sub-dominant in this case, the holonomy effects are
absent in this region. The solution to the above equation can be found in
closed form, given by
\begin{eqnarray}\label{Eq: Region II nonminimal}
    \psi_{0}^{(2)}=A_{II}+\frac{B_{II}}{2M}\ln\left(\frac{x-2M}{x}\right)\,,
\end{eqnarray}
which equals the classical solution. After matching, however, $A_{II}$ and $B_{II}$ are nonclassical due to the quantum corrections we found in region I. Near the horizon, the above solution can be approximated as
\begin{eqnarray}
    \psi_{0}^{(2)}\simeq A_{II}-\frac{B_{II}}{2M}\ln2M+\frac{B_{II}}{2M}\ln\left(x-2M\right)\,.
\end{eqnarray}
Matching this solution with Eq.~(\ref{Eq: Region I Near the horizon nonminimal coupling}), we obtain
\begin{eqnarray}
A_{II}&=&A_{I}\left[1+2iM\omega\left(\gamma_{\tilde{\lambda}}+\chi_{0}^{-1}\ln\left(\frac{2}{\left(1+\chi_{0}^{2}\left(1-\tilde{\lambda}^{2}\right)\right)}\right)\right)\right]\,, \\
    B_{II}&=&\frac{4iM^{2}\omega}{\chi_{0}}A_{I}\,.
\end{eqnarray}

The asymptotic form of the solution (\ref{Eq: Region II nonminimal}) can be approximated as 
\begin{eqnarray}\label{Eq: Region II asymptotic nonminimal}
    \psi_{0}^{(II)}\simeq A_{II}-\frac{B_{II}\omega}{\rho}\,,
\end{eqnarray}
where the new radial coordinate is $\rho=\omega x$. The asymptotic region
(region III) has the same form as the minimally coupled and classical case,
thanks to Eq.~(\ref{Eq: Asymptotic approximation}), hence the solution is also
the same as (\ref{Eq: Region III}):
\begin{eqnarray}\label{Eq: Region III nonminimal}
     \psi_{0}^{(III)}=\frac{e^{-i\rho}A_{III}}{\rho}-\frac{ie^{i\rho}B_{III}}{2\rho}\,.
\end{eqnarray}
We match this expression with the asymptotic form of the solution in region II, Eq.~(\ref{Eq: Region II asymptotic nonminimal}) after performing the expansion 
\begin{eqnarray}\label{Eq: Region III match with II nonminimal}
    \psi_{0}^{(III)}\simeq -i\left(A_{III}+\frac{iB_{III}}{2}\right)+\frac{1}{\rho}\left(A_{III}-\frac{iB_{III}}{2}\right)\,,
\end{eqnarray}
with the result
\begin{eqnarray}\label{Eq: Asymprotic coefficients}
    A_{III}&=&\frac{iA_{I}}{2\chi_{0}}\left[\chi_{0}-4M^{2}\omega^{2}+2iM\omega\left(\gamma_{\tilde{\lambda}}(M)\chi_{0}+\ln\left(\frac{2}{1+\chi_{0}\left(1-\frac{\tilde{\lambda}^{2}\chi_{0}}{2}\right)}\right)\right)\right]\,,\\
     B_{III}&=&\frac{A_{I}}{\chi_{0}}\left[\chi_{0}+4M^{2}\omega^{2}+2iM\omega\left(\gamma_{\tilde{\lambda}}(M)\chi_{0}+\ln\left(\frac{2}{1+\chi_{0}\left(1-\frac{\tilde{\lambda}^{2}\chi_{0}}{2}\right)}\right)\right)\right]\,.
\end{eqnarray}
These expressions agree with the minimally coupled form because the asymptotic form of the solutions in region II agree.

Just as in the minimally coupled case, we need to check the conservation law for this case as well. The current flux (\ref{eq:Current - nonminimal scalar complex}) in the Schwarzschild gauge reads
\begin{eqnarray}
    J^{x}_{({\rm NMC})}&=&iN\chi_{0}\frac{\left(E^{x}\right)^{3/2}}{E^\varphi}\left(\psi^{*}_{0}\psi'_{0}-\psi_{0}\phi^{*'}_{0}\right) \nonumber \\
    &=&i\chi_{0}x^{2}\left(1-\frac{2M}{x}\right)\left(\psi^{*}_{0}\psi'_{0}-\psi_{0}\phi^{*'}_{0}\right)\,.
\end{eqnarray}
Computing the flux near the horizon gives
\begin{eqnarray}
    J_{\mathcal{H}}&=&-8\pi \chi_{0} M^{2}\omega \left|A_{I}\right|^{2}\,.
\end{eqnarray}
The asymptotic waves consist of ingoing and outgoing modes (\ref{Eq: Region III nonminimal}), with the total flux
\begin{eqnarray}
    J_{\infty}&=&\frac{2\chi_{0}}{\omega}\left(\left|A_{III}\right|^{2}-\frac{\left|B_{III}\right|^{2}}{4}\right)=J_{\mathcal{J}^{+}}-J_{\mathcal{J}^{-}}\nonumber \\
    &=&-8\pi \chi_{0} M^{2}\omega \left|A_{I}\right|^{2}=J_{\mathcal{H}}\,,
\end{eqnarray}
where $J_{\mathcal{J}^{+}}$ denotes the outgoing modes reaching future null infinity, while $J_{\mathcal{J}^{-}}$ stands for the ingoing modes from past null infinity. 

The greybody factor, defined as the transmission coefficient, is then given by
\begin{eqnarray}\label{Eq: nonminimally coupled transmission coefficient constant holonomy}
    \mathcal{T}_{0}(\omega)&=&\frac{J_{\mathcal{H}}}{J_{\mathcal{J}^{-}}} \nonumber \\
    &=&\frac{16M^2\omega^2}{\left(\chi_{0}^{2}-4M^2\omega^2\right)^2+M^2\omega^2 \chi_{0}^{2}\alpha_{\tilde{\lambda}}^2(M)}\,.
\end{eqnarray}
This result  differs from the minimally coupled case, (\ref{Eq: Minimally coupled transmission coefficient constant holonomy}), due to the presence of the global factor $\chi^{2}$. Recall that $\chi$ and $\alpha_{\tilde{\lambda}}$ have been defined in (\ref{Eq: Definition for gamma constant holonomy}) and (\ref{Eq: Definition for alpha constant holonomy}), respectively.

\subsubsection{Decreasing holonomy function}
The line element for a decreasing holonomy function of $\bar{\mu}$ type  is given by Eq.~(\ref{Eq: Line-element decreasing holonomy function}). The potential (\ref{Eq: nonminimally coupling potential}) and damping factor (\ref{Eq: nonminimally coupling damping term}) simplify to 
\begin{eqnarray}
    V_{l}(x)&=&\frac{\beta(x)}{x^{2}}\left(1-\frac{2M}{x}\right)\left(l(l+1)+\frac{2M}{x}\right)\,, \label{Eq: potential term in nonminimally coupled} \\
    \zeta(x)&=&-\frac{\Delta}{2x^{2}\sqrt{\beta(x)}}\left(1-\frac{2M}{x}\right)\left(\frac{2M}{x^2}+\left(1-\frac{2M}{x}\right)\left(\ln \frac{\Delta}{x^2}\right)'\right)\,,\label{Eq: damping term in nonminimally coupled}
\end{eqnarray}
with
\begin{eqnarray}
    \beta(x)=1+\frac{\Delta}{x^2}\left(1-\frac{2M}{x}\right)\,.
\end{eqnarray}
As shown earlier for a decreasing holonomy function, it is necessary to
perform an expansion in order  to integrate the tortoise coordinate in the
near-horizon region.

Near the horizon (region I), both the potential (\ref{Eq: potential term in
  nonminimally coupled}) and the damping terms (\ref{Eq: damping term in
  nonminimally coupled}) are negligible once again. We therefore obtain 
solutions similar to the minimally-coupled case in region I, given by (\ref{Eq:
  Region I decreasing holonomy function}). The intermediate
region (region II) is dominated by the potential and the damping terms
($V,\zeta\gg\omega^2$). Accordingly, Eq.~(\ref{Eq: Reduce nonminimal
  KG-equation in radial coordinates 1}) reduces to
\begin{eqnarray}
    \frac{1}{x}\beta\left(1-\frac{2M}{x}\right)\partial_{x}\left[x^{2}\left(1-\frac{2M}{x}\right)\psi_{0}^{(II)}\right]=0
\end{eqnarray}
with solution
\begin{eqnarray}\label{Eq: solution Region II NMC}
\psi_{0}^{(II)}&=&A_{II}+\int\frac{{\rm d}x}{x^2\left(1-\frac{2M}{x}\right)} \nonumber \\
&=&A_{II}+\frac{B_{II}}{2M}\ln\left(1-\frac{2M}{x}\right)\,.
\end{eqnarray}
Near the horizon, this function can be written as
\begin{eqnarray}
\psi_{0}^{(II)}&\approx&A_{II}+\frac{B_{II}}{2M}\ln\left(\frac{x-2M}{2M}\right)
\end{eqnarray}
in order to match with Eq.(\ref{Eq: Region I decreasing holonomy function}) in
region I, determining
\begin{eqnarray}
    A_{II}&=&A_{I}\left(1+2iM\omega \left(1+\frac{\Delta}{8M^2}-\frac{\Delta^{2}}{512M^4}\right)\right)\,, \\
    B_{II}&=&4iM^2\omega A_{I}\,.
\end{eqnarray}
Asymptotically, the solution ({\ref{Eq: solution Region II NMC}) has the form
\begin{eqnarray}\label{Eq: Asymptotic form of solution Region II NMC}
    \psi_{0}^{(II)}&\approx&A_{I}\left[1+2iM\omega \left(1+\frac{\Delta}{8M^2}-\frac{\Delta^{2}}{512M^4}\right)-\frac{4iM^2\omega^2}{\rho}\right]\,.
\end{eqnarray}
where we have used the usual redefinition $\rho=\omega x$. 

This function must match with the asymptotic solution in region III. Since
both the potential and damping terms are asymptotically suppressed, the
solution is again similar to the minimally coupled one, given by (\ref{Eq: Region III decreasing holonomy function}). After some algebra, we obtain 
\begin{eqnarray}\label{Eq: Region III coefficients in NMC}
    A_{III}&=&\frac{iA_{I}}{2}\left[1+2iM\omega-4M^{2}\omega^{2}-iM\omega\left(-\frac{\Delta}{4M^{2}}+\frac{3\Delta^{2}}{256M^{4}}\right)\right]\,, \nonumber \\
    B_{III}&=&A_{I}\left[1+2iM\omega-4M^{2}\omega^{2}+iM\omega\left(\frac{\Delta}{4M^{2}}-\frac{3\Delta^{2}}{256M^{4}}\right)\right]\,,
\end{eqnarray}
relating the coefficients in region III to $A_{I}$.

Proceeding as in the previous subsection, the conservation laws can be seen to
hold with the decreasing holonomy function for a nonminimally coupled scalar. Using (\ref{Eq: Region III coefficients in NMC}), the transmission coefficient in this case is given by
\begin{eqnarray}\label{Eq: Transmission coefficient exact decreasing holonomy function NMC}
     \mathcal{T}_{0}(\omega)&=&\frac{J_{\mathcal{H}}}{J_{\mathcal{J}^{-}}}\nonumber \\
     &=&\frac{16M^{2}\omega^{2}}{\left(1-4M^{2}\omega^{2}\right)^{2}+4M^{2}\omega^{2}\left(1+\frac{\Delta}{8M^{2}}-\frac{\Delta^{2}}{512M^{4}}\right)^{2}}\,,
\end{eqnarray}
which provides information about the greybody factors and retains the universality of black hole emission rates.


\section{Black hole evaporation}


We are now ready to explore the thermal properties of LQG black holes modeled
by the EMG formalism. As is the case for Schwarzschild black holes, we will
see that an asymptotic observer has different notions of vacua before and
after gravitational collapse. If the observer waits long enough, they will see
the collapse of the star until a stable black hole is formed. Having initially
been in
their standard Minkowski vacuum, they will then detect a swarm of particles,
the distribution of which is given by (\ref{Eq: Thermal
  distribution}). Consequently, the spacetime can no longer be assumed to be a
vacuum solution of the field equations, and the radiation will backreact on
the geometry. In this section, we will develop the covariant stress-energy
tensor due to the backreaction for our emergent line element, utilizing the techniques developed in \cite{Davies-Fulling-Unruh, Fulling}. For a review, see \cite{Brout, Fabbri-Salas}. 

The significance of introducing this covariant stress-energy tensor lies in
its ability to quantify the extent to which backreaction influences black hole
thermodynamics. In the case of the Schwarzschild solution, using this
covariant tensor as a matter source in the Einstein field equations leads to
the ADM energy getting smaller in magnitude. In LQG, the total stress-energy
momentum of the system, is the sum of the matter contribution and the
gravitational contribution through the effective Einstein tensor. This concept of net stress-energy tensor was introduced in \cite{Idrus1}, and we will use it here to study the black hole emission rates. 

The construction of a consistent stress-energy tensor relies on the covariance
of the physical fields. Due to this covariant nature of our line-element, we can proceed with the machinery of \cite{Davies-Fulling-Unruh, Fulling} to evaluates it. The difficulty of solving the Klein--Gordon equation in
curved spacetimes makes it necessary to employ the near-horizon regime for an
approximate definition of the stress-energy tensor. Covariance allows us to
use coordinate transformations and to define the stress-energy tensor at any
local point in spacetime.

Near the horizon, the nonangular part of the spacetime metric is conformally flat,
\begin{eqnarray}\label{Eq: Conformal 2d metric}
    {\rm d}s^{2}_{(2)}=-C(u,v){\rm d}u{\rm d}v
\end{eqnarray}
where $C(u,v)$ is the conformal factor and the line element  is written in null coordinates $\left(u,v\right)$. 
In this two-dimensional reduction, the covariant stress-energy tensor is given by
\cite{Fulling, Fabbri-Salas}
\begin{eqnarray}
\braket{\Psi\left|T_{\pm\pm}\right|\Psi}&=&\frac{1}{24\pi}\left(\frac{1}{C}\frac{\partial^{2}C}{\partial \left(x^{\pm}\right)^{2}}-\frac{3}{2C^{2}}\left(\frac{\partial C}{\partial x^{\pm}}\right)^{2}\right)+\braket{\Psi\left|:T_{\pm\pm}(x^{\pm}):\right|\Psi} \label{Eq: Covariant energy tensor 1} \\
\braket{\Psi\left|T_{+-}\right|\Psi}&=&-\frac{R_{(2)}}{96\pi}C\,. \label{Eq: Covariant energy tensor 2}
\end{eqnarray}
For the sake of convenience, we defined $x^{+} = u$ and $x^{-} = v$ in these
equations. The quantity $R_{(2)}$ denotes the Ricci scalar associated with the
metric in Eq.~(\ref{Eq: Conformal 2d metric}). The state $\ket{\Psi}$
represents an arbitrary quantum state of the scalar field. Among the
components of the covariant stress-energy tensor, Eq.~(\ref{Eq:
  Covariant energy tensor 2}) is purely geometrical. On the other hand,
Eq.~(\ref{Eq: Covariant energy tensor 1}) contains not only geometric
contributions but also an additional term that arises from the normal-ordering
of the stress-energy tensor, the details of which can be found in the
Appendix~\ref{Appendix B}. The latter depend on the chosen vacuum state, while
the geometry contribution, usually referred to as the vacuum-polarization term, is independent of the matter contribution and appears
even for a vacuum state.

In what follows, we review the commonly used vacuum states in the context of
the Schwarzschild solution. We then adopt the same choice of vacuum in order to investigate the black hole evaporation in the EMG scenario.


\subsection{2d Covariant stress-tensor in EMG background}
The two dimensional conformally flat line element depends on the global factor $\chi$.
Different choices have an effect on the vacuum polarization. In this
subsection we provide the derivation for a constant global factor
$\chi=\chi_{0}$. The case with $\chi=1/\sqrt{1+\lambda^{2}(x)}$ requires a
renormalization, for which a detailed discussion is provided in Appendix~\ref{Appendix: Covariant tensor components}.

The nonangular line element corresponding to $\chi=\chi_{0}$ is given by
\begin{eqnarray}\label{Eq: Conformal metric nonEuclidean}
    {\rm d}s^{2}_{(2)}=-\left(1-\frac{2M}{x}\right)\chi_{0}^{2}{\rm d}u{\rm d}v\equiv -C_{2}(u,v){\rm d}u{\rm d}v
\end{eqnarray}
where the null coordinates $(u, v)$ are defined by Eq.~(\ref{Eq: Null coordinates definition for arbitrary mass}). The vacuum state naturally associated with these coordinates is the Boulware state, which is the vacuum most appropriate for studying the exterior region of a stable, spherically symmetric, and static star. In this background, the scalar field can be expanded in terms of the mode basis $(e^{-i\omega u}, e^{i\omega u})$ for the right-moving sector and $(e^{-i\omega v}, e^{i\omega v})$ for the left-moving sector. The components of the vacuum polarization tensor for this geometry are given by
\begin{eqnarray}
    \braket{0_{\rm B}|T_{uu}|0_{\rm B}}&=&\braket{0_{\rm B}|T_{vv}|0_{\rm B}}=\frac{M\chi_{0}^{2}}{48 \pi x^{5}}\left[x\left(3M-2x\right)+\lambda\left(x-2M\right)^{2}\left(-2\lambda+x\lambda'\right)\right],\, \\
    \braket{0_{\rm B}|T_{uv}|0_{\rm B}}&=&-\frac{M\left(x-2M\right)\chi_{0}^{2}}{48\pi x^{5}}\left[2x+\lambda^{2}\left(2x-5M\right)+x\lambda \lambda'\left(2M-x\right)\right]\,,
\end{eqnarray}
which reduce to their classical values in the limit $\lambda \to 0$ and
$\chi_{0} \to 1$. In the limit $M \to 0$, the covariant tensor components
vanish altogether. In all cases, the components vanish at infinity, such that
an asymptotic observer will not detect any particles.

Although the Boulware vacuum is a proper vacuum state for the exterior, it is
not well-defined at the horizon; see \cite{Fabbri-Salas} for a detailed
discussion. This behavior is due to the divergence of the null coordinates $u$ and
$v$ at the horizon, which in turn is a consequence of $x_{*}$ diverging at the
horizon.  A vacuum state that is well defined throughout spacetime can be
obtained by utilizing the Kruskal coordinates $\left(U,V\right)$, given by
exponentiated versions of $\left(u,v\right)$. The scalar field in this set of
coordinates can then be expanded in the basis
$\left(e^{i\omega U},e^{-i\omega U}\right)$ for the right-moving modes and
$\left(e^{i\omega V},e^{-i\omega V}\right)$ for the left-moving modes. The
vacuum state annihilated by these sets of modes is known as the Hartle--Hawking
state $\ket{0_{\rm H}}$. Using the definition of normal ordering given in
Appendix~\ref{Appendix B} and the relation (\ref{Eq: Krsukal coordinates for
  flat space}) between $\left(U,V\right)$ and $\left(u,v\right)$, the
normal-ordered component $T_{uu}$ computed in the vacuum $\ket{0_{\rm H}}$ is
obtained as
\begin{eqnarray}\label{Eq: HH normal ordering}
    \braket{0_{\rm H}\left|:T_{uu}:\right|0_{\rm H}}&=&\left(\frac{{\rm d}U}{{\rm d}u}\right)^{2}\braket{0_{\rm H}\left|:T_{UU}:\right|0_{\rm H}}-\frac{1}{24\pi}\left(\frac{{\rm d}^{3}U/{\rm d}u^3}{{\rm d}U/{\rm d}u}-\frac{3}{2}\left(\frac{{\rm d}^{2}U/{\rm d}u^{2}}{{\rm d}U/{\rm d}u}\right)^{2}\right) \nonumber \\ 
    &=&\frac{\kappa^{2}}{48 \pi }\,.
\end{eqnarray}
Following similar steps, one can show that
\begin{eqnarray}
    \braket{0_{\rm H}\left|:T_{vv}:\right|0_{\rm H}}&=&\frac{\kappa^{2}}{48 \pi }\label{Eq: HH normal ordering 2}\,, \\
    \braket{0_{\rm H}\left|:T_{uv}:\right|0_{\rm H}}&=&0\,.\label{Eq: HH normal ordering 3} 
\end{eqnarray}
Adding the normal-ordered stress tensor  (\ref{Eq: HH
  normal ordering})--(\ref{Eq: HH normal ordering 3}) to the diagonal
components
\begin{eqnarray}\label{Eq: HH EMG nonEuclidean}
    \braket{0_{\rm H}\left|T^{(2)}_{uu}\right|0_{\rm H}}=\braket{0_{\rm H}\left|T^{(2)}_{vv}\right|0_{\rm H}}&=&\left(1-\frac{2M}{x}\right)^{2}\left[\frac{\kappa^2}{48 \pi }\left(1+\frac{4M}{x}+\frac{12M^{2}}{x^{2}}\right)\right.\nonumber \\&&\left.+\frac{M\chi_{0}^{2}}{48\pi x^{3}}\lambda^{2}\left(x\left(\ln\lambda\right)'-2\right)\right]
\end{eqnarray}
results in the complete stress-energy components in the Hartle--Hawking
state. They vanish at the horizon, while an observer at infinity
will detect thermal equilibrium with incoming and outgoing fluxes of energy at
rates $\kappa^{2}/(48\pi)$. One, thus, concludes that this vacuum state can be
associated with an eternal black hole.

The two vacua recalled here correspond to static black hole solutions. To
study the evaporation process, we consider a black hole as produced
by a dynamical process and study the propagation of a scalar field on its exterior region. As discussed in detail in Sec.~\ref{Sec: LQGBH Hawking distribution}, the scalar field is prepared in its vacuum state in the far past $\mathcal{I}^{-}$, and hence
\begin{eqnarray}
    {\rm d}s^{2}_{\mathcal{I}^{-}}=-{\rm d}u_{\rm in}{\rm d}v_{\rm
  in}+x^{2}(u_{\rm in},v_{\rm in}){\rm d}\Omega^{2}\,.
\end{eqnarray}
After some time, the matter collapse will produce a black hole with mass $M$,
and hence the line element is given by our emergent metric (\ref{Eq: Null
  line-element of constant chi}). The relation between late time coordinates $(u,v)$ and
early time coordinates $(u_{\rm in},v_{\rm in})$ is given by (\ref{Eq: Typical logarithmic
  relation}).

The vacuum state $\ket{0_{\rm in}}$ described in Sec.~\ref{Subsection:
  Bogoliubov} as the vacuum state at early times at $\mathcal{J}^{-}$ is
annihilated by the basis modes
$\left(e^{-i\omega u_{\rm in}},e^{i\omega u_{\rm in}}\right)$ and
$\left(e^{-i\omega v_{\rm in}},e^{i\omega v_{\rm in}}\right)$. The
normal-ordered contributions to the stress-energy tensor in this vacuum are
given by
\begin{eqnarray}
    \braket{0_{\rm in}\left|:T_{uu}:\right|0_{\rm in}}&=&-\frac{1}{24\pi}\left(\frac{{\rm d}^{3}v/{\rm d}u^3}{{\rm d}v/{\rm d}u}-\frac{3}{2}\left(\frac{{\rm d}^{2}v/{\rm d}u^{2}}{{\rm d}v/{\rm d}u}\right)^{2}\right)= \frac{\kappa^{2}}{48\pi}\,, \label{Eq: Normal ordering uu in-state}\\
     \braket{0_{\rm in}\left|:T_{vv}:\right|0_{\rm in}}&=&0\,. \label{Eq: Normal ordering vv in-state}
\end{eqnarray}
The $vv$-component vanishes because the late-time coordinate $v$ is linear in
terms of the early coordinates $v_{\rm in}$, while the normal-ordered terms consist of second and third order derivatives. 
The vacuum associated with the far past is the in-state vacuum $\ket{0_{\rm in}}$. In this vacuum state, the outgoing flux is given by 
\begin{eqnarray}\label{Eq: Collapse uu component}
    \braket{0_{\rm in}\left|T^{(2)}_{uu}\right|0_{\rm in}}=\braket{0_{\rm H}\left|T^{(2)}_{uu}\right|0_{\rm H}}
\end{eqnarray}
and the ingoing flux by
\begin{eqnarray}
    \braket{0_{\rm in}\left|T^{(2)}_{vv}\right|0_{\rm in}}=\braket{0_{\rm B}\left|T^{(2)}_{vv}\right|0_{\rm B}}.
\end{eqnarray}
An asymptotic observer at $\mathcal{J}^{+}$ will detect a net flux of energy 
\begin{eqnarray}\label{Eq: uu in components EMG}
    \lim_{x\to\infty}\braket{0_{\rm in}\left|T_{uu}\right|0_{\rm in}}=\frac{\kappa^{2}}{48\pi} \quad {\rm and}\quad
    \lim_{x\to\infty}\braket{0_{\rm in}\left|T_{vv}\right|0_{\rm in}}&=&0
\end{eqnarray}
of thermal outgoing particles, showing the black hole is losing energy. Energy conservation  dictates that this energy must be extracted from the black hole, which can be seen by instead taking the limit at the horizon
\begin{eqnarray}
\lim_{x\to2M}\braket{0_{\rm in}\left|T_{uu}\right|0_{\rm in}}=0\quad {\rm and}\quad \nonumber 
\lim_{x\to2M}\braket{0_{\rm in}\left|T_{vv}\right|0_{\rm in}}=-\frac{\kappa^{2}}{48\pi}\,.
\end{eqnarray}
which gives an ingoing flux of negative energy compensating (\ref{Eq: uu
  in components EMG}), that is, to be associated with the black hole losing energy.

\subsection{The covariant backscattering tensor: s-wave approximation}\label{sub 6.2}

The backreaction can likewise be derived via the net stress-energy tensor formalism. Starting from the  functional \eqref{eq:EH functional} and the Klein--Gordon functional \eqref{eq:KG functional}, one obtains the conserved stress-energy tensor \eqref{eq:Net stress-energy tensor}, modulo the ambiguity encoded in the constant parameter $\alpha$.

As a minimal model for the backreaction, one may retain the background spacetime form while allowing the mass parameter to vary $M\to M(u)$. The flux of energy detected at $\mathcal{I}^{+}$ can be computed through the semiclassical net stress energy tensor (\ref{eq:Net stress-energy tensor})
\begin{eqnarray}\label{Eq: Net stress energy tensor}
\bar{T}_{\mu\nu}=\braket{T^{(4)}_{\mu\nu}}-\frac{\alpha}{8\pi}G_{\mu\nu}\,,
\end{eqnarray}
where $G_{\mu\nu}$ is the Einstein tensor of the emergent metric, $\alpha$ is
the free constant parameter allowed by EMG, see Eq.~(\ref{eq:Net stress-energy
  tensor}). Moreover, $\braket{T^{(4)}_{\mu\nu}}$ is the covariant 4-dimensional stress-energy tensor defined as \cite{Brout, Fabbri-Salas}
\begin{eqnarray}\label{Eq: 4d covariant tensor}
\braket{T^{(4)}_{\mu\nu}}=\frac{\delta^{A}_{\mu}\delta^{B}_{\nu}}{4\pi q_{\theta\theta}}\braket{T^{(2)}_{AB}}
\end{eqnarray}
which relates the four-dimensional tensor $\braket{T^{(4)}_{\mu\nu}}$ to the
two-dimensional tensor $\braket{T^{(2)}_{AB}}$ if the line element is decomposed as
\begin{eqnarray}
    {\rm d}s^{2}=g_{\mu\nu}{\rm d}x^{\mu}{\rm d}x^{\nu}=g_{AB}(y){\rm d}y^{A}{\rm d}y^{B}+q_{\theta\theta}(y){\rm d}\Omega^{2}\,.
\end{eqnarray}
The covariant tensor in Eq.~(\ref{Eq: 4d covariant tensor}) is conserved by construction, as a consequence of metric compatibility, $\nabla_{\mu} q_{\theta\theta}=0$, and the conservation of $\braket{T^{(2)}_{AB}}$, which has no angular components.

One may exploit the covariant nature of the stress-energy tensor
(\ref{Eq: 4d covariant tensor}) and derive the corresponding components in
coordinate charts
$\{\tilde{x}^{A}\}=\left(\tilde{x}^{1},\tilde{x}^{2}\right)=\left(t,x\right)$,
using the tensor transformation
\begin{eqnarray}
\tilde{T}^{(2)}_{CD}=\left(\frac{\partial x^{A}}{\partial \tilde{x}^{C}}\right)\left(\frac{\partial x^{B}}{\partial \tilde{x}^{D}}\right)T^{(2)}_{AB}\,.
\end{eqnarray}
A coordinate transformation from $\{x^{A}\}=\left(u,v\right)$ to $\{\tilde{x}^{A}\}=\left(t,x\right)$ implies
\begin{eqnarray}
\braket{T^{(2)}_{tt}}&=&\frac{1}{\chi_{0}^{2}}\left(\braket{T^{(2)}_{uu}}+\braket{T^{(2)}_{vv}}+\braket{T^{(2)}_{uv}}\right) \\
\braket{T^{(2)}_{xx}}&=&\frac{1}{\chi_{0}^{2}}\left(\braket{T^{(2)}_{uu}}+\braket{T^{(2)}_{vv}}-\braket{T^{(2)}_{uv}}\right)\left(\frac{{\rm d}x_{*}}{{\rm d}x}\right)^{2} \\
\braket{T^{(2)}_{tx}}&=&\frac{1}{\chi_{0}^{2}}\left(\braket{T^{(2)}_{vv}}-\braket{T^{(2)}_{uu}}\right)\left(\frac{{\rm d}x_{*}}{{\rm d}x}\right).
\end{eqnarray}
We are not interested in their complete local expressions, but only in their
asymptotic values for $x\to\infty$ because we are considering a black hole as
the final outcome of a collapsing star. Hence, we will compute the energy flux in the "in" vacuum state $\ket{0_{\rm in}}$.

The amount of energy flux at infinity can be computed through the
$tx$-component of the net stress-energy tensor. Classically, when Einstein
equations are satisfied, the net stress-energy tensor vanishes, $\bar{T}_{tx}=0$, and the Einstein tensor yields
\begin{eqnarray} 
    G_{tx}=-\frac{2}{x^{2}}\left(1-\frac{2M(t)}{x}\right)^{-1}\dot{M}(t)\,.
\end{eqnarray}
Therefore, the $tx$-component of Eq.~(\ref{Eq: Net stress energy tensor}) when $\bar{T}_{tx}=0$ can be rewritten as
\begin{eqnarray}\label{Eq: Energy flux classical in tx}
    \dot{M}(t)=-\frac{\kappa^{2}}{48\pi\alpha}=-\frac{1}{768\pi\alpha M^{2}}
\end{eqnarray}
which matches the classical result when $\alpha=1$. 

When $\lambda\neq0$, the local components of the stress-tensor are given by
\begin{eqnarray}
\braket{T^{(2)}_{tt}}&=&\frac{1}{\chi_{0}^{2}}\left(\braket{T^{(2)}_{uu}}+\braket{T^{(2)}_{vv}}+\braket{T^{(2)}_{uv}}\right)\,, \\
\braket{T^{(2)}_{xx}}&=&\frac{\left(\braket{T^{(2)}_{uu}}+\braket{T^{(2)}_{vv}}-\braket{T^{(2)}_{uv}}\right)}{\chi_{0}^{2}\left(1-2M/x\right)^2\left(1+\lambda^{2}(x)\left(1-2M/x\right)\right)}\,,\\
\braket{T^{(2)}_{tx}}&=&\frac{\braket{T^{(2)}_{vv}}-\braket{T^{(2)}_{uu}}}{\chi^{3}_{0}\left(1-2M/x\right)\left(1+\lambda^{2}(x)\left(1-2M/x\right)\right)},
\end{eqnarray}
while the Einstein tensor is given by
\begin{eqnarray} \label{Gtx}
    G_{tx}=\frac{2\left(1+2\lambda^{2}(x)\left(1-2M/x\right)\right)}{x^{2}\left(1-2M/x\right)\left(1+\lambda^{2}(x)\left(1-2M/x\right)\right)}\dot{M}(t).
\end{eqnarray}
Computed in the ``in'' vacuum state at infinity, the $tx$-components give us a conservation relation
\begin{eqnarray}
    \frac{{\rm d}E_{\rm T}}{{\rm d}t}=-\frac{\kappa^{2}}{48\pi \chi_{0}^{2}}-\frac{1+2\lambda_{\infty}^{2}}{1+\lambda^{2}_{\infty}} \alpha \dot{M}(t)\,.
\end{eqnarray}
where we have use the fact that
\begin{eqnarray}\label{Eq: tx-component at infinity}
   \lim_{x\to\infty} \braket{0_{\rm in}|T_{tx}^{(2)}|0_{\rm
  in}}=-\frac{\kappa^{2}}{48\pi\chi_{0}^{2}}=-\frac{1}{768\pi M^{2}}\,. 
\end{eqnarray}
Energy conservation $E_{\rm T}\big|_{\infty}\to {\rm const.}$ implies
\begin{eqnarray} \label{Mdotalpha}
    \frac{1+2\lambda_{\infty}^{2}}{1+\lambda^{2}_{\infty}} \alpha \dot{M}(t)=-\frac{\kappa^{2}}{48\pi \chi_{0}^{2}}=-\frac{1}{768\pi M^{2}}\,.
\end{eqnarray}
Using (\ref{Eq: ADM mass in general}) for the relationship between the
parameter $M$ and the ADM mass, we obtain
\begin{eqnarray}\label{eq:Net stress-energy backreaction}
   \alpha \dot{M}_{\rm ADM}=-\frac{\kappa^{2}}{48\pi \chi_{0}^{2}}=-\frac{1}{768\pi M^{2}}\,.
\end{eqnarray}

\subsection{Canonical backreaction}

While the net stress-energy approach offers an intuitive way of understanding
the evaporation process, the freedom to choose a constant $\alpha$ prevents us from deriving an unambiguous evaporation rate.
Here, we present a canonical approach based on Hamilton's equations of motion, which generate the fundamental time evolution of the mass as a phase-space function and are free of ambiguities.

In the Schwarzschild gauge
\begin{equation}\label{eq:Schwarzschild gauge}
    N^x=0
    \quad,\quad
    E^x=x^2\,,
\end{equation}
the vacuum solution to the equations of motion is given by
\begin{eqnarray}
    \bar{K}_\varphi=\bar{K}_x=0
    \quad,\quad
    \bar{E}^\varphi = \frac{x}{\sqrt{1-\frac{2 M_0}{x}}}
    \quad,\quad
    \bar{N} = \sqrt{1-\frac{2 M_0}{x}}
    \\
    \bar{\phi} = 0 \quad,\quad \bar{P}_\phi = 0
    \quad,\quad
    \bar{\phi}^* = 0 \quad,\quad \bar{P}_\phi^* = 0
    \,,
\end{eqnarray}
with constant $M_0$.
We used the bar to identify these functions as representing the background vacuum.
In the presence of a test field
\begin{equation}
    \phi = \epsilon \delta \phi
    \quad,\quad
    P_\phi = \epsilon \delta P_\phi
    \quad,\quad
    \phi^* = \epsilon \delta \phi^*
    \quad,\quad
    P_\phi^* = \epsilon \delta P_\phi^*
\end{equation}
where we use the constant $\epsilon$ for book-keeping of the perturbative
order, the dynamical solution to the gravitational variables receives
corrections of second order. We parametrize them as
\begin{equation}
    K_\varphi = \epsilon^2 \delta K_\varphi
    \quad,\quad K_x = \epsilon^2 \delta K_x\,,
\end{equation}
\begin{eqnarray}
    E^\varphi = \frac{x}{\sqrt{1-\frac{2 \left(M_0+\epsilon^2\delta M\right)}{x}}}
    \quad,\quad
    \bar{N} = \sqrt{1-\frac{2 \left(M_0+\epsilon^2\delta M\right)}{x}}
    \,,
\end{eqnarray}
while we preserve (\ref{eq:Schwarzschild gauge}) as the gauge condition.

To second order in $\epsilon$, the constraints and equations of motion for nonminimal coupling determine
\begin{eqnarray}\label{eq:d K_phi}
    \delta K_\varphi &=& 0
    \\
    \delta K_x &=& \frac{\delta P_\phi \delta \phi' + \delta P_\phi^* (\delta \phi^*)'}{2 x}
    \label{eq:d K_x}
\end{eqnarray}
and require
\begin{eqnarray}
    \delta M' = x^2 \left(1-\frac{2 M_0}{x}\right) |\delta \phi'|^2
    - \left(1+\lambda^2 \left(1-\frac{2 M_0}{x}\right)\right) \left(1-\frac{2 M_0}{x}\right) \frac{|\delta P_\phi|^2}{x^2}
    \,,
\end{eqnarray}
assuming $V,\partial_\phi V,\partial_{\phi^*} V\to 0$ as well as $\phi,\phi^*\to0$.
Furthermore, using $\dot{\cal M}=\{{\cal M},H[N]+H_x[N^x]\}$, where ${\cal M}$ is the vacuum mass observable (\ref{eq:Weak observable in simple case - nonperiodic}), and setting ${\cal M}=M_0+\epsilon^2 \delta M$, we obtain
\begin{equation}
    \delta \dot{M} = \chi_0 \left(1-\frac{2 M_0}{x}\right)^2 \left(1+\lambda^2 \left(1-\frac{2 M_0}{x}\right)\right) \left(\delta P_\phi \delta \phi'+\delta P_\phi^* (\delta \phi^*)'\right)
    \,.
\end{equation}

Now setting $\epsilon\to1$, as well as $\delta \phi\to \phi$ and $\delta P_\phi\to P_\phi$ since they are test fields, we obtain two equations for the mass,
\begin{equation}\label{eq:M' minimal}
    M' = \left(1-\frac{2 M}{x}\right) \left[\left(1+\lambda^2 \left(1-\frac{2 M}{x}\right)\right) \frac{|P_\phi|^2}{x^2}
    + x^2 |\phi'|^2\right]
    \,,
\end{equation}
and
\begin{equation}\label{eq:Mdot canonical}
    \dot{M} = \chi_0 \left(1-\frac{2 M}{x}\right)^2 \left(1+\lambda^2 \left(1-\frac{2 M}{x}\right)\right) \left(P_\phi \phi'+P_\phi^* (\phi^*)'\right)
    \,.
\end{equation}
Asymptotically, we can focus on the simpler equation
\begin{equation}
    \lim_{x\to\infty} \dot{M} = \chi_0^{-1} \lim_{x\to\infty} \left(P_\phi \phi'+P_\phi^* (\phi^*)'\right)
    \propto T_{tx}
    \,.
\end{equation}
Evaluating the equations of motion for nonminimal coupling, this equation reads
\begin{eqnarray}\label{eq:dot M nonminimal}
    \lim_{x\to\infty} \dot{M} &=& x^2 \left((\phi^*)' \dot{\phi}+\phi' \dot{\phi}^*\right)
    \nonumber\\
    &=& x^2\phi' \dot{\phi} = 4 \pi x^2 T_{tx}
    \,,
\end{eqnarray}
where we used the reality conditions for the scalar field, which requires the
replacement $\phi\to\phi/\sqrt{2}$, in the second line. (The factor of $4\pi$
results from restoring coefficients of $\sqrt{4\pi}$ absorbed in the canonical
variables $\phi$ and $P_\phi$ after angular integrations in the
spherically symmetric system. After absorbing these coefficients, the basic brackets (\ref{eq:Basic
  brackets -
  scalar}) follow.)
The semiclassical equation uses the expectation value
$\braket{T_{tx}^{(4)}}$ for the stress-energy contribution, asymptotically given by
(\ref{Eq: tx-component
  at infinity}) and (\ref{Eq: 4d covariant tensor}), yielding
\begin{eqnarray}\label{eq:dot M nonminimal - semiclassical}
    \lim_{x\to\infty} \dot{M} &=& 4\pi x^2 \braket{T_{tx}^{(4)}}
    \nonumber\\
    &=& - \frac{\kappa^{2}}{48\pi\chi_0^2}
    \,.
\end{eqnarray}

Using the minimally coupled system, the same equations (\ref{eq:d K_phi}), (\ref{eq:d K_x}), and (\ref{eq:Mdot canonical}) are obtained, while Eq.~(\ref{eq:M' minimal}) is replaced by
\begin{equation}\label{eq:M' nonminimal}
    M' = \left(1-\frac{2 M_0}{x}\right) \sqrt{1+\lambda^2 \left(1-\frac{2 M_0}{x}\right)} \left(\frac{|P_\phi|^2}{x^2}+x^2 |\phi'|^2\right)\,.
\end{equation}
Therefore, evaluating the equations of motion for minimal coupling, we obtain
\begin{equation}\label{eq:dot M minimal}
    \lim_{x\to\infty} \dot{M} = \chi_0^{-1} x^2 \left((\phi^*)' \dot{\phi}+\phi' \dot{\phi}^*\right)\,,
\end{equation}
and the semiclassical equation for a real scalar field yields
\begin{equation}\label{eq:dot M minimal - semiclassical}
    \lim_{x\to\infty} \dot{M} = - \frac{\kappa^{2}}{48\pi\chi_0^3}\,.
\end{equation}

Inserting the results
(\ref{eq:dot M nonminimal - semiclassical}) and (\ref{eq:dot M minimal -
  semiclassical})  into Eq.~\eqref{Mdotalpha}
allows us to fix the parameter $\alpha$ as
\begin{eqnarray}
    \alpha=\frac{1}{\chi_{0}(1+2\lambda_{\infty}^{2})}
\end{eqnarray}
for minimal coupling and
\begin{eqnarray}
    \alpha=\frac{1}{\chi^{2}_{0}(1+2\lambda_{\infty}^{2})}
\end{eqnarray}
for nonminimal coupling. In both cases,
$\chi_0=1/\sqrt{1+\lambda_{\infty}^2}$.  This factor approaches unity in the
case of monotonically decreasing holonomy function with $\lambda_\infty\to0$, which includes the
$\bar{\mu}$-scheme as a special case. 


\subsection{The black hole energy loss}
As shown in the previous sections, our model implies a thermal distribution with a temperature given by the surface gravity.
Consequently, we can associate the quanta of Hawking radiation to a thermal density matrix
\begin{eqnarray}
\rho=\sum_{N_{\omega}=0}^{\infty}\frac{e^{-2\pi N\omega \kappa^{-1}}}{1-e^{2\pi\omega \kappa^{-1}}}\ket{N_{\omega}}\bra{N_{\omega}}\,,
\end{eqnarray} 
where $\ket{N_{\omega}}$ is the Fock state for $N_{\omega}$ Hawking quanta with frequency $\omega$. In what follows, we sketch the main results while the detailed derivation is provided in Appendix (\ref{Appendix B}).

We now consider modes that escape from the black hole horizon and get transmitted
to $\mathcal{I}^{+}$, given by the right-moving modes of the expansion (See Eq.\eqref{Eq: Scalar field expansion to right and left modes} in the Appendix (\ref{Appendix B})):
\begin{eqnarray}
    \phi_{\rm R}=\int\frac{{\rm d}\omega }{\sqrt{4\pi \omega}}\left(\mathcal{T}_{l}(\omega) c_{\omega}e^{-i\omega u}+\mathcal{T}^{*}_{l}(\omega) c^{\dagger}_{\omega}e^{i\omega u}\right)\,.
\end{eqnarray}
Using (\ref{Eq: momentum-energy tensor components}), the normal-ordered stress-energy tensor (\ref{Normal ordered correlator}) is given by
\begin{eqnarray}
\braket{N_{\omega}|:T_{uu}:|N_{\omega}}=\frac{N_{\omega} \ \omega \left|\mathcal{T}_{0}(\omega)\right|^2}{2\pi}\,.
\end{eqnarray}
The energy flux can then be computed as
\begin{eqnarray}\label{Eq: General energy flux}
    \frac{{\rm d}\bar{E}}{{\rm d}u}=-{\rm Tr}\left[:T_{uu}:\rho\right]=-\frac{1}{2\pi}\int_{0}^{\infty}\frac{\left|\mathcal{T}_{0}(\omega)\right|^{2}\omega{\rm d}\omega}{e^{2\pi \omega \kappa^{-1}}-1}\,.
\end{eqnarray}
The total energy loss is derived by using the $tx$-component of the stress-energy tensor,
\begin{eqnarray}\label{Eq: Generic energy loss integral}
    \frac{{\rm d}E}{{\rm d}t}=:\alpha \frac{{\rm d}M_{\rm ADM}}{{\rm d}t}=-{\rm Tr}\left[:T_{tx}:\rho\right]=-\frac{1}{2\pi}\int_{0}^{\infty}\frac{\left|\mathcal{T}_{0}(\omega)\right|^{2}\omega{\rm d}\omega}{e^{2\pi \omega \kappa^{-1}}-1} \,.
\end{eqnarray}
where $\mathcal{T}_{0}(\omega)$ is the greybody factor. Below, we will analyze the flux of ADM energy asymptotically for both constant and decreasing holonomy functions.

In the constant holonomy case, the line element is given by Eq.~(\ref{Eq:
  Asier-Brizuela line-element}), implying a spacetime geometry determined by
$M$ and $\tilde{\lambda}$ as independent parameters. Fig.~\eqref{Fig: Emission rates 1} shows the energy loss for the minimally
coupled case with various values of the parameter $\tilde{\lambda}$, while
Fig.~\eqref{fig: Emission rates 2} shows a comparison between the minimal,
non-minimal, and classical cases. The transmission coefficients for the minimal
and nonminimal cases are given in
Eqs.~\eqref{Eq: Minimally coupled transmission coefficient constant holonomy} and
\eqref{Eq: nonminimally coupled transmission coefficient constant holonomy},
respectively.
 The qualitative behavior of the amount of energy loss is very
similar to the classical case: The evaporation process will continue
indefinitely, until the mass is completely depleted.  

According to the derivation in EMG, the line element in Eq.~(\ref{Eq: Asier-Brizuela line-element}) is parametrized by $M$ and $\tilde{\lambda}$. Recent proposals 
discuss the dynamical evaporation scenario in the case of constant holonomy $\lambda(x)$, with matter minimally coupled to the gravitational degrees of freedom. They postulates
the minimum radius $x_{\tilde{\lambda}}$ as the independent parameter instead
of $\tilde{\lambda}$, using it as a system scale that is also independent of
$M$. It is then possible to consider a backreaction process in which
$x_{\tilde{\lambda}}$ stays constant while $M$ decreases. Eventually, the
Schwarzschild radius $2M$ will
then approach the minimal radius, and evaporation must stop completely. This
proposal led the authors to suggest a remnant scenario.

While such a change of parameterization is mathematically possible because the
defining equation (\ref{Eq: Minimum radius for constant holonomy}) of $x_{\tilde{\lambda}}$ can be
solved for $\tilde{\lambda}$, it is physically misleading because $M$ and
$\tilde{\lambda}$ play different roles: The modification parameter
$\tilde{\lambda}$ appears in the Hamiltonian constraint that defines a
covariant gravitational theory, while $M$ appears in a given solution of this
theory. Alternatively, one may express $M$ as the observable (\ref{eq:Weak
  observable in simple case - nonperiodic}) without considering specific
solutions. But this role of $M$ is also different from $\tilde{\lambda}$
because it turns $M$ into a phase-space function rather than a constant
parameter.

Considering $x_{\tilde{\lambda}}$ and $M$
independent parameters requires $\tilde{\lambda}$ to be a function of the former,
\begin{eqnarray}\label{eq:lambda function of M}
    \tilde{\lambda}=\sqrt{\frac{x_{\tilde{\lambda}}}{2M-x_{\tilde{\lambda}}}}\,.
\end{eqnarray}
Therefore,
\begin{equation}\label{eq:chi0 function of M}
    \chi_0=\frac{1}{\sqrt{1+\tilde{\lambda}^2}}=\sqrt{1-\frac{x_{\tilde{\lambda}}}{2M}}
\end{equation}
is a also function of $M$ and $x_{\tilde{\lambda}}$. Depending on how one
interprets $M$, as a parameter on the solution space or as a phase-space
function, the $M$-dependence of $\tilde{\lambda}$ and $\chi_0$ is problematic
because these parameters appear in the Hamiltonian constraint. Therefore, one
would either deal with a solution-dependent constraint, which is not
meaningful, or the Hamiltonian constraint would be a mixture of different
derivative terms to higher polynomial orders through the spatial derivatives
in (\ref{eq:Weak observable in simple case - nonperiodic}). Since these terms
include only a specific form of higher polynomials in derivatives, they are
not complete in the sense of effective field theory. It is then questionable
that any effects derived from such a theory may be considered reliable because
they could easily change if other derivative terms of the same order are
included. By contrast, the parameterization used here, based on EMG, makes sure that all
covariant terms up to a given derivative order are included in the Hamiltonian constraint (at least in the non-minimal coupling).

Moreover, if $M$ is interpreted as the dynamical value of the phase-space
function (\ref{eq:Weak observable in simple case - nonperiodic}), then
$\tilde{\lambda}$, according to the proposal of
\cite{Asier-radiation,Asier-gray, Menezes}, should not be constant but a function on the phase
space.
As argued in \cite{Norbert-polymerisation}, this requires that the
postulate be made at the kinematical level where the phase space is
available, which in turn redefines the Hamiltonian vector field and hence the
dynamics: Therefore, a system with phase-space dependent $\tilde{\lambda}$ is
not equivalent to  the original one. In addition, there is a pathology of such a
system because $\tilde{\lambda}$ increases unboundedly as $M$ decreases, and it
diverges as $2M\to x_{\tilde{\lambda}}$. This limiting value is
defined as the remnant mass in \cite{Asier-radiation,Asier-gray,Menezes}, which is
approached in the asymptotic future.

Furthermore, an assumption in \cite{Asier-radiation, Asier-gray,Menezes} is that, to
zeroth order (imposing a unit greybody factor), the black hole satisfies the
Stefan--Boltzmann law with no modification.  As we show below, this is not
consistent with their postulate that the minimum radius $x_{\tilde{\lambda}}$
is held fixed while $M$ and $\tilde{\lambda}$ vary.

To analyze the arguments of \cite{Asier-radiation, Asier-gray, Menezes}, let us assume
that the holonomy parameter $\tilde{\lambda}$ follows the dynamical trajectory
implied by the backreaction effect, \textit{i.e.,} the holonomy parameter $\tilde{\lambda}$ evolves due to the time-changing mass $M(t)$.
In this case, the Einstein tensor component $G_{tx}$ of the line element (\ref{Eq: Asier-Brizuela line-element}) remains classical,
\begin{eqnarray}
    G_{tx}=\frac{2\dot{M}(t)}{x^{2}\left(1-\frac{2M}{x}\right)}\,.
\end{eqnarray}

By analogy with the classical case, the \cite{Asier-radiation, Asier-gray,Menezes}
then assume that the radiated power is proportional to the surface gravity,
\begin{eqnarray}
    \dot{M}=-\frac{\kappa^{2}}{48\pi}
\end{eqnarray}
where in their case
\begin{eqnarray}
    \kappa=\frac{\chi_{0}(M)}{4M}=\frac{1}{4M}\sqrt{1-\frac{x_{\tilde{\lambda}}}{2M}}\,,
\end{eqnarray}
using (\ref{eq:chi0 function of M}).
Therefore, it was claimed that an inflection point appears at
$x_{\tilde{\lambda}} = 2M$. This statement relies on the assumption that the
Stefan--Boltzmann law, which is derived from the Einstein equations, remains
unmodified.

A careful treatment, however, reveals inconsistencies in this line of
arguments. First, the parameter $\alpha$ in (\ref{eq:Net stress-energy
  tensor}) cannot be assumed constant in this case since it depends on
$\tilde{\lambda}$ for both types of coupling. Treating $\alpha$ as constant
leads to a violation of energy conservation. Second, even if one proceeds
under this assumption, the asymptotic $tx$-component of Hawking radiation is given
by (\ref{Eq: tx-component at infinity}), and thus the energy flux becomes
\begin{eqnarray}\label{Eq: Asier-David energy loss}
    \dot{M}=-\frac{1}{1536\pi M^{3}}\frac{2M+x_{\tilde{\lambda}}}{\left(1-x_{\tilde{\lambda}}/2M\right)^{3/2}}
\end{eqnarray}
for the minimal coupling, while
\begin{eqnarray}\label{Eq: Asier-David energy loss 2}
    \dot{M}=-\frac{1}{1536\pi M^{2}}\frac{2M+x_{\tilde{\lambda}}}{2M-x_{\tilde{\lambda}}}
\end{eqnarray}
for the nonminimal coupling. Both expressions are computed by assuming that
$\tilde{\lambda}=\tilde{\lambda}(M)$ while $x_{\tilde{\lambda}}$ is held constant. 
Equations~(\ref{Eq: Asier-David energy loss})--(\ref{Eq: Asier-David energy
  loss 2}) represent the modified Stefan--Boltzmann relation, incorporating
the holonomy effects together with the assumption of a time-dependent holonomy
parameter $\tilde{\lambda}$. Both relations exhibit a divergence as $2M \to
x_{\tilde{\lambda}}$, and therefore no remnant can be formed. This behavior resembles the classical case: The black hole energy diverges, with the singular point shifted to $x_{\tilde{\lambda}}$. 

\begin{figure}[h]\label{Fig: Constant holonomy energy loss}
    \begin{subfigure}{0.35\textwidth}
        \centering
\includegraphics[width=7.5cm]{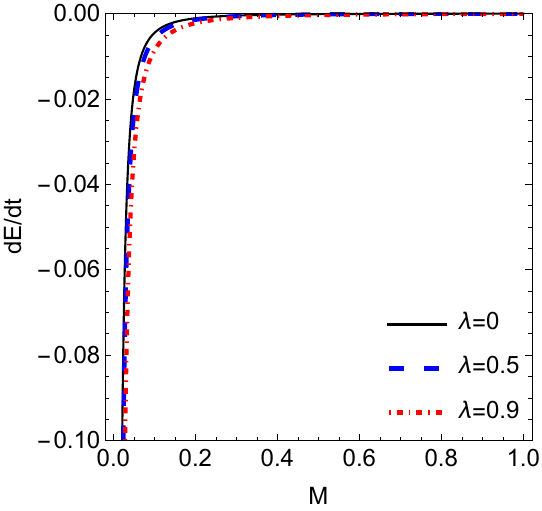}
        \caption{}
        \label{Fig: Emission rates 1}
    \end{subfigure}
    \hfill
    \begin{subfigure}{0.35\textwidth}
        \centering
  \includegraphics[width=7.5cm]{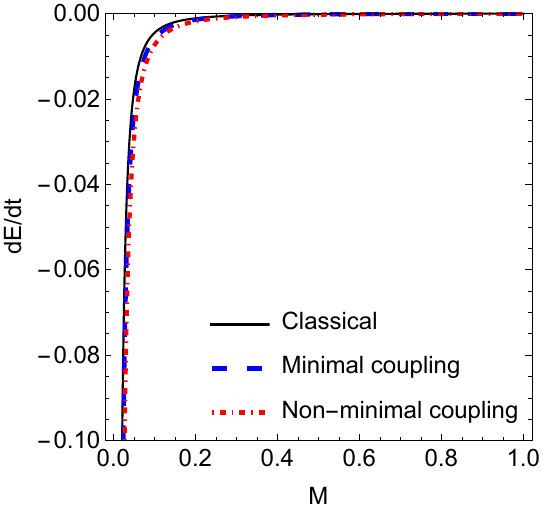}
        \caption{}
        \label{fig: Emission rates 2}
    \end{subfigure}
    \hfill
    \caption{(a) Energy emission for a classical black hole represented by a
      black solid line ($\tilde{\lambda}=0$) and the minimally coupled constant
      holonomy correction by a dashed line ($\tilde{\lambda}=0.5$) and dot-dashed line
      $\tilde{\lambda}=0.9$. (b) The energy emission for a constant holonomy function with $\tilde{\lambda}=0.5$ for minimal coupling (solid line) and nonminimal coupling (dashed line). The nonminimal coupling implies  slower rates in comparison with the minimal coupling. However, both are faster in comparison with the classical evaporation ($\tilde{\lambda}=0$).}
\end{figure}

Rather than promoting the minimum radius $x_{\tilde{\lambda}}$ to the
fundamental scale, adopting a decreasing holonomy function analogous to the
$\bar{\mu}$-scheme of loop quantum cosmology implies a consistent slow-down of
the evaporation process. In this approach, we do not need to impose an \textit{ad hoc}
assumption that the minimum radius remains constant.  Instead, we use the length scale that naturally appears
  when we write a unitless parameter $\lambda$ as a decreasing function of the
  length parameter $x$. For instance, one may heuristically refer to the area
  gap $\Delta$ in models of loop quantum gravity. The holonomy
function is then given by $\lambda(x)=\sqrt{\Delta}/x$, where
$\Delta$ is a constant.  The corresponding energy loss is shown in
Fig.~\ref{Fig: Decreasing holonomy function energy flux}. The transmission
coefficients are provided in Eq.~(\ref{Eq: Transmission coefficient exact
  decreasing holonomy function}) for the minimally coupled case and in
Eq.~(\ref{Eq: Transmission coefficient exact decreasing holonomy function
  NMC}) for the nonminimally coupled case. These results were previously
discussed in \cite{Idrus-short thermal BH}.

The evaporation rates (\ref{Eq: Generic energy loss integral}) can be computed
using (\ref{Eq: Transmission coefficient exact decreasing holonomy function})
and (\ref{Eq: Transmission coefficient exact decreasing holonomy function
  NMC}) as the transmission coefficient for the minimal and nonminimal
coupling case, respectively. The numerical plot of the radiated ADM energy is
shown in Fig.~\ref{Fig: Decreasing holonomy function energy flux}. As opposed
to the constant-holonomy case, here the black hole is undergoing a change of
thermodynamical properties for both the minimal and the nonminimal coupling
case. As can be seen, the black hole energy flux reaches a minimum value at a
sub-Planckian scale around $M_{r}\sim 0.15 \sqrt{\Delta}$, before
transitioning to a phase of  slowed-down evaporation. The minimum point
indicates a stable remnant. However, as shown in \cite{EMGscalarQNM} and
mentioned in \cite{Idrus-short thermal BH}, in the case of nonminimal coupling there will be a gravitational instability due to the changing sign of the imaginary component of the quasinormal $s$-mode frequency at the critical mass $M_{c}\sim 0.57 \sqrt{\Delta}>M_{r}$. Hence, near this critical mass, the nonminimally coupled system becomes unstable, jump-starting a white-hole emission \cite{Idrus-short thermal BH}.

\begin{figure}[h]
    \centering
\includegraphics[width=10cm]{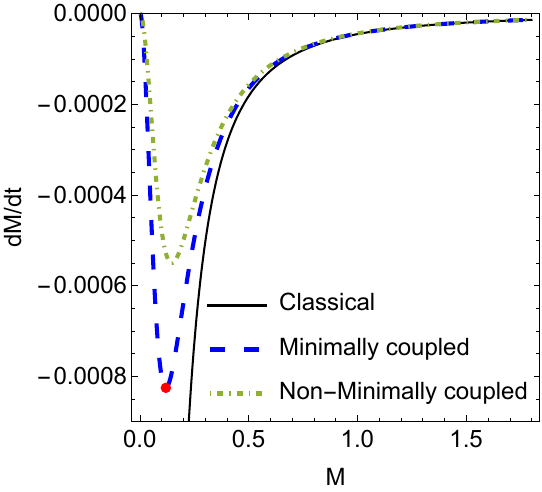}
    \caption{Energy flux for the classical case (solid line), the minimally
      coupled (dashed line), and the nonminimally coupled (dot-dashed) scalar
      field with a decreasing holonomy function, using $\Delta=1$. The minimum appears for both minimally and nonminimally coupled
      cases, with the latter evaporating more slowly and reaching a smaller value of energy extraction.}
    \label{Fig: Decreasing holonomy function energy flux}
\end{figure}

\section{Conclusion}
Emergent modified gravity (EMG) provides a tool to incorporate corrections
motivated by loop quantum gravity into a consistent black-hole line element by
requiring the constraint brackets to be free of anomalies and their gauge
transformations to be on-shell equivalent to coordinate transformations of the
spacetime metric. The model can be solved generically for arbitrary holonomy
function $\lambda(x)$, showing the robustness of the model \cite{Idrus1}. Even
though current evaluations of EMG start by restricting
the derivatives to be up to second order, as in the classical case, 
covariant modifications appear that are not of higher-curvature form.

In the present work, we primarily investigated the Hawking thermal
spectrum in a holonomy-corrected black hole background and examined how such
corrections affect the fate of black hole evaporation. These holonomy
corrections can be viewed as effective nonperturbative contributions arising
from loop quantum gravity. To address these questions, we treat a scalar field
as a quantum field propagating on the EMG background. In EMG, however, matter
fields do not couple directly to the classical geometry but rather to the
underlying gravitational degrees of freedom, formulated canonically. As shown in \cite{Bojowald-Duque
  scalar}, the coupling between matter fields and gravity is not unique:
Anomaly freedom and general covariance permit both minimal and nonminimal
couplings, even when the order of spatial derivatives in the Hamiltonian
matches that of general relativity. Here, we demonstrated that black
hole thermality is preserved in both coupling scenarios.

In the classical Schwarzschild solution, the black hole radiates away all its
energy. An asymptotic observer at future lightlike infinity will detect
particle creation due to the initial vacuum state at past infinity
$\ket{0_{\rm in}}$ no longer being the vacuum state in the far future, once
the black hole is formed. Another way to perceive this process is by
considering the tunneling process of particles from the black hole
interior. The probability of an $s$-wave  propagating through the
classically forbidden region around the horizon, turns out to correspond to the Boltzmann
factor, where the temperature is given by the particle-creation process. We
found that the Hawking distribution remains unchanged, using
several routes just as in general relativity. This agreement can be traced
back to the strict preservation of general covariance.

The LQG background (\ref{Eq: Effective line-element in the full case}) shares
the same horizon location with the classical Schwarzschild geometry. Since
thermal properties of a black hole are determined by the horizon, both
backgrounds yield the same temperature up to a global factor $\chi_{0}$. This
factor approaches unity if $\lambda(x)$ decreases
monotonically. Consequently, the Hawking temperature receives only a slight
modification when the holonomy parameter is constant, taking the form
$T_{\rm H} = \frac{\chi_0^{2}}{8\pi M}$, where
$\chi_0^{-1} = \sqrt{1 + \tilde{\lambda}^{2}}$ and $\tilde{\lambda}$ is the
holonomy length that parameterizes modifications in models of loop quantum
gravity.  We therefore conclude that, to leading order in the computation of
the Hawking distribution, holonomy corrections do not contribute
directly. Instead, these corrections enter through the scalar-field
potential. The effects of this modified potential can be probed in at least
two ways: through black-hole quasinormal modes and through the greybody
factor. The former has been analyzed in detail in \cite{EMGscalarQNM}, while
in this work we focussed on the latter and its implications for the
black-hole evaporation process.

Due to the Schrodinger-like nature of the wave equation for radial modes of the scalar field, the greybody factors can be computed using the standard procedure of matching conditions in quantum mechanics, as was initially developed by Unruh \cite{Unruh2}.
In this approach, spacetime is divided into three regions where the
wavefunction can be solved explicitly. Requiring that the solution be
continuous at the junctions, one is able to obtain the solutions in those
regions parametrized by a single coefficient. The solution in each region can
be solved explicitly by restricting to the modes with energy less than the
background thermal energy. Consequently, modes with the lowest multipole
($l=0$) dominate. We study the greybody factors both for the minimally coupled
and nonminimally coupled scalar field as well as for both $\mu_{0}$ and $\bar{\mu}$
schemes corresponding to constant and decreasing holonomy functions,
respectively. Unlike the latter, the $\mu_{0}$ scheme does not result in significant
deviations to the final expression of the greybody factors from the classical
behavior, which can be traced back to the fact that the constant $\tilde{\lambda}$ parameter is
dimensionless and  does not scale with the mass of the black hole. On the
other hand, in the $\bar{\mu}$-scheme, the area gap parameter $\Delta$ in the holonomy function $\lambda(x)=\sqrt{\Delta}/x$
has the same dimension as $M^2$, and hence it is always accompanied by $M^2$ in
equations such as (\ref{Eq: Transmission coefficient exact decreasing holonomy
  function}) and (\ref{Eq: Transmission coefficient exact decreasing holonomy
  function NMC}) for the minimally and nonminimally coupled scalar
field. Despite this difference, both maintain the universality of the black
hole absorption rates. 

Finally, we discuss the backreaction effects in the emergent spacetime. We treat the backreaction within a semiclassical framework, employing both a canonical analysis and an approach based on the net stress-energy tensor. The former is obtained by considering perturbations around the vacuum solution, while the latter is derived from an analysis of the action functional. This action should not be interpreted as generating the equations of motion; rather, it is introduced as a purely kinematical functional constructed from the emergent spacetime metric.
Such a definition introduces an ambiguity, since matter is not coupled directly to the metric (which is emergent), but instead to the gravitational degrees of freedom through the Hamiltonian constraint. However, this ambiguity can be resolved by imposing consistency with the canonical analysis. Within the net stress-energy tensor formalism, the distinction between minimal and non-minimal coupling is encoded in a coefficient $\alpha$, which appears as an arbitrary constant in the net stress–energy tensor \eqref{eq:Net stress-energy tensor}. This constant depends on the asymptotic value of $\lambda$ and approaches unity for both types of coupling in the case of a decreasing holonomy function, including the $\bar{\mu}$-scheme.

Using the semiclassical approach described above, we analyze the behavior of black hole evaporation by computing the associated energy loss. In the $\mu_{0}$-scheme, the evaporation exhibits a qualitatively classical behavior, leading to a complete depletion of the black hole mass. By contrast, in the $\bar{\mu}$-scheme the evaporation process is halted as the mass approaches the Planck scale, potentially resulting in the formation of a sub-Planckian black hole remnant with mass $M_{r}\sim 0.15\,M_{\mathrm{Pl}}$.
In the presence of a non-minimally coupled scalar field, however, the quasinormal modes become unstable at a critical mass $M_{c}\sim 0.67\,M_{\mathrm{Pl}}$, with $M_{c}>M_{r}$, which instead opens up the possibility of a transition to a white-hole phase. Details of this behavior can be found in Refs.~\cite{EMGscalarQNM, Idrus-short thermal BH}. From the perspective of the information loss paradox, it is important to note that, within this semiclassical framework, the dominant portion of the black hole mass and correspondingly of the semiclassical entropy is radiated away before the onset of this Planckian regime. As a result, neither the formation of a sub-Planckian remnant nor a possible late-time white-hole transition, as described here, by itself constitutes a definitive resolution of the information loss problem.
Moreover, as discussed in \cite{Idrus-short thermal BH} following the approach in \cite{Parikh:2024zmu}, quantum fluctuations of the horizon geometry are expected to become comparable to the mean geometric quantities in the vicinity of the Planck scale.  This outcome suggests that the
effective spacetime description employed in our analysis may no longer be sufficient. The behavior identified here should therefore be interpreted as signaling the limitations of the semiclassical and effective framework, rather than as a complete characterization of the ultimate endpoint of black hole evaporation.

\section*{Acknowledgments}
IHB is supported by the Indonesia Endowment Fund for Education (LPDP) grant from the Ministry of Indonesia. The work of MB and ED was supported in part by NSF grant PHY-2206591. SB is supported in part by the start-up grant from the Indian Statistical Institute. SB acknowledges support from the Blaumann Foundation for a grant on ‘Black Holes in Loop Quantum Gravity’.

\appendix

\section{Flat space in the zero-mass limit}\label{Apendix}

\subsection{Spacetime structure with nonconstant $\chi$}
The line element (\ref{Eq: Effective line-element in the full case}) has  ambiguities when $\lambda$ is not constant. In this appendix we discuss $\chi(E^{x})=1/\sqrt{1+\lambda^{2}(E^{x}})$. This choice corresponds to the case with flat spatial metric in the $M\to0$ limit.
The line element then has the form
\begin{eqnarray}\label{Eq: line-element with Euclidean space}
    {\rm d}s^{2}=-\left(1-\frac{2M}{x}\right)\frac{1+\lambda^{2}}{1+\lambda_{\infty}^{2}}{\rm d}t^2+\frac{{\rm d}x^{2}}{\left(1-\frac{2M}{x}\right)\left(1-\frac{2M\lambda^2}{(1+\lambda^2)x}\right)}+x^{2}{\rm d}\Omega^2
\end{eqnarray}
The line element can be expressed in terms of Kruskal coordinates. For this, it is useful to first define the Eddington--Finkelstein coordinates and determine the surface gravity.

The four-acceleration $a_{\mu}=\nabla_{\mu}\ln \sqrt{-g_{tt}}$ required to
keep a particle  at a constant radial position $x$ is given by 
\begin{eqnarray}\label{Eq: Four acceleration component in flat space coordinates}
    a_{\mu}=\delta_{\mu}^{x}\left(\frac{M}{x^{2}}\left(1-\frac{2M}{x}\right)^{-1}+\frac{\lambda \lambda'}{1+\lambda^{2}}\right)\,.
\end{eqnarray}
Using the definition of the surface gravity as (\ref{Eq: General Surface
  Gravity}), we obtain
\begin{eqnarray}\label{Eq: Surface gravity flat space}
    \kappa=\frac{\chi_{0}}{4M}\,,
\end{eqnarray}
the same  as Eq.~(\ref{Eq: Surface gravity flat time}).

We now define the null coordinates
\begin{eqnarray}
    u=\chi_{0}^{-1}\left(t-x_{*}\right)\quad {\rm and}\quad v=\chi_{0}^{-1}\left(t+x_{*}\right)
\end{eqnarray}
where 
\begin{eqnarray}
    {\rm d}x_{*}=\frac{\sqrt{1+\lambda^{2}_{\infty}}{\rm d}x}{\sqrt{1+\lambda^{2}(x)\left(1-2M/x\right)}\left(1-2M/x\right)}
\end{eqnarray}
is the radial tortoise coordinate.
In these coordinates, the line element takes the form
\begin{eqnarray}\label{Eq: Null line-element of nonconstant chi}
    {\rm d}s^{2}=-\left(1-\frac{2M}{x}\right)\left(1+\lambda^2\right)\chi_{0}^{4}{\rm d}u{\rm d}v+x^{2}{\rm d}\Omega^{2}.
\end{eqnarray}
Using the Kruskal coordinates defined in (\ref{Eq: Krsukal coordinates for flat space}), the line element becomes
\begin{eqnarray}\label{Eq: Kruskal metric for flat space}
    {\rm d}s^{2}=-\left(1-\frac{2M}{x}\right)\left(1+\lambda^2\right)\chi_{0}^{2}{\rm exp}\left(-\frac{2\kappa}{\chi_{0}}x_{*}\right){\rm d}U{\rm d}V+x^{2}(x_{*}(u,v)){\rm d}\Omega^{2}.
\end{eqnarray}
The global factor $\chi_0 = 1/\sqrt{1 + \lambda_\infty^2}$, appearing in the conformal factor of the nonangular part of the metric ensures the preservation of asymptotic flatness as a consequence of the construction.

\subsection{Scalar field coupling}\label{Appendix Scalar Field Coupling}
In the case of minimal coupling, the Klein--Gordon equation (\ref{Eq: General KG in radial coordinates}) couples to the metric (\ref{Eq: line-element with Euclidean space}) to gives the scalar field dynamics
\begin{eqnarray}
    0&=&-\frac{1+\lambda_{\infty}^{2}}{1+\lambda^{2}}\left(1-\frac{2M}{x}\right)^{-1}\ddot{\phi}_{lm}+\left(1-\frac{2M\lambda^{2}}{\left(1+\lambda^2\right)x}\right)\left(1-\frac{2M}{x}\right)\partial_{x}^{2}\phi_{lm}\nonumber \\ &&+\left(\frac{2\left(x-M\right)}{x^2\left(1+\lambda^2\right)}+\lambda^2\frac{\left(x-2M\right)\left(M-2x\right)}{x^{3}\left(1+\lambda^2\right)}+\lambda\lambda'\frac{\left(x-2M\right)^2}{x^2\left(1+\lambda^2\right)}\right)\partial_{x}\phi_{lm}\,.
\end{eqnarray}
Written in the tortoise coordinate (\ref{Eq: Radial EMG tortoise coordinates}), the expression of Eq~.(\ref{Eq: General KG in tortoise coordinates}) reduces to 
\begin{eqnarray}
    \left[-\partial_{t}^{2}+\partial^{2}_{x_{*}}-V_{l}(x)\right]\Psi_{lm}(t,x)=0
\end{eqnarray}
and the potential is given by
\begin{eqnarray}\label{Eq: Potential with Euclidean space}
    \bar{V}_{l}(x)=\left(1-\frac{2M}{x}\right)\chi_{0}^{2}\left[\frac{2M}{x^3}+\frac{l(l+1)}{x^2}+\lambda^{2}\left(\frac{3M}{x^3}-\frac{6M}{x^4}+\frac{l(l+1)}{x^2}+\frac{\lambda\lambda'}{x}\left(1-\frac{2M}{x}\right)^{2}\right)\right] \nonumber \\
\end{eqnarray}
which again vanishes on the horizon and asymptotically. 

In the case of $\chi=1/\sqrt{1+\lambda^{2}}$, the equation of motion (\ref{Eq: Generic NMC Coupled scalar field}) gives
\begin{eqnarray}
   0&=& -\beta^{-1}\left(1-\frac{2M}{x}\right)^{-1}\ddot{\phi}_{lm}+\frac{\chi}{x^{2}}\partial_{x}\left[\frac{x^{2}}{\chi}\left(1-\frac{2M}{x}\right)\phi_{lm}'\right]-\frac{l(l+1)}{x^{2}}\phi_{lm}+\left(1-\frac{2M}{x}\right)\phi_{lm}'\left(\ln\chi\right)'\nonumber \\
   &=&-\beta^{-1}\left(1-\frac{2M}{x}\right)^{-1}\ddot{\phi}_{lm}+\frac{2}{x}\left(1-\frac{M}{x}\right)\partial_{x}\phi_{lm}+\left(1-\frac{2M}{x}\right)\partial_{x}^{2}\phi_{lm}-\frac{l(l+1)}{x^{2}}\phi_{lm}\nonumber \\
\end{eqnarray}
which recovers the nonminimally coupled Klein--Gordon equation as in (\ref{Eq: Reduce nonminimal KG-equation in radial coordinates 1}). Hence, unlike in the minimally coupled case, the nonconstant $\chi$ gives the same dynamics as $\chi=\chi_{0}$. 

\subsection{Euclidean space in zero mass limit}\label{Appendix: Covariant tensor components}
Recall that the 2d line-element for the case of $\chi=1/\sqrt{1+\lambda^{2}(x)}$ is given by
\begin{eqnarray}\label{Eq: Conformal metric Euclidean}
    {\rm d}s^{2}_{(2)}=-\left(1-\frac{2M}{x}\right)\left(1+\lambda^{2}\right)\chi_{0}^{4}{\rm d}u{\rm d}v\equiv -C_{1}(u,v){\rm d}u{\rm d}v\,.
\end{eqnarray}
The vacuum polarization associated to the (\ref{Eq: Conformal metric Euclidean}) equals
\begin{eqnarray}\label{Eq: EMG on diagonal stress tensor Euclidean}
    \braket{0_{\rm B}\left|T^{(1)}_{uu}\right|0_{\rm B}}&=&\braket{0_{\rm B}\left|T^{(1)}_{vv}\right|0_{\rm B}}\nonumber \\
    &=&-\frac{\chi_{0}^{2}}{48\pi x^{5}}\left[2Mx^{2}\left(1-\frac{3M}{2x}\right)+2M\lambda^{6}x^{2}\left(1-\frac{2M}{x}\right)^{2}-x^{5}\left(1-\frac{2M}{x}\right)^{2}\left(\lambda'\right)^2\right.\nonumber \\ &&\left.+\lambda^{4}\left(M\left(16M^2-19Mx+6x^2\right)-\left(2M-x\right)^{3}\left(x\lambda'\right)^{2}\right)\right. \nonumber \\ &&\left. +2\lambda^{2}M\left(\left(4M-3x\right)\left(M-x\right)+2x^{2}\left(x-2M\right)^{2}\left(\lambda'\right)^{2}\right)-x\left(x-2M\right)^{2}\lambda\left(M\lambda'+x^{2}\lambda''\right)\right.\nonumber \\
    &&\left. +\lambda^{3}x\left(x-2M\right)^{2}\left(-3M\lambda'+2\left(M-x\right)x\lambda''\right)\right.\nonumber \\ &&\left.-x\left(x-2M\right)^{2}\lambda^{5}\left(2M\lambda'+x\left(x-2M\right)\lambda''\right)\right]
\end{eqnarray}
for the diagonal components, while for the off-diagonal component,
\begin{eqnarray}\label{Eq: EMG off diagonal stress tensor Euclidean}
    \braket{0_{\rm B}\left|T^{(1)}_{uv}\right|0_{\rm B}}&=&\frac{\chi_{0}^{2}\left(x-2M\right)}{48\pi x^{5}\left(1+\lambda^{2}\right)}\left[-2Mx+2M\lambda^{4}\left(5M-3x\right)+M\lambda^{6}\left(5M-2x\right)+x^{3}\left(\lambda'\right)^{2}\left(x-2M\right)\right.\nonumber \\ &&\left. +\lambda^{2}\left(M\left(5M-6x\right)+\left(x\lambda'\right)^{2}\left(x-4M\right)\left(x-2M\right)\right)\right. \nonumber \\ &&\left.+x\lambda^{5}\left(2M-x\right)\left(-4M\lambda'+\left(2M-x\right)x\lambda''\right)\right.\nonumber \\
    &&\left.+\lambda \left(Mx\left(3x-2M\right)\lambda'+x^{3}\left(x-2M\right)\lambda''\right)\right. \nonumber \\ &&\left. +x\lambda^{3}\left(M\left(7x-10M\right)\lambda'+2x\left(x-2M\right)\left(x-M\right)\lambda''\right)
    \right]\,.
\end{eqnarray}
Both Eqs.~(\ref{Eq: EMG off diagonal stress tensor Euclidean}) and (\ref{Eq: EMG on diagonal stress tensor Euclidean}) reduce to the classical expressions in the limit when $\chi_{0}\to1$ and $\lambda\to 0$. They also approach zero in the asymptotic limit for either constant $\lambda$ or monotonically decreasing $\lambda$. Thus, the vacuum state $\ket{0_{\rm B}}$ serves as the good Minkowski state at infinity. 

Unlike the classical counterpart, the covariant stress tensor for this scheme survives in the zero mass limit:
\begin{eqnarray}
    \lim_{M\to 0}\braket{0_{\rm B}\left|T^{(1)}_{uu}\right|0_{\rm B}}&=&\lim_{M\to 0}\braket{0_{\rm B}\left|T^{(1)}_{uu}\right|0_{\rm B}} \nonumber \\
    &=&\frac{\chi_{0}^{2}}{48\pi \left(1+\lambda^{2}\right)}\left(\left(\lambda'\right)^{2}\left(1-\lambda^{2}\right)+\lambda\lambda''\left(1+\lambda^{2}\right)\right)
\end{eqnarray}
and 
\begin{eqnarray}
    \lim_{M\to 0}\braket{0_{\rm B}\left|T^{(1)}_{uv}\right|0_{\rm B}}=\frac{\chi_{0}^{2}}{48\pi \left(1+\lambda^{2}\right)}\left(\left(\lambda'\right)^{2}+\lambda\lambda''\left(1+\lambda^{2}\right)\right).
\end{eqnarray}
They trivially vanish when $\lambda$ is constant because the line-element
(\ref{Eq: Conformal metric Euclidean}) reduces to Minkowski spacetime for
this particular choice in the zero mass limit. Due to our definition (\ref{Eq:
  Null coordinates definition for arbitrary mass}), we have $\left(u,v\right)\to \chi^{-1}_{0} \left(u_{(c)},v_{(c)}\right)$ in this limit (for constant $\lambda$), where $\left(u_{(c)},v_{(c)}\right)$ are the Minkowski null coordinates. Note also that these contributions vanish asymptotically when $\lambda$ decreases monotonically.

The covariant components of the stress-energy tensor are nonvanishing for
arbitrary $\lambda(x)$ in the zero mass limit because the line element is conformally flat with nontrivial conformal factor, 
\begin{eqnarray}
    \lim_{M\to 0}{\rm d}s^{2}=-\left(1+\lambda^{2}\right)\chi_{0}^{4}{\rm d}u{\rm d}v=-\left(1+\lambda^{2}\right)\chi_{0}^{2}{\rm d}u_{(c)}{\rm d}v_{(c)}
\end{eqnarray}
The nontriviality of the vacuum polarization arises from its construction: it
is computed using local inertial coordinates near the horizon and extended to
the full spacetime using the covariance of the stress-energy tensor. In EMG,
the subtlety lies in the fact that while the near-horizon geometry resembles
flat Minkowski spacetime, the spacetime does not become Minkowski in the $M
\to 0$ limit. This leads to residual artifacts in the covariant components of
the stress-energy tensor when considering the limit in which no black hole is present. In particular, the tensor components diverge when $x\to0$. 
To eliminate these artifacts and divergences, the tensor is redefined as
\begin{eqnarray}
    \braket{0_{\rm B}\left|\tilde{T}_{uu}^{(1)}\right|0_{\rm B}}&=&\braket{0_{\rm B}\left|T_{uu}^{(1)}\right|0_{\rm B}}-\lim_{M\to 0}\braket{0_{\rm B}\left|T^{(1)}_{uu}\right|0_{\rm B}}\label{Eq: Bare uu tensor} \nonumber \\
    \braket{0_{\rm B}\left|\tilde{T}_{vv}^{(1)}\right|0_{\rm B}}&=&\braket{0_{\rm B}\left|T_{vv}^{(1)}\right|0_{\rm B}}-\lim_{M\to 0}\braket{0_{\rm B}\left|T^{(1)}_{vv}\right|0_{\rm B}}\label{Eq: Bare vv tensor} \nonumber \\
    \braket{0_{\rm B}\left|\tilde{T}_{uv}^{(1)}\right|0_{\rm B}}&=&\braket{0_{\rm B}\left|T_{uv}^{(1)}\right|0_{\rm B}}-\lim_{M\to 0}\braket{0_{\rm B}\left|T^{(1)}_{uv}\right|0_{\rm B}}\label{Eq: Bare uv tensor}
\end{eqnarray}
such that the components vanish in the absence of mass. The renormalized tensor remains covariant, as it is the difference of covariant tensors.

In the case of the Hartle--Hawking vacuum, the normal-order contributions
(\ref{Eq: HH normal ordering})--(\ref{Eq: HH normal ordering 3}) are still
valid here, because the global Kruskal coordinates have a similar relation with $\left(u,v\right)$ up to a constant $\chi_{0}$. Therefore the renormalized covariant stress-energy tensor in this vacuum state is 
\begin{eqnarray}
    \braket{0_{\rm H}|\tilde{T}^{(1)}_{uu}|0_{\rm H}}&=& \braket{0_{\rm B}|\tilde{T}^{(1)}_{uu}|0_{\rm B}}+\frac{\kappa^{2}}{48\pi} \\
     \braket{0_{\rm H}|\tilde{T}^{(1)}_{vv}|0_{\rm H}}&=& \braket{0_{\rm B}|\tilde{T}^{(1)}_{vv}|0_{\rm B}}+\frac{\kappa^{2}}{48\pi} \\
      \braket{0_{\rm H}|\tilde{T}^{(1)}_{uv}|0_{\rm H}}&=& \braket{0_{\rm B}|\tilde{T}^{(1)}_{uv}|0_{\rm B}}.
\end{eqnarray}
For constant and monotonically decreasing holonomy function $\lambda(x)$, the flux of incoming and outgoing energy at infinity is constant
\begin{eqnarray}
    \lim_{x\to \infty}\braket{0_{\rm H}|T^{(1)}_{uu}|0_{\rm H}}=\lim_{x\to \infty}\braket{0_{\rm H}|T^{(1)}_{vv}|0_{\rm H}}=\frac{\kappa^{2}}{48\pi}
\end{eqnarray}
which means that this state corresponds to a black hole in thermal equilibrium with the environment.

Describing the collapse, the classical normal ordered stress-energy tensor is
(\ref{Eq: Normal ordering uu in-state})--(\ref{Eq: Normal ordering vv
  in-state}). Thus, the correction to the covariant stress-energy  tensor comes purely from the renormalized vacuum polarization contribution (\ref{Eq: Bare uu tensor})--(\ref{Eq: Bare uv tensor}),
\begin{eqnarray}
    \braket{0_{\rm in}|\tilde{T}^{(1)}_{uu}|0_{\rm in}}&=&\braket{0_{\rm B}|\tilde{T}^{(1)}_{uu}|0_{\rm B}}+\frac{\kappa^{2}}{48\pi} \\
    \braket{0_{\rm in}|\tilde{T}^{(1)}_{vv}|0_{\rm in}}&=&\braket{0_{\rm B}|\tilde{T}^{(1)}_{vv}|0_{\rm B}}\\
    \braket{0_{\rm in}|\tilde{T}^{(1)}_{uv}|0_{\rm in}}&=&\braket{0_{\rm B}|\tilde{T}^{(1)}_{uv}|0_{\rm B}}\,.
\end{eqnarray}
Here it is interesting to see that for an observer at $\mathcal{J}^{+}$, the flux of energy is
\begin{eqnarray}
    \lim_{x\to\infty}\braket{0_{\rm in}|\tilde{T}_{uu}^{(1)}|0_{\rm in}} &\to& \frac{\kappa^{2}}{48\pi} \\
    \lim_{x\to\infty}\braket{0_{\rm in}|\tilde{T}_{uu}^{(1)}|0_{\rm in}}&\to& 0\,.
\end{eqnarray}
At the horizon, the incoming flux is given by
\begin{eqnarray}
    \lim_{x\to 2M}\braket{0_{\rm in}|T_{vv}|0_{\rm in}}&=&-\frac{\chi_{0}^{2}}{768\pi M^{2}\left(1+\lambda_{\rm H}^{2}\right)}\left(1+16M^{2}\left(\lambda_{\rm H}'\right)^{2}+\lambda_{\rm H}^{2}\left(1-16M^{2}\left(\lambda_{\rm H}'\right)^{2}\right)\right.\nonumber \\ &&\left.+16M^{2}\lambda_{\rm H}\lambda_{\rm H}''\left(1+\lambda_{\rm H}^{2}\right)\right)
\end{eqnarray}
which gives a negative flux, while the outgoing flux is
\begin{eqnarray}
    \lim_{x\to 2M}\braket{0_{\rm in}|T_{uu}|0_{\rm in}}&=&-\frac{\chi_{0}^{2}}{48\pi\left(1+\lambda^{2}_{\rm H}\right)}\left(\left(\lambda_{\rm H}'\right)^{2}\left(1-\lambda_{\rm H}^{2}\right)+\lambda_{\rm H}\lambda_{\rm H}''\left(1+\lambda_{\rm H}^{2}\right)\right)\\
\end{eqnarray}
and is nonvanishing, unlike its classical counterpart. The net flux is given by 
\begin{eqnarray}
    \Delta =  \lim_{x\to 2M}\braket{0_{\rm in}|T_{vv}|0_{\rm in}}-\lim_{x\to 2M}\braket{0_{\rm in}|T_{uu}|0_{\rm in}}=-\frac{\kappa^{2}}{48\pi}
\end{eqnarray}
which match the amount of energy reaching $\mathcal{J}^{+}$.

\section{CFT approach to particle creation}\label{Appendix B}
In \cite{Fabbri-Salas-Olmo}, the authors proposed a method using conformal field techniques to derive the Hawking distribution.  The near horizon expansion of the Schwarzschild metric can be cast into the Rindler form
\begin{eqnarray}\label{Eq: near-horizon expansion}
    {\rm d}s^{2}\big|_{\rm NH}=-\left(\kappa \rho\right){\rm d}t^{2}+{\rm d}\rho^{2}+\left(2M\right)^2{\rm d}\Omega^{2}\,,
\end{eqnarray}
where $x=2M\left(1+\kappa^{2}\rho^{2}\right)$, with the surface gravity
$\kappa=1/4M$.  A similar form exists for the emergent
line-element (\ref{Eq: Effective line-element in the full case}), the only
difference being that the surface gravity is then $\kappa=\chi_{0}/4M$. 

The transformation from Rindler form to a conformally flat line elemny is achieved by introducing a new radial coordinate $\zeta$ by
\begin{eqnarray}
    {\rm d}\rho=\kappa\rho{\rm d}\zeta\quad\mbox{such that} \quad \rho= \kappa^{-1}e^{\kappa \zeta}\,.
\end{eqnarray}
The nonangular part of the line element (\ref{Eq: near-horizon expansion}) transforms to 
\begin{eqnarray}\label{Eq: Near horizon conformal metric}
    {\rm d}s_{(2)}^{2}=e^{2\kappa\zeta}\left(-{\rm d}t^2+{\rm d}\zeta^{2}\right)\,.
\end{eqnarray}
The near horizon radial coordinates $\zeta$ is the leading order of the radial tortoise coordinate near the horizon:
\begin{eqnarray}
    {\rm d}x_{*}
    &\approx&4M\frac{{\rm d}\rho}{\rho}\left(1+\lambda_{\rm H}^{2}\frac{\kappa^2 \rho^2}{2M}\right)^{-1} \nonumber \\
    &\approx&\frac{{\rm d}\rho}{\kappa \rho}={\rm d}\zeta
\end{eqnarray}
which allows us later to substitute $x_{*}\to \zeta$ in the near horizon scalar field equation. 

As proposed by the authors of \cite{Fabbri-Salas-Olmo}, any 2-dimensional spacetimes related by a conformal transformation that is not of the M\"obius type in general have different notions of the vacuum state.
Near the horizon, both the minimally coupled, (\ref{Eq: General KG in tortoise coordinates}), and nonminimally coupled scalar field, (\ref{Eq: nonminimally coupled reduced scalar field}), reduce to
\begin{eqnarray}\label{Eq: Reduced CFT minimally coupled scalar dynamics}
    e^{2\kappa\zeta}\left(-\partial_{t}^{2}+\partial_{\zeta}^{2}\right)\Psi_{lm}(t,\zeta)=0
\end{eqnarray}
which means that the dynamics of the reduced 2-dimensional field $\Psi_{lm}$ can be deduce from the 2-dimensional CFT as
\begin{eqnarray}\label{Eq: Reduced CFT scalar action}
    S^{(2d)}_{\Psi}=\frac{1}{2}\int{\rm d}t{\rm d}\eta \sqrt{g^{(2)}}g^{\alpha\beta}_{(2)}\nabla_{\alpha}\Psi_{lm}\nabla_{\beta}\Psi_{lm}
\end{eqnarray}
We introduce null-coordinates $u=t-\eta$ and $v=t+\eta$, which is  convenient for massless scalar fields. Consequently, the dynamics (\ref{Eq: Reduced CFT minimally coupled scalar dynamics}) reduces to a simpler form 
\begin{eqnarray}\label{Eq: CFT in null coordinates}
    \partial_{u}\partial_{v}u_{lm}=0
\end{eqnarray}
which means the right-moving and left-moving modes are linearly independent. The late time solution can be decomposed into modes that reach $\mathcal{H}^{+}$ and $\mathcal{I}^{+}$
\begin{eqnarray}\label{Eq: Right and Left Modes}
    \Psi_{lm}(u,v)&=&\Psi_{\rm R}(u)+\Psi_{\rm L}(v)\nonumber \\
    &=&\int_{0}^{\infty} {\rm d}\omega \left[a_{\omega}g_{lm\omega}(u)+a^{\dagger}g_{lm\omega}^{*}(u)+b_{\omega}h_{lm\omega}(v)+b_{\omega}^{\dagger}h^{*}_{lm\omega}(v)\right] 
\end{eqnarray}
where the outgoing (right moving) modes $\Psi_{\rm R}$ are the modes that reach $\mathcal{I}^{+}$, with the inner product given by (\ref{Eq: Orthonormality condition of g}).
It is important to note here that the dynamics (\ref{Eq: CFT in null coordinates}) is invariant under a conformal transformation 
$(u,v)\rightarrow (\tilde{u}(u),\tilde{v}(v))$,
\begin{eqnarray}
    e^{2\kappa(v-u)}{\rm du}{\rm d}v\rightarrow  e^{2\kappa(v(\tilde{v})-u(\tilde{u}))}\frac{{\rm d}u}{{\rm d}\tilde{u}}\frac{{\rm d}v}{{\rm d}\tilde{v}}{\rm d}\tilde{u}{\rm d}\tilde{v}
\end{eqnarray}

An important ingredient in this approach is the normal ordering of the stress-energy tensor. The reduced stress-energy tensor can be derived from the reduced action (\ref{Eq: Reduced CFT scalar action}) as \( T_{\alpha\beta} \sim \delta S/(\sqrt{-g^{(2)}} \delta g^{\alpha\beta}_{(2)}) \), yielding:
\begin{eqnarray}\label{Eq: Momentum energy tensor}
    T_{\alpha\beta} = \partial_{\alpha}u_{lm}\partial_{\beta}u_{lm} - \frac{1}{2}g_{\alpha\beta}\partial_{\mu}u_{lm}\partial^{\mu}u_{lm}\,.
\end{eqnarray}
In null coordinates, the components are:
\begin{eqnarray}\label{Eq: momentum-energy tensor components}
    T_{uu} &=& \partial_{u}\phi_{l}\partial_{u}\phi_{l}, \nonumber \\
T_{vv} &=& \partial_{v}\phi_{l}\partial_{v}\phi_{l}, \nonumber \\
T_{uv} &=& T_{vu} = 0\,.
\end{eqnarray}
For simplicity in later calculations, we redefine the notation for the right- and left-moving null coordinates as:
\begin{eqnarray}
    u&:=&\eta_{+}=t-\eta \nonumber \\
     v&:=&\eta_{-}=t+\eta
\end{eqnarray}

The operators for components of the stress-energy tensor are thus quadratic, leading to potential UV divergences if not properly addressed. These divergences can be mitigated by renormalizing the stress-energy tensor using the point-splitting method. Along with the normal ordering operator, the normalized momentum energy tensor reads
\begin{eqnarray}\label{Eq: normal ordering momentum energy tensor}
    :T_{\pm\pm}(\eta^{\pm}):=\lim_{\eta^{\pm'}\rightarrow\eta^{\pm}}\left[\partial_{\pm}\Psi_{lm}(\eta^{\pm})\partial_{\pm'}\Psi_{lm}(\eta^{\pm'})+\frac{\hbar}{4\pi}\frac{1}{\left(\eta^{\pm'}-\eta^{\pm}\right)^2}\right]
\end{eqnarray}
Let us define null coordinates \( y^{\pm} \), which are conformally related to the original coordinates \( \eta^{\pm} \) through the relation \( \eta^{\pm} = \eta^{\pm}(y^{\pm}) \). In these new coordinates, the field can be expanded as
\begin{eqnarray}\label{Eq: Scalar field expansion to right and left modes}
   \phi_{l} (y^{\pm})&=& \int \frac{{\rm d}\Omega}{\sqrt{4\pi \Omega}} \left[ c_{\Omega} e^{-i\Omega y^{-}} +  c^{\dagger}_{\Omega} e^{i\Omega y^{-}} + d_{\Omega} e^{-i\Omega y^{+}} +  d^{\dagger}_{\Omega} e^{i\Omega y^{+}} \right],  \nonumber \\
   &=&\phi_{\rm R}(y^{\pm})+\phi_{\rm L}(y^{\pm})\,.
\end{eqnarray}
 The quantum normal ordered stress-energy tensor in this coordinate system
 therefore equals
\begin{eqnarray}\label{Normal ordered correlator}
    :T_{\pm\pm}(y^{\pm}):&=&\lim_{y^{\pm'}\rightarrow y^{\pm}}\left[T_{\pm\pm}(y^{\pm},y^{\pm'})+\frac{\hbar}{4\pi}\frac{1}{\left(y^{\pm}-y^{\pm'}\right)^{2}}\right] \nonumber \\
    &=&\lim_{y^{\pm'}\rightarrow y^{\pm}}\left[\frac{{\rm d}\zeta^{\pm}(y^{\pm})}{{\rm d}y^{\pm}}\frac{{\rm d}\zeta^{\pm}(y^{\pm'})}{{\rm d}y^{\pm'}}\partial_{\pm}\phi_{l}(y^{\pm})\partial_{\pm'}\phi_{l}(y^{\pm'})+\frac{\hbar}{4\pi}\frac{1}{\left(y^{\pm'}-y^{\pm}\right)^2}\right]\,.
\end{eqnarray}
The conformal transformation $\zeta^{\pm}=\zeta^{\pm}(y^{\pm})$ is
continuous. Therefore, $y^{\pm'}\rightarrow y^{\pm}$ implies
$\zeta^{\pm'}\rightarrow \zeta^{\pm}$ and thus 
\begin{eqnarray}\label{Eq: Virasoro anomaly}
   :T_{\pm\pm}(y^{\pm}):&=&\left(\frac{{\rm d}\zeta^{\pm}}{{\rm d}y^{\pm}}\right)^{2}:T_{\pm\pm}(\zeta^{\pm}):-\frac{\hbar}{4\pi}\lim_{y^{\pm'}\rightarrow y^{\pm}}\left[\frac{\frac{{\rm d}\zeta^{\pm}(y^{\pm})}{{\rm d}y^{\pm}}\frac{{\rm d}\zeta^{\pm}(y^{\pm'})}{{\rm d}y^{\pm'}}}{\left(\zeta^{\pm'}-\zeta^{\pm}\right)^2}-\frac{1}{\left(y^{\pm'}-y^{\pm}\right)^2}\right]\nonumber \\
   &=&\left(\frac{{\rm d}\zeta^{\pm}}{{\rm d}y^{\pm}}\right)^{2}:T_{\pm\pm}(\zeta^{\pm}):-\frac{\hbar}{24\pi}\left(\frac{{\rm d}^{3}\zeta^{\pm}/{\rm d}(y^{\pm})^3}{{\rm d}\zeta^{\pm}/{\rm d}y^{\pm}}-\frac{3}{2}\left(\frac{{\rm d}^{2}\zeta^{\pm}/{\rm d}(y^{\pm})^2}{{\rm d}\zeta^{\pm}/{\rm d}y^{\pm}}\right)^{2}\right)
\end{eqnarray}
The terms in parentheses are the well-known (quantum) \textit{Virasoro anomaly} contributions. 

Eq.~(\ref{Eq: Virasoro anomaly}) is the main ingredient in this approach. We will now follow the discussion of \cite{Fabbri-Salas-Olmo} in deriving the particle number distribution through the above transformation. For simplicity, let us restrict our attention to the right-moving modes
\begin{eqnarray}
    \phi_{R}(y^{-})=\int \frac{{\rm d}\Omega}{\sqrt{4\pi \Omega}}\left[c_{\Omega}e^{-i\Omega y^{-}}+c^{\dagger}_{\Omega}e^{i\Omega y^{-}}\right]
\end{eqnarray}
Wwe compute the $\partial_{-}\phi_{\rm R}(y^{-})$-correlator in the $\ket{0_{\zeta}}$-vacuum:
\begin{eqnarray}
    &&\braket{0_{\zeta}\left|:\partial_{-}\phi_{\rm R}(y^{-})\partial_{-'}\phi_{\rm R}(y^{-'}):\right|0_{\zeta}}=\int \frac{{\rm d}\Omega {\rm d}\tilde{\Omega}}{4\pi\sqrt{\Omega\tilde{\Omega}}}\nonumber \\
    &&\times\braket{0_{\zeta}|\left(-i\Omega c_{\Omega}e^{-i\Omega y^{-}}+i\Omega c_{\Omega}^{\dagger}e^{i\Omega y^{-}}\right)    
    \left(-i\tilde{\Omega} c_{\tilde{\Omega}}e^{-i\tilde{\Omega} y^{-'}}+i\tilde{\Omega}c_{\tilde{\Omega}}^{\dagger}e^{i\tilde{\Omega}y^{-'}}\right)|0_{\zeta}}\,.
\end{eqnarray}
In order to isolate the spacetime integration, such that we can investigate
the short-distance physics, we multiplied the right-hand side with
$e^{-i\left(wy^{-}-\tilde{w}y^{-'}\right)}$ which makes all the terms vanish,
except for $c_{\Omega}^{\dagger}c_{\Omega}$. Hence,
\begin{eqnarray}
    \int_{-\infty}^{\infty}{\rm d}y^{-}{\rm d}y^{-'}\frac{e^{-i\left(wy^{-}-\tilde{w}y^{-'}\right)}}{4\pi\sqrt{w\tilde{w}}}  &&\braket{0_{\zeta}\left|:\partial_{-}\phi_{\rm R}(y^{-})\partial_{-'}\phi_{\rm R}(y^{-'}):\right|0_{\zeta}}=\int\frac{{\rm d}\Omega {\rm d}\tilde{\Omega}}{16\pi^2}\sqrt{\frac{\Omega\tilde{\Omega}}{w\tilde{w}}}\nonumber \\
    &&\left(\int_{-\infty}^{\infty}{\rm d}y^{-}e^{i\left(\Omega-w\right)y^{-}}\right)\left(\int_{-\infty}^{\infty}{\rm d}y^{-'}e^{i\left(-\tilde{\Omega}+\tilde{w}\right)y^{-'}}\right)\braket{0_{\zeta}\left|c_{\Omega}^{\dagger}c_{\tilde{\Omega}}\right|0_{\zeta}}\nonumber \\
    &=&\frac{1}{4}\braket{0_{\zeta}\left|c_{w}^{\dagger}c_{\tilde{w}}\right|0_{\zeta}}\,.
\end{eqnarray}
The right-hand side implies the particle number operator by setting $w=\tilde{w}$:
\begin{eqnarray}\label{Eq: General particle production}
\braket{0_{\zeta}\left|N_{\omega}\right|0_{\zeta}}&=&\frac{1}{\pi\hbar}\int_{-\infty}^{\infty}\int_{-\infty}^{\infty}{\rm d}y^{-}{\rm d}y^{-'}e^{-i\omega\left(y^{-}-y^{-'}\right)}\braket{0_{\zeta}|:\partial_{-}\phi_{R}(y^{-})\partial_{-'}\phi_{R}(y^{-'}):|0_{\zeta}}\nonumber \\
&=&-\frac{1}{\pi^{2}}\int_{-\infty}^{\infty}\int_{-\infty}^{\infty}{\rm d}\tilde{u}{\rm d}\tilde{u}'e^{-i\omega\left(\tilde{u}-\tilde{u}'\right)}\left[\frac{{\rm d}\zeta^{+}(\tilde{u})}{{\rm d}\tilde{u}}\frac{{\rm d}\zeta^{+}(\tilde{u}')}{{\rm d}\tilde{u}'}\frac{1}{\left(\zeta^{+}(\tilde{u})-\zeta^{+}(\tilde{u}')\right)^{2}}-\frac{1}{\left(\tilde{u}-\tilde{u}'\right)^{2}}\right]\nonumber \\
\end{eqnarray}
where $\zeta^{+}$ corresponds to outgoing null-coordinates in the new coordinate system. The relation indicates that for any linear (Poincar\'e) or M\"obius transformation, particle creation is inevitable. As we have discussed in the Hawking set-up, the late-time coordinate $u$ can be written in terms of the early-time coordinate $v$ through Eq.~(\ref{Eq: Typical logarithmic relation}), and hence by substituting $\tilde{u}=v$ in the above relation one will end up with
\begin{eqnarray}
\braket{0_{\eta}\left|N_{\omega}\right|0_{\eta}} =\frac{1}{e^{2\pi \omega/\kappa}-1}\,,
\end{eqnarray}
which is consistent with the rest of the method.

\end{document}